\documentclass[]{aa}

\usepackage{graphicx}
\usepackage{txfonts}
\usepackage{natbib}
\usepackage{graphicx}
\usepackage{txfonts}
\usepackage{hyperref}
\usepackage{booktabs}
\usepackage{amsmath}
\usepackage{amssymb}
\usepackage{mathtools}
\usepackage{txfonts}
\usepackage{makecell}
\usepackage{indentfirst}
\usepackage{longtable}
\usepackage[table, dvipsnames]{xcolor}
\usepackage{nicematrix}
\NiceMatrixOptions{cell-space-top-limit=2pt,cell-space-bottom-limit=1.5pt}

\usepackage[normalem]{ulem}
\usepackage{multirow,booktabs}
\usepackage[flushleft]{threeparttable} 
\usepackage{caption}

\newcommand\gaia{\textit{Gaia}\xspace}

\newcommand\gdrtwo{\gaia DR2\xspace}
\newcommand\gdrthree{\gaia DR3\xspace}

\newcommand{\allcand}{358\xspace}
\newcommand{\bestcand}{67\xspace}
\newcommand{\besttaxocand}{156\xspace}

\definecolor{mulberry}{rgb}{0.77, 0.29, 0.55}

\makeatletter
\renewcommand*\aa@pageof{, page \thepage{} of \pageref*{LastPage}}
\makeatother

\setcitestyle{citesep={,}}

\newcommand{\orcit}[1]{\protect\href{https://orcid.org/#1}{\protect\includegraphics[width=8pt]{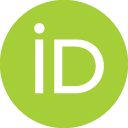}}}

\makeatletter
\renewcommand*\maketitle{%
  \thispagestyle{firstpage}
\begingroup
    \if@wideboxfn
    \setlength\bibindent{1.4\parindent}
    \else
    \setlength\bibindent{\parindent}
    \fi
    \renewcommand*\thefootnote{\@fnsymbol\c@footnote}%
    \renewcommand\@makefntext[1]{%
    \ifaa@longfn\hsize\textwidth\fi
    \noindent
    \hb@xt@\bibindent{\hss\@makefnmark\enspace}##1}
  \ifaa@twocolumn
  \begingroup
    \begin{aa@strip}
          \aa@maketitle
    \end{aa@strip}
    \@thanks            
  \endgroup
  \else
    \begingroup
      \let\thanks\footnote
      \aa@maketitle
    \endgroup
  \fi
\endgroup
  \setcounter{footnote}{0}%
}
\makeatother

\begin{document}
\title{Binary asteroid candidates in \gdrthree astrometry\thanks{Full Table~\ref{tab:all_candidates} will be available at the CDS and at \url{https://lagrange.oca.eu/fr/gaiamoons}}}

\author{
L. Liberato     \orcit{0000-0003-3433-6269}\inst{\ref{inst:0001},\ref{inst:0002}}
\and
P. Tanga        \orcit{0000-0002-2718-997X}\inst{\ref{inst:0001}}
\and
D. Mary         \orcit{0000-0002-9047-5768}\inst{\ref{inst:0001}}
\and
K. Minker      \orcit{0009-0002-6435-9453}\inst{\ref{inst:0001}}
\and
B. Carry        \orcit{0000-0001-5242-3089}\inst{\ref{inst:0001}}
\and
F. Spoto        \orcit{0000-0001-7319-5847}\inst{\ref{inst:0003}}
\and
P. Bartczak     \orcit{0000-0002-3466-3190}\inst{\ref{inst:0005},\ref{inst:0008}}
\and
B. Sicardy      \orcit{0000-0003-1995-0842}\inst{\ref{inst:0004}}
\and
D. Oszkiewicz   \orcit{0000-0002-5356-6433}\inst{\ref{inst:0005}}
\and
J. Desmars      \orcit{0000-0002-2193-8204}\inst{\ref{inst:0006},\ref{inst:0007}}
}

\institute{ 
Université Côte d’Azur, Observatoire de la Côte d’Azur, CNRS, Laboratoire Lagrange, Bd de l’Observatoire, CS 34229, 06304 Nice Cedex 4, France\relax 
\label{inst:0001}
\and São Paulo State University, Grupo de Dinâmica Orbital e Planetologia, CEP 12516-410, Guaratinguetá, SP, Brazil\relax 
\label{inst:0002}
\and Harvard-Smithsonian Center for Astrophysics, 60 Garden St., MS 15, Cambridge, MA 02138, USA\relax 
\label{inst:0003}
\and LESIA, Paris Observatory, PSL University, CNRS, Sorbonne University, Univ. Paris Diderot, Sorbonne Paris Cité, 5 place Jules Janssen, 92195 Meudon, France\relax
\label{inst:0004}
\and Astronomical Observatory Institute, Faculty of Physics, Adam Mickiewicz University, Słoneczna 36, 60-286 Poznań, Poland\relax
\label{inst:0005}
\and Polytechnic Institute of Advanced Sciences-IPSA, 63 Boulevard de
Brandebourg, 94200 Ivry-sur-Seine, France\relax
\label{inst:0006}
\and Institute of Celestial Mechanics and Ephemeris Calculation (IMCCE), Paris Observatory, PSL Research University, CNRS, Sorbonne University, UPMC Univ Paris 06, Univ. Lille, 77, Av. Denfert-Rochereau, 75014 Paris, France\relax
\label{inst:0007}
\and Instituto Universitario de Física Aplicada a las Ciencias y las Tecnologías (IUFACyT). Universidad de Alicante, Ctra. San Vicente del Raspeig s/n. 03690 San Vicente del Raspeig, Alicante, Spain\relax
\label{inst:0008}}

\date{Accepted}

\abstract
{Asteroids with companions constitute an excellent sample for studying the collisional and dynamical evolution of minor planets. The currently known binary population were discovered by different complementary techniques that produce, for the moment, a strongly biased distribution, especially in a range of intermediate asteroid sizes ($\approx$ 20 to 100 km) where both mutual photometric events and high-resolution adaptive optic imaging are poorly efficient.}
{A totally independent technique of binary asteroid discovery, based on astrometry, can help to reveal new binary systems and populate a range of sizes and separations that remain nearly unexplored.} 
{In this work, we describe a dedicated period detection method and its results for the \gdrthree data set. This method looks for the presence of a periodic signature in the orbit post-fit residuals.}
{After conservative filtering and validation based on statistical and physical criteria, we are able to present a first sample of astrometric binary candidates, to be confirmed by other observation techniques such as photometric light curves and stellar occultations.}
{}
\maketitle

\section{Introduction}
Binary asteroids have attracted the attention of the scientific community due to their interesting properties and the significant impact they have on our understanding of the Solar System. Unlike single asteroids, binary systems offer unique insights into many fundamental processes, including the formation and evolution of planetary bodies, collision dynamics, and gravitational interactions.

Numerical simulations suggest the existence of nearly equal-sized binaries as a byproduct of catastrophic collisions \citep{durda2004formation} but they remain essentially unknown. On the other hand, multiple craters on Mars demonstrate that asteroid binaries with properties not represented by the known sample should exist, or have existed \citep{vavilov2022evidence}. More generally, the formation of asteroid companions by fragment ejection and in-orbit re-accumulation \citep{walsh2006binary,walsh2008rotational,cuk2007formation,pravec2010formation,jacobson2013formation,madeira2023dynamical} seems to be common among near-Earth asteroids (and probably small main belt objects) but the collisional evolution is also expected to play a role \citep{doressoundiram1997formation,michel2001collisions,durda2007size,jutzi2019shape}, with an undefined boundary between the two creation mechanisms. However, some open questions on the origin of binaries and the evolution of asteroids cannot be answered without a greater number of known objects and a better knowledge of the sample.

The presence of asteroid companions has been revealed by different techniques, such as high-resolution imaging from ground-based and space-based telescopes, photometry, radar ranging, and stellar occultations. When taken together, these different approaches are complementary and cover a wide range of separations and size ratios for the system components. However, the known population of binary asteroids is still strongly biased. 

While imaging favours large separations of the primary and the secondary, and bright primaries in the case of adaptive optics, photometric studies, based on detecting mutual events, strongly favour compact systems \citep{merline2002asteroids, pravec2006photometric,richardson2006review}. Radar techniques are strongly limited in range and are efficient for near-Earth asteroids \citep{ostro2002asteroid}. 

Some satellites have also been discovered by space probes and studied during close encounters. While they constitute a precious sample of detailed information, in statistical terms their contribution to the knowledge of the global population of binaries is limited. 
Astrometric detection of binaries alone cannot solve all these problems, but can very usefully fill an unexplored space of exploration \citep{pravec2012small}.
The astrometric signature of the presence of asteroid satellites is expected to show up in the residuals of the asteroid motion (on the sky or, equivalently, concerning the heliocentric orbit) due to the difference between the position of the photocentre (the one usually provided by the observations) and the centre of mass of the system. This difference, which in principle can also be due to the irregular shape of a single object, is enhanced by the presence of a satellite. 

\begin{figure*}[htbp]
    \centering
    \includegraphics[width=0.86\linewidth]{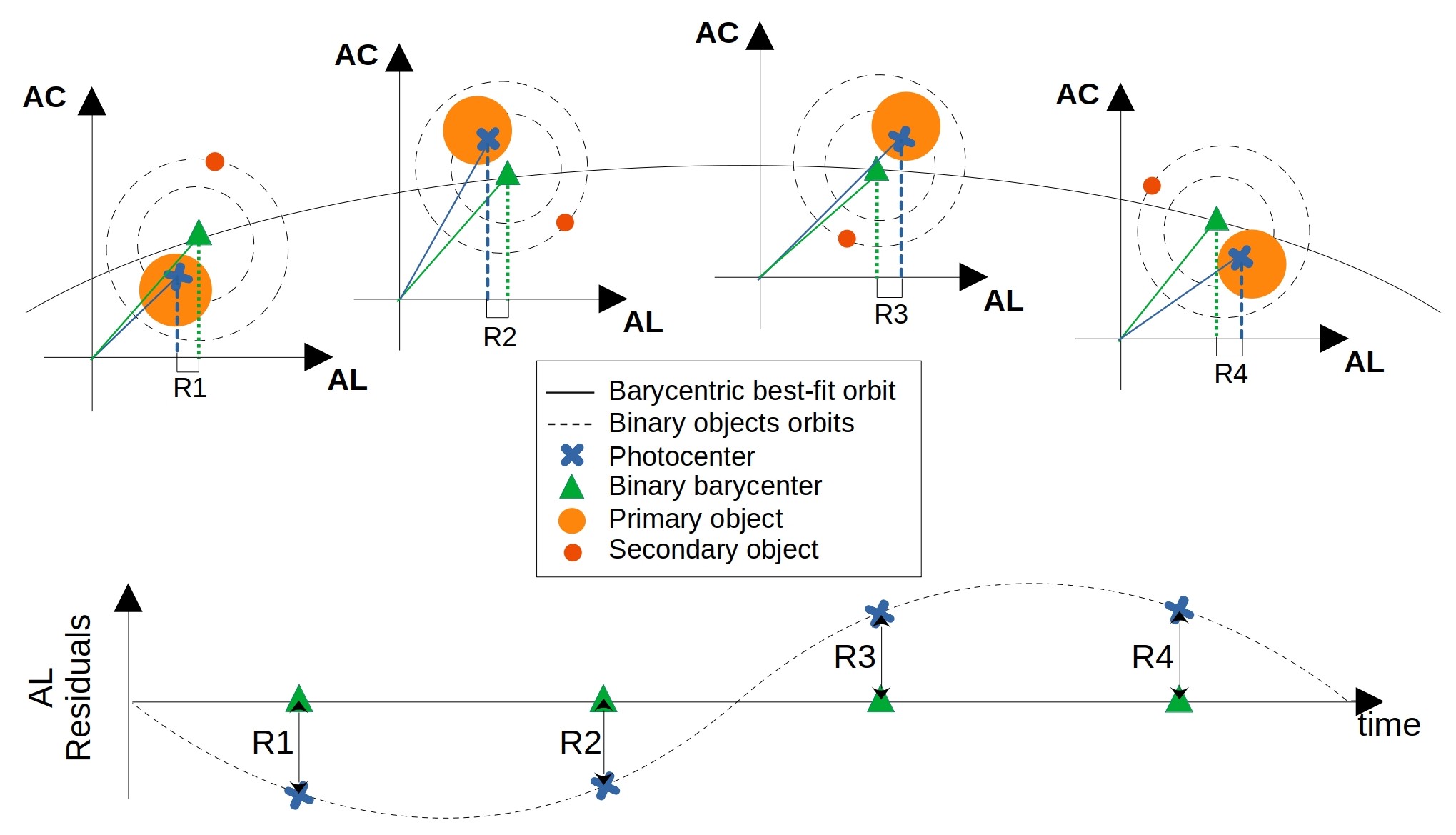}
    \caption{Representation of a simplified case where the \gaia satellite would be observing a binary system perpendicularly to the plane of rotation of the bodies, along with the projections of the positions of the barycentre and the photocentre on the AL direction and the representation of the orbital fit residuals' projection in the along-scan direction. AL direction is represented horizontally for simplicity, but in reality it can take any orientation, slowly rotating by some degrees on a scale of several hours. }
    \label{fig:wob_rep}
\end{figure*}

Anomalous residuals in the astrometry, linked to the presence of a satellite, have been revealed in the case of (90482) Orcus \citep{ortiz2011mid}. For trans-Neptunian objects, the effect is amplified by the large separation of the components, particularly in cases of high size ratio (secondary/primary). 

The astrometric signature is generally much harder to detect for main belt asteroids since it requires much more precise astrometry to reveal a perturbed motion due to a companion that is not spatially resolved. A rather solid detection, the first using \gaia astrometry, was obtained for (4337) Arecibo \citep{tanga2022gaia}. 
Also in this case, however, the presence of a satellite was already known. A systematic search of astrometry for new binary systems is a much more complex task. Only the exploration of a large data set of homogeneous measurements of extremely high quality can potentially be successful. The data produced by the \gaia mission of ESA are the main sources available today for this search. A first attempt on the very limited data set (in the time range and the number of objects) provided by \gdrtwo was unsuccessful  \citep{segev2023astrometric}.

In the following we present a blind search of astrometric binaries in the \gdrthree asteroid astrometry. Our exploration represents a first step only where we focus on the detection of periodic signals (wobble, in the following) in the residuals obtained from orbit adjustment to the data. For the moment this approach limits our investigation to sets of astrometric positions collected in short time spans. As a consequence, short-period systems are favoured. The result of our search is also filtered on the basis of rather strict criteria. These choices produce results that are certainly not comprehensive of what can be obtained from \gaia data. Nonetheless, it provides a first consolidated set of candidate detections that can be fruitfully explored by other techniques for confirmation.

The article is organised as follows. In Section~\ref{S:data} we recall the main properties of the data set that we explore and the goal of the search. Section~\ref{S:psearch} presents the method of period search. The selection, filtering, and validation of the candidates found, by statistical and by physical criteria, are presented in Section~\ref{S:selection}. Section~\ref{S:final} presents the content of the final list of binary candidates. We put our findings in a broader context in Section~\ref{S:Discussion}, and resume the work, with future perspectives, in Section~\ref{S:Conclusions}.


\section{Astrometry and binary signatures: Data set and model}
\label{S:data}

The asteroid astrometry in \gdrthree for a large number of asteroids provides an outstanding opportunity for a search of astrometric signatures produced by possible satellites. The properties of \gaia astrometry are illustrated in detail by \cite{tanga2022gaia}. Here we recall a few properties specifically relevant to our search.

The first property is the time distribution of observations. The two \gaia telescopes scan the sky continuously with a six-hour rotation period. The two fields of view (FOV) sweep the same area of the sky 106 minutes apart. The spin axis precesses in such a way that a large field overlap is granted every subsequent rotation. The combination of this rotation--precession with the revolution around the Sun results in a sparse series of detections (also called transits in the FOV) for any given source, fixed or moving. However, when the motion on the scanning path in the sky combines favourably with the position and displacement of a target source, a short consecutive series of detections over a few satellite rotations is possible. 

Over the coverage of \gdrthree (34 months), it is rather common to obtain sequences of a few consecutive observations spanning a few rotations (12h, four detections) for each object. A few occurrences of longer sequences with several rotations up to 10-12 detections (five or six revolutions) can also happen, although less frequently.

As is discussed in the following section, the length of time windows of consecutive data and the presence of constant periods in the data sampling control, respectively, the longest detectable period in our analysis, and the possible emergence of spurious frequencies. At the same time, irregular sampling allows the exploration of a large range of frequencies \citep{eyer1999variable,VanderPlas_2018_GLS}.

In addition, the nearly mono-dimensional astrometry provided for each transit brings accurate information only in the direction of the scan (along-scan, AL). The only measurable residual from orbital fitting is the projection of the total residual onto the AL direction, whose direction is known from the attitude data of the satellite. The direction of the total post-orbital fit residual in the sky, however, is not directly measurable from \gaia astrometry alone at a given epoch (or in an interval of time in which the geometry of the system concerning the scan direction does not evolve). Constraints on the total wobble in the astrometry, and the possible primary-secondary separation, are the result of applying some physical constraints to the measured wobble projection and period. 

For this first search for astrometric binaries, we use the simplest first-order model for the wobbling motion, assuming systems composed of uniformly illuminated spheres, of equal bulk densities and albedos. The other important assumption is that the components are not resolved so that the wobble is generated by the difference in position between the barycentre of the system (to which the orbit is adjusted) and the photocentre. Given the size of the \gaia pixel ($60~$mas) in the AL direction, this should be the case for a wide range of binaries in a reasonable range of sizes.

Figure~\ref{fig:wob_rep} illustrates the principles discussed, where at the top of the plot it is shown a representation of an ideal case where \gaia is observing a binary asteroid perpendicularly to the plane of rotation of the objects around the barycentre, the along-scan direction (AL) of the \gaia satellite is on the x-axis and the across-scan direction is on the y-axis. In principle, at any given epoch, the projection of the instantaneous wobble amplitude $\alpha$ for the satellite position on its orbit, projected along AL, is provided by the residual of \gaia astrometry to the orbital fit. The blue and green dashed lines represent the projection of the positions of the photocentre and the barycentre of the binary, respectively. The difference between these projections represents the AL residuals, as shown at the bottom of Fig.~\ref{fig:wob_rep}.

With the assumptions of uniform illumination and unresolved components, it is possible to derive the wobble amplitude as the difference between the position of the barycentre and the position of the photocentre. This is a function of the mass ratio $q$ only \citep{hestroffer2010gaia}, scaled for the separation $a$:

\begin{equation}
    \alpha= a \left|\frac{1}{1+q^{-1}} - \frac{1}{1+q^{-\frac{2}{3}}}\right|  \quad, q \in\ ]0,1]
    \label{eq:sep}
\end{equation}

\begin{figure}[htpb]
    \centering
    \includegraphics[width=0.9\linewidth]{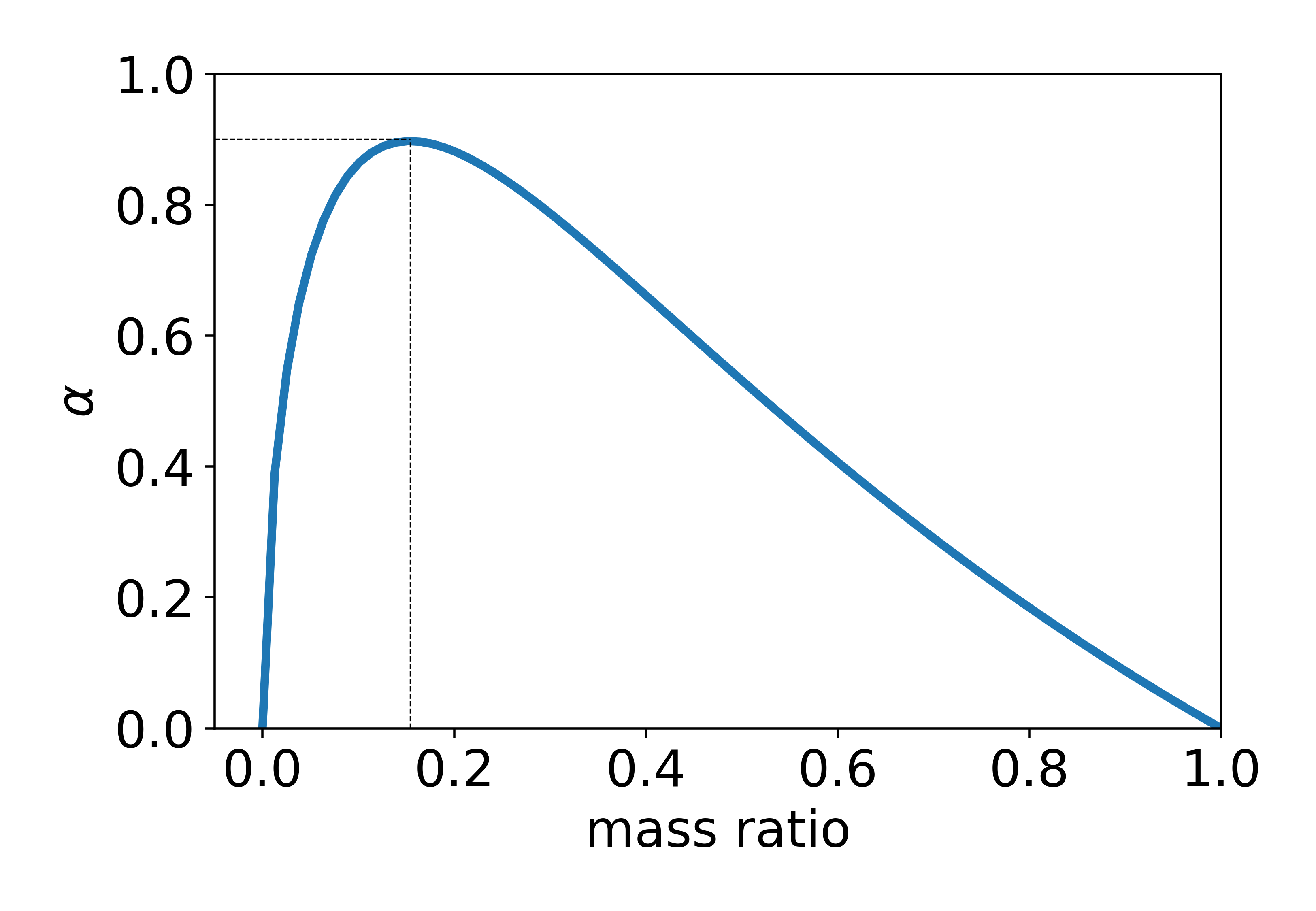}
    \caption{Plot of the amplitude of the photocentre wobbling as a function of the binary mass ratio, as described in Eq.~\eqref{eq:sep}. In this example we use a separation $a=10~$mas, which means that we would observe a maximum photocentre wobbling of $0.9~$mas for a mass ratio $q\approx 0.1538$. }
    \label{fig:alpha}
\end{figure}

This function (Fig.~\ref{fig:alpha}) has a maximum at $q_{max}\approx0.154$ (corresponding to a size ratio $k=\sqrt[3]{q_{max}}=0.536$) and tends to zero at the extremes of the range of $q \in ]0,1]$, that is, for a vanishingly small satellite or equal-sized components. During a single transit in the Astrometric Field of \gaia (about 40 seconds), up to nine consecutive positions are obtained. 
We fit all \gdrthree astrometry for each asteroid by a version of the \textsc{orbfit} software, \footnote{See http://adams.dm.unipi.it/orbfit} optimised to exploit the full accuracy of \gaia, following the same procedure as in \citet{gaiacollaborationDR2_2018}.
From the orbital fit, for each astrometric point, a residual is obtained in the AL direction. 

Given that the observation geometry over that time span is nearly constant (and a relative displacement of a satellite negligible) we average the residuals from all transits from one observation into a single transit. Their standard deviation also provides uncertainties on the measured residual itself. The average AL residuals per transit, and their uncertainty, are the data on which we base our analysis.
To prevent future confusion, throughout the work we refer to the residuals from the orbital fitting of the average transit observations in the along-scan direction simply as AL residuals.


\section{Wobbling detection}
\label{S:psearch}

For this first data exploration, our search targets intervals of time in which several consecutive astrometric measurements are available, sufficiently short to neglect any change of orientation of the asteroid with respect to \gaia. On the observations available over that interval, a period search is performed. 


\subsection{Sample selection}
\label{Ss:sample_filter}
\begin{figure*}[htbp]
    \centering
    \includegraphics[width=0.45\linewidth]{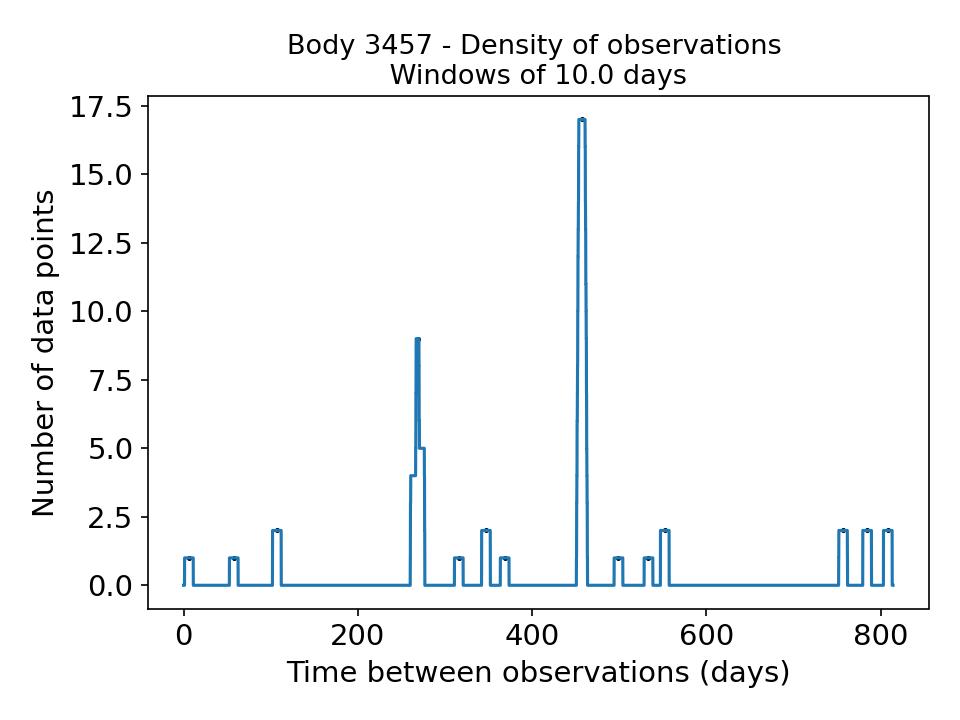}
    \includegraphics[width=0.45\linewidth]{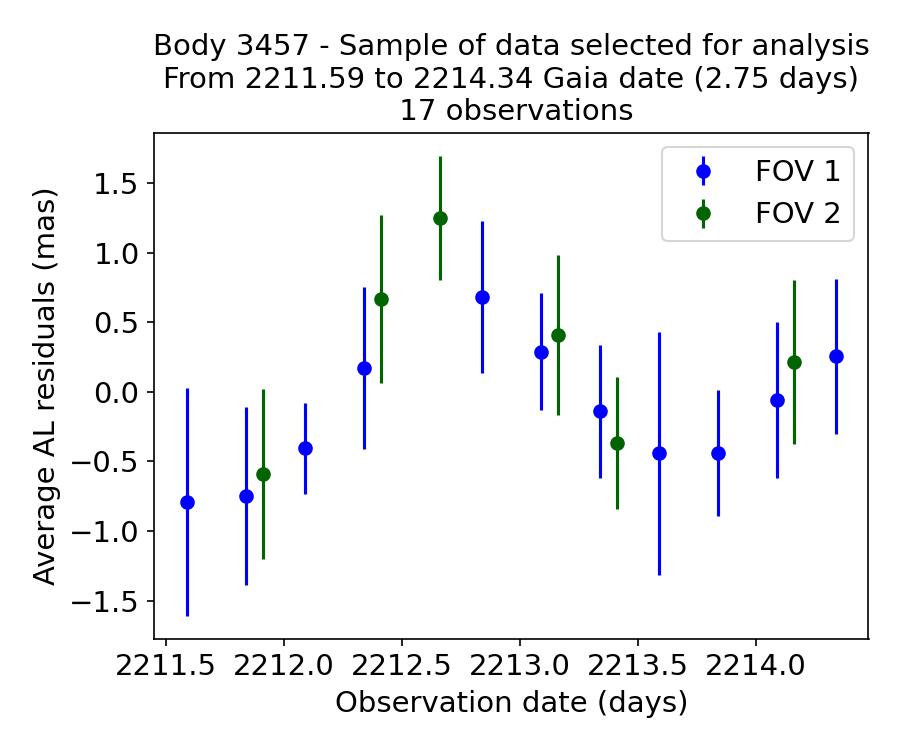}
    \caption{Plots of the density of observations (number in ten consecutive days) for asteroid (3457) Arnenordheim and selected sample of data. Left: Resulting convolution of the observation time span in days vs the number of observations in each ten-day window. Right: Sample of data at the peak of the ten-day window of search combining 17 observations. The x-axis is the observation date and the y-axis is the average AL residual from the orbital fit. Each dot represents one averaged set of transits observed in the FOV 1 in blue and FOV 2 in green.}
    \label{fig:windows}
\end{figure*}

 To easily identify the longest sequences of consecutive astrometric measurements, we perform a convolution of the observation epochs with a rectangular function and find the peaks of large data density, as shown in Fig.~\ref{fig:windows}. In the left plot, we show the density of AL residuals over time for asteroid (3457) Arnenordheim, and the peak close to 450 days represents the 2.75-day window of time with 17 observations shown in the right plot. The rectangular function has a unity value over an interval of ten days and zero elsewhere.
 
The considered threshold of ten days is a compromise between the length of the time series and the resulting number of candidates, as longer time series allow for better frequency analysis but reduce the number of targets. It should be noted that this selection method allows for gaps in the data sequence, where the observations are absent due to some technical reasons (missed observation, rejection for low quality). Over the whole DR3 data set of 156,801 asteroids, our selection extracts 30,030 objects with sequences satisfying the above-mentioned criteria. 

Next, we identify the observations in each one of the windows and perform an outlier check, removing the points with nominal values or standard deviations larger than $3\sigma$ of the dispersion of AL residuals in the window of search.


\subsection{Period search}
\label{Ss:periodogram}

The Generalised Lomb-Scargle Periodogram (GLSP) \citep{zechmeister2009generalised,vanderplas2015periodograms,VanderPlas_2018_GLS} is a statistical technique for analysing irregularly sampled time series data and identifying periodic signals in the data. The GLSP method is a generalisation of the classical Lomb-Scargle Periodogram (LSP) \citep{lomb1976,scargle1982}. 

As described in \cite{VanderPlas_2018_GLS}, the GLSP corresponds to a weighted least-squares problem where a sinusoidal signal of unknown amplitude and phase plus an unknown constant are jointly estimated at each test frequency. Formally, using the notation from \cite{VanderPlas_2018_GLS}, the GLSP $P(f)$ can be defined from the following equations:

\begin{equation}
    y_{\rm{model}}(t;f) \equiv y_0(f)+A_f \sin (2 \pi ft+ \phi_f) \quad,
    \label{eq:ymodel}
\end{equation}
\begin{equation}
    \chi^2(f) \equiv \sum_{ n=1}^{N} \left( \frac{y_n-y_{ \rm{model}}(t_n;f)}{\sigma_n} \right)^2 \quad,
    \label{eq:chi2}
\end{equation}
\begin{equation}
    P(f) \equiv \frac{ \widehat{\chi}_0^2-\widehat{\chi}^2(f)}{\widehat{\chi}_0^2} \quad .
    \label{eq:peri}
\end{equation}

Here, $y_{\rm{model}}$ Eq.~\eqref{eq:ymodel} is the assumed true model: a sinusoid (unknown frequency $f$, amplitude $A_f$, and phase $\phi_f$) plus a constant offset term $y_0$ that may depend on the considered frequency. If the $N$ data points of a time series {$y$}, $y(t_n) \equiv y_n$ for $n=1,\cdots, N$, correspond to the sum of the true model of  Eq.~\eqref{eq:ymodel} plus uncorrelated Gaussian noise of variance $\sigma^2_n$ for each of the $n$ samples, then the weighted sum in Eq.~\eqref{eq:chi2} is a $\chi ^2$ random variable with $N$ degrees of freedom. 

The GLSP computes the residual sum of squares (RSS) of the weighted residuals (noted $\widehat{\chi}^2(f) $) obtained when adjusting a sinusoid with unknown amplitude and phase for each test frequency plus a constant and compares it to the RSS obtained when fitting only a constant ($\widehat{\chi}^2_0 $).
Precisely, $\widehat{\chi}^2(f) $ (resp., $\widehat{\chi}^2_0 $) is obtained by replacing $y_{\rm{model}} $ by the estimated sinusoid plus constant (resp.,  the estimated constant) in Eq.\eqref{eq:chi2}.
The GLSP $P(f)$ in Eq.~\eqref{eq:peri} is thus the relative reduction in the sum of square brought by the adjusted model at considered frequency $f$ (with some abuse of notation on $f$). The higher $P(f)$, the smaller $\widehat{\chi}^2(f)$, which means that the model and the data are in greater agreement.

The GLSP is probably the most common technique to detect periodic signals in time series with irregular sampling. We opted for this approach because it has proven efficiency and is computationally simple enough to allow for large-scale Monte Carlo simulations (this is an important aspect of our study, see next section). The GLSP in Eq.~\eqref{eq:peri} is implemented with the \textsc{astropy} package \citep{0astropy2013,1astropy2018,2astropy2022}, with error bars $\sigma_n$ estimated as explained in the last paragraph of Sec.~\ref{S:data}.

Given the sampling frequency and the time range of ten days for our data sequences, we limit the search to periods between 3 hours and 3 times the length of the observations sample and apply our period search to all the asteroids in the 30,030 objects selected in the previous step.

\begin{figure*}[htbp]
\centering
\includegraphics[width=0.45\linewidth]{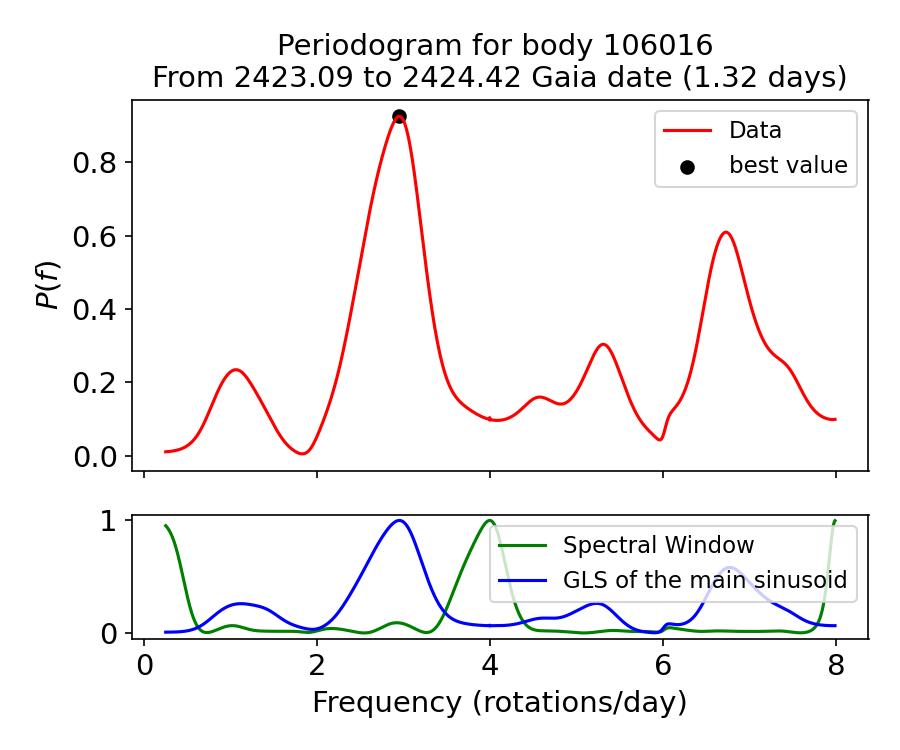} 
\includegraphics[width=0.45\linewidth]{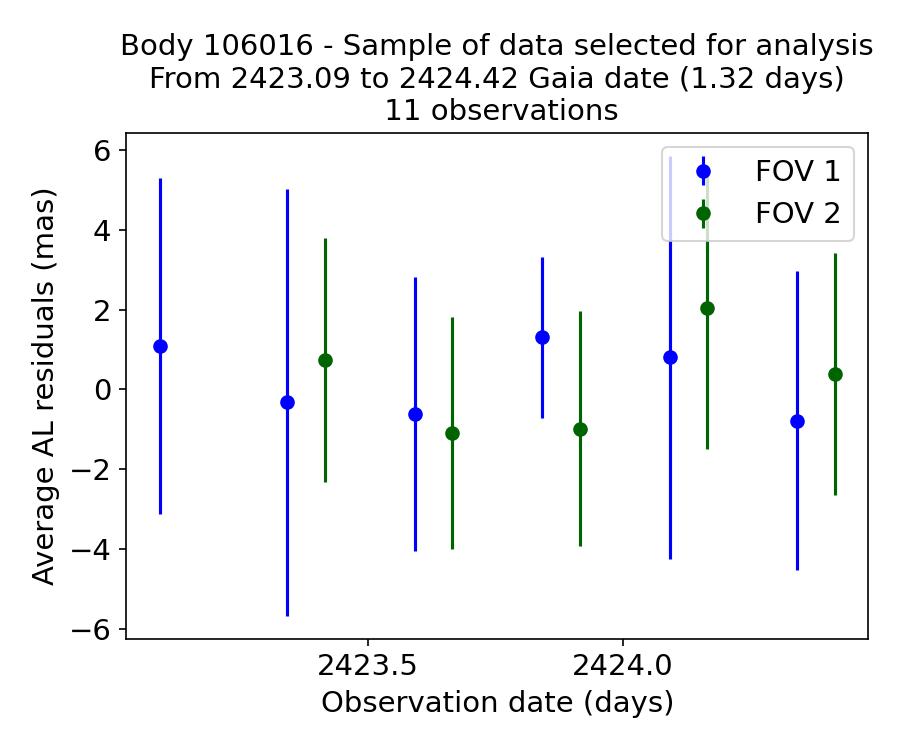}
\caption{Examples of periodograms obtained for asteroid (106016) 2000 SS293  (left plot), with residuals shown on the right plot. The y-axis shows the GLSP $P(f)$ for each frequency shown on the x-axis. The red line represents the periodogram obtained for the data analysed. The blue line represents the GLSP obtained for a purely sinusoidal signal whose parameters correspond to the largest peak of $P(f)$ (about 3 cycles/day here), with the same time sampling as the data and the same error bars for the weights. Finally, the green line represents the spectral window of the time sampling. The right plot shows the AL residual time series that reproduces the periodogram on the left plot. }
\label{fig:periodogram}
\end{figure*}

The period search provides a periodogram for all the studied bodies, and they are very different depending on the time sampling of observations. The highest peak in the periodogram represents the period that best fits the sample. The left plot of Fig.~\ref{fig:periodogram} shows, as an example, the GLSP obtained for the asteroid (106016) 2000 SS293 from the sample of AL residuals shown in the right plot of the same figure. The red line is the GLSP from the fitting on the data of AL residuals in the current window of search. Using the frequency with the highest power from the GLSP on the data we obtain the sinusoidal fit with amplitude $A$ and, along with the same time sampling and uncertainties, we run the GLSP to obtain the blue line that represents the profile of the periodogram only due to the sinusoidal signal. Finally, the spectral window (green line) is obtained by computing the Lomb-Scargle periodogram of a time series with constant unit amplitude.\footnote{Precisely, the spectral window indicates, at each frequency, the reduction of the residual sum of squares (relative to the initial sum of squares) obtained when fitting a pure sinusoid at that frequency to a constant signal sampled like the data.}

It is interesting to compare the shapes of the GLSP of the data (red curve in the left panel of Fig.~\ref{fig:periodogram}), of the main estimated sinusoid (blue curve) and spectral window (green curve). The spectral window indicates the specific frequency pattern related to the specific time sampling in a time series.  Similarities in the structures of the spectral window and the GLSP are often a clear sign that some peaks are fake (sampling-induced) rather than genuine signal periodicities. Similarly, the pure sinusoid analysis (blue curve) encodes the same information but focuses on the analysis of a particular frequency. As an example, we see that the peak at almost 7 cycles/day in the data GLSP (red curve) is probably, from the blue curve, a spurious sidelobe created by the specific time sampling. This is also visible from the green curve, which shows that an alias peak appears at four rotations/days after the main peak.

From the considerations explained above, it is clear that a detection method based on the highest peak of the periodogram must evaluate the ``statistical significance'' of a given maximum peak's height. One way to do this is to measure the probability of obtaining a value of the largest peak as high as the one observed in the data when there is only noise in the time series. This is investigated in the next section. 


\section{Binary Candidates Selection Method}
\label{S:selection}

\subsection{Statistical selection}
\label{Ss:selection_statistic}

For each window of search of each object, we 
computed the GLSP as described in Sect.~\ref{Ss:periodogram} and obtained the value (say, $v$) of the highest peak. A common proxy for the ``statistical significance'' of $v$ is the $p$-value \citep[see e.g.][]{efron_2010},
defined as follows. Let $V$ be the random variable defined as the maximum value of the periodogram, conditional on the time sampling grid and error bars of the considered data. The $p$-value of $v$ is the probability of obtaining a maximum peak value at least as large as $v$ under the signal-absent hypothesis, with the same time sampling and noise error bars (denoted by ${\mathcal{H}}_0$). Formally this probability is

\begin{equation}
    p(v) \equiv \textrm{Pr}\;(V>v \; |\; {\mathcal{H}}_0)\quad .
    \label{eq:pval}
\end{equation}

In practice, the $p$-value $p(v)$ of a particular time series can be evaluated by Monte Carlo simulations by estimating empirically this probability as follows. We generate 10,000 simulated time series sampled with the same time sampling as the time series under investigation. Each time series is composed of a Gaussian noise consistent with the error bars plus a random constant (an offset drawn randomly from a zero-mean Laplacian distribution, which is consistent with our data, see Appendix \ref{app:offset}). The GLSPs of the 10,000 time series are computed and we count how often their highest peak has an equal or larger power than the one from the GLSP under investigation ($v$).

\begin{figure}[htbp]
    \centering
    \includegraphics[width=\linewidth]{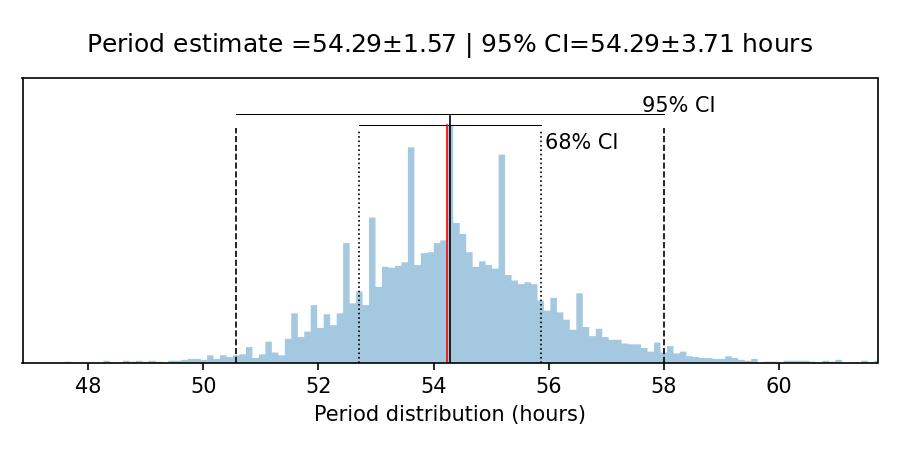}\\
    \includegraphics[width=\linewidth]{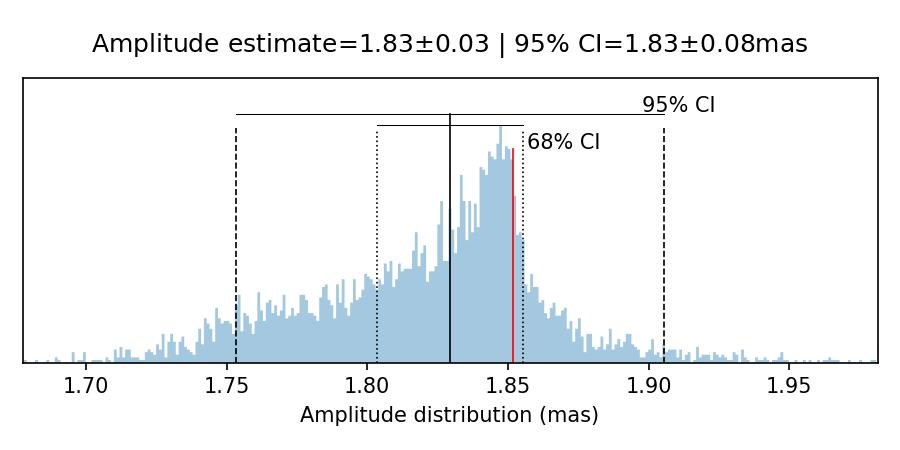}
    \caption{Examples of uncertainty determination from MC simulations under ${\widehat{\mathcal{H}}}_1$. The top and bottom plots show, respectively, the distribution of periods and amplitudes obtained from the best fit in each MC simulation (in blue). The red lines show the value estimated from the first GLSP run only on \gaia data, and the solid black line represents the median obtained from the distribution of amplitude and period values from the MC under ${\widehat{\mathcal{H}}}_1$. The dotted vertical lines show the 68\% distribution interval while the dashed lines represent the 95\% confidence interval of the distribution. The final results are given as the first estimate from the \gaia data (red line), and the uncertainty is the 95\% interval. For instance, in this example, the final amplitude value is $1.85\pm 0.08$.}
    \label{fig:uncertainties}
\end{figure}

 A $p$-value is computed in this way for each of the selected 30,030 time series. To be selected as a possible candidate, the time series must satisfy three conditions. The first condition is that the  $p$-value estimated from the MC under ${{\mathcal{H}}}_0$ must be smaller than $0.05$.  The second condition is a check of the 'robustness' of the periodicity which we call ``quality factor'' below, and which also provides a method to estimate uncertainties for the amplitude and frequency associated with the main peak of the GLSP. \\
 For this second condition, we perform MC simulations under the hypothesis that there is a sinusoidal signal in the data with the frequency and amplitude found from the GLSP of the candidate time series (plus noise with the considered error bars, plus an unknown offset, and with same time-sampling; we note this hypothesis ${\widehat{\mathcal{H}}}_1)$. We then perform another series of 5,000 MC simulations of synthetic time series under ${\widehat{\mathcal{H}}}_1$.

 A robust period detection should repeat nearly identically several times with different realisations of the synthetic data. To obtain the quality factor of the signal, denoted by $Q$ below, we estimate the fraction of time series for which the GLSP highest peak is found $\pm$ 0.15 cycles/day around the injected frequency. Our second condition is that $Q$ should be larger than $0.5$, meaning that the initially injected period should be retrieved more than half of the time in nominal noise and sampling conditions.
 
 We underline that a large maximum peak caused by noise in the GLSP (i.e. a false alarm) will lead to a high value of the quality factor, because the wobbling amplitude estimated from the data will be large, so any strong but `fake' detection will appear as robust concerning the $Q$ analysis. In the opposite direction, a maximum peak with low amplitude will lead to a low estimated wobbling amplitude and a low value of the quality factor. Hence, while a large value of $Q$ is not a guarantee that the detection is robust, the cross-analysis brought by the quality factor essentially provides a supplementary flag that some detections may not be robust (this will be further investigated in Sec.~\ref{Ss: validation}). 

Finally, we turn to the third condition. An important aspect is that we are looking for astrometric binaries, which means that \gaia satellite would not have been able to observe the asteroid and the satellite separately. A wobble amplitude of $20~$mas translates to a minimum separation of about $220~$mas, which is above the \gaia resolution limit of about $200~$mas. Hence, the third condition is that the wobble amplitude estimated $\widehat{A}$ must be smaller than $20~$mas.

\begin{figure*}[htpb]
    \centering
    \includegraphics[width=0.9\linewidth]{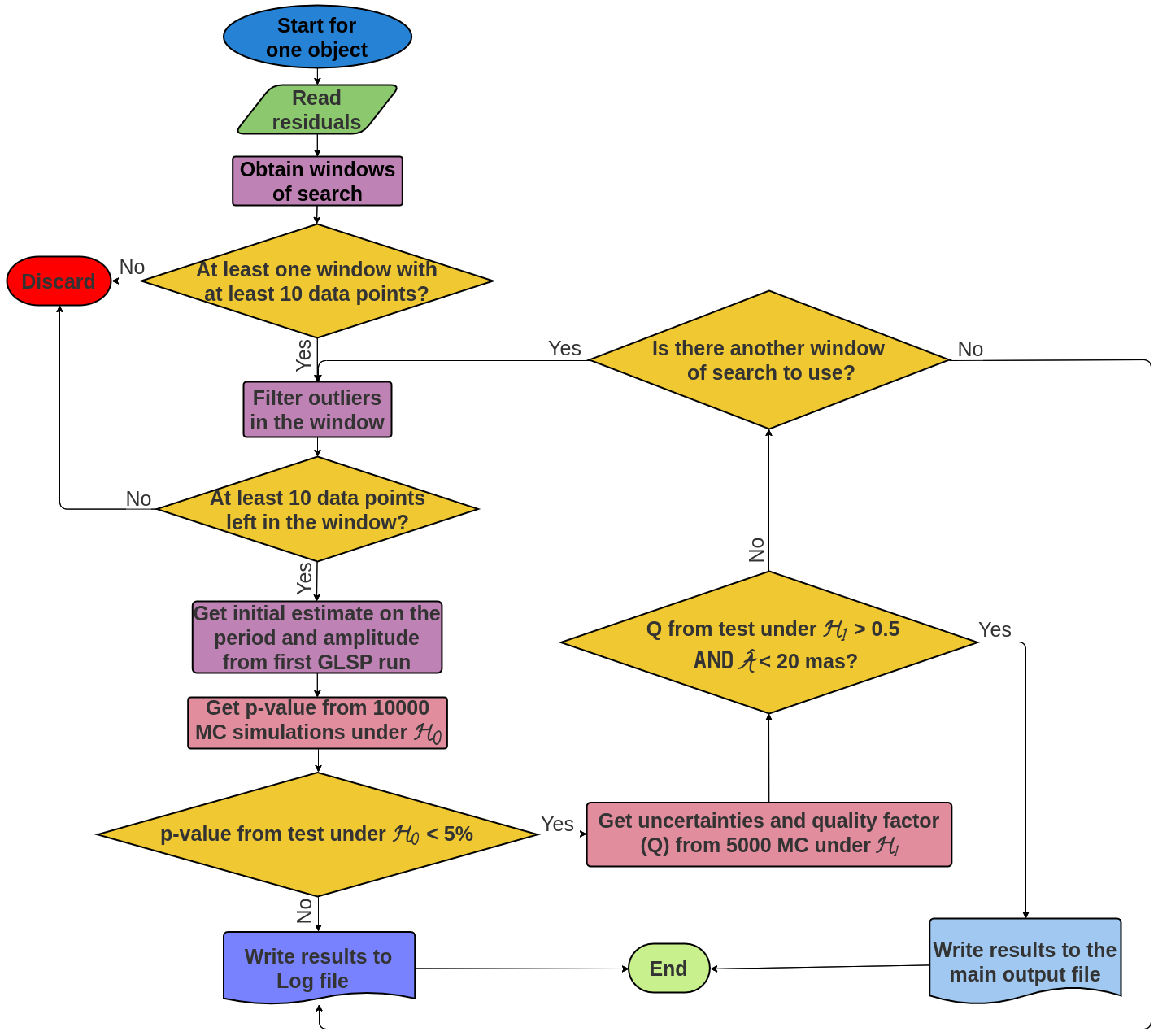}
    \caption{Flowchart showing the period search and statistical selection processes discussed in Sections \ref{S:psearch} and \ref{S:selection}. It represents part of the full detection process, which is coupled with physical characterisation tests that are discussed in the following sections.}
    \label{fig:flowchart}
\end{figure*}

In summary, if the signal detected in the current window has a $p$-value$<5\%$ as estimated from MC simulations under ${{\mathcal{H}}}_0$, a quality factor $Q>0.5$ as estimated from MC simulations under ${\widehat{\mathcal{H}}}_1$ and estimated wobble amplitude smaller than $20~$mas, then the object is selected as a preliminary candidate and the values estimated in the current window are stored. The threshold of $p$-value$<5\%$ is not very conservative since the expected number of false positives is about 1,500 (5\% of 30,030). In our case, the number of candidates that pass the triple test above is 3,038. This number (larger than 1,500) can be explained by two reasons. First, by a population of true binaries that lead to smaller $p$-values. Second, by possible modelling errors (for instance, some un-modelled, low-order polynomial trends would lead to GLSP with larger peaks at low frequencies).\footnote{While we do have some hints that a model with low-order polynomials could be more accurate than the model in Eq. \eqref{eq:ymodel}, this would require to compute a much more computationally expensive least-squares fitting at each frequency than the current GLSP. This improvement will be implemented in a subsequent version of our detection procedure.} However, as we discuss in Sect.~\ref{Ss: validation}, the latter seems unlikely. 

To obtain estimates of the period and amplitude uncertainties, we use the results from the MC simulations under ${\widehat{\mathcal{H}}}_1$. In the case of mean period determination, we look for frequencies that are no more than 0.25 cycles/day away from the best frequency found with the first GLSP and scan along the distribution to find the mean value, the 68\% confidence interval, which corresponds to the mean $ \pm 1 \sigma $ for a Gaussian distribution, and the 95\% confidence interval. The value estimated from the first GLSP run is selected as the wobble period and the 95\% CI is the uncertainty on the period. The procedure is the same for the amplitude, but in this case, we limit the interval for amplitudes up to $2~$mas away from the first amplitude estimate. Figure ~\ref{fig:uncertainties} shows an example of uncertainty determination for the period on the top plot and the amplitude on the bottom plot. 

If the object under the period search has more than one window of data, as discussed in Sec.~\ref{Ss:sample_filter}, we repeat the same procedure described in all the other windows as well. Figure~\ref{fig:flowchart} shows a flowchart that summarises the statistical selection processes discussed above.


\subsection{Analysis of the detection performance of the statistical selection process}
\label{Ss: validation}

To evaluate the detection performance of our method, in this section, we conduct a general study based on the analysis of two cases. In the first case, there is indeed a signal in each time series analysed and we look at the effectiveness of the method at detecting and characterising it. In the second case, the time series have no signal and we examine how incorrect the method can be in falsely selecting residuals with only noise. A supplementary investigation into the detection of eccentric binary systems is discussed in Appendix \ref{app:ecc}.


\subsubsection{Signal detection}

In this analysis, we study the signal detection performance by grouping the simulated time series into classes of signal-to-noise ratio and looking at the results in terms of detections (at the right period or not), missed detections, associated estimated $p$-values and $Q$-factors.
We use two signal-to-noise ratio definitions:
\begin{align}
    &\textrm{``Base~S/N''}:~S/N_{\rm base}\equiv {A}/\sigma \quad \textrm{and}  \label{eq:base_snr}  \\
    &\textrm{``Post-fit~S/N''}: ~\widehat{S/N}\equiv \widehat{A}/\bar{\sigma}  \label{eq:postfit_snr}  \quad,
\end{align}

where $A$ is the true amplitude of the time series and $\sigma$ is the nominal average uncertainty of the data set, $\widehat{A}$ is the amplitude estimate from the data time series and $\bar{\sigma}$ is the estimated average uncertainty of the data set. To obtain the performance analysis we perform the period search and selection method shown in Fig.~\ref{fig:flowchart} for six different $S/N_{\rm base}$ values, with Monte Carlo simulations based on 1,000 time series of synthetic AL residual. 

First, we choose a set of different values of $S/N_{\rm base}$ between 0.5 and 3.0. We create each set of synthetic data from a sinusoid with a known amplitude $A$ of 1 mas, a period of 23 hours (a frequency close to but not exactly 1 cycle/day to avoid possible bias) and a random phase between 0 and $2\pi$. The nominal average noise $\sigma$ for each set of synthetic samples is estimated using Eq.~\eqref{eq:base_snr}.

For the simulations of the time series, we draw 12 observation epochs, randomly distributed in 6 days. The small number of points, the randomness in the data and the time sampling are aimed not only at representing the data from \gaia, which is usually modestly sampled over a few days but also at providing conservative results that cover most of the non-ideal cases. 
 
For each time series, the actual standard deviations of the errors $\sigma_n$ are obtained by a random perturbation of $\sigma$ (using a zero-mean, normal distribution of std $0.8$; this setting is aimed at creating some scatter in the actual S/N within each S/N class), and we add one random Laplacian offset as described in Appendix \ref{app:offset}. Then, for each one of the 6,000 synthetic time series simulated in this way, we run the statistical selection (Sect.~\ref{Ss:selection_statistic}). 

\begin{figure}[htbp]
    \centering
    \includegraphics[width=0.9\linewidth]{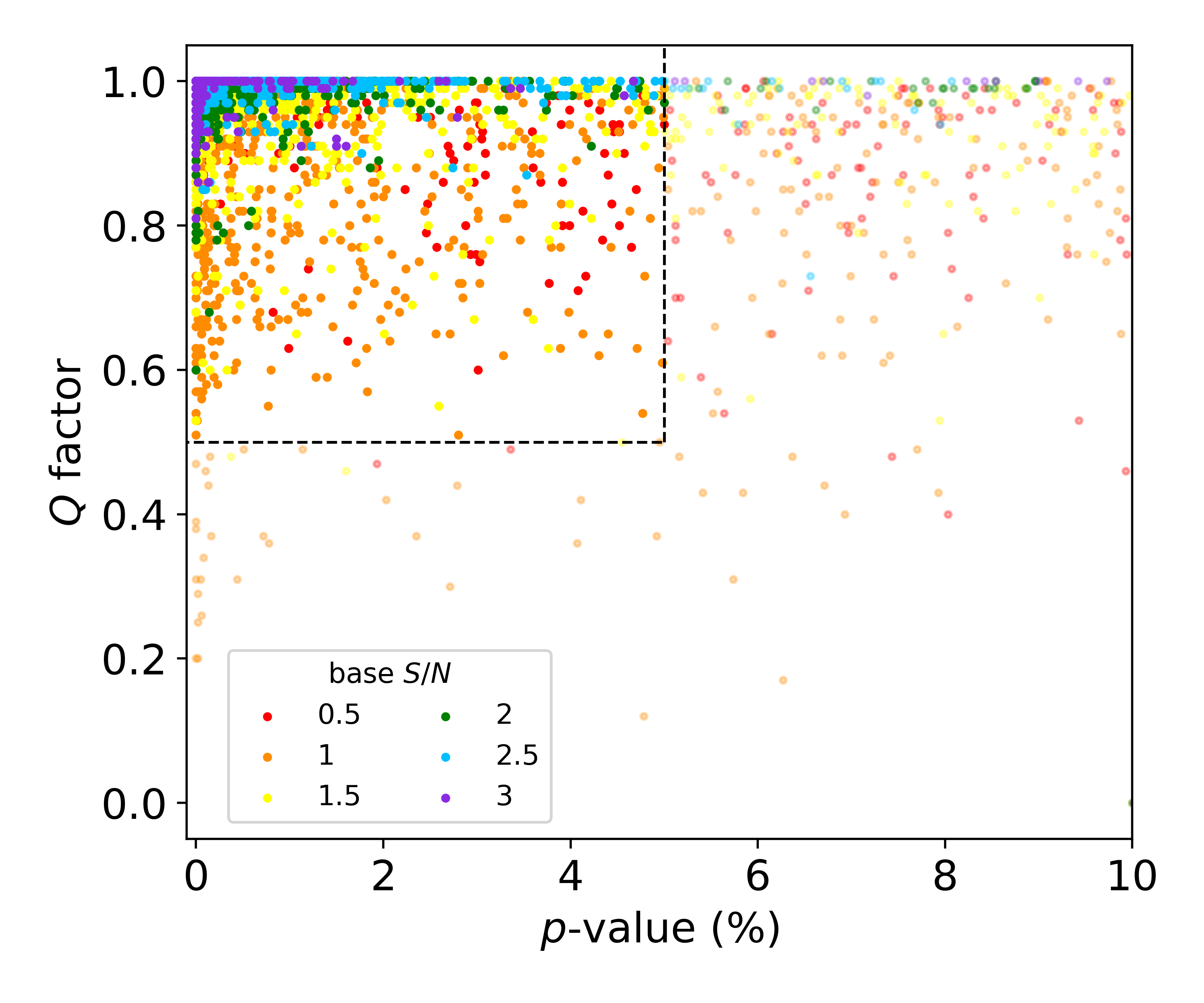}

    \caption{Distribution of pairs ($p$, $Q$) for each of the 6,000 sets of synthetic data simulated. The partially transparent dots represent the cases where the detections do not meet the requirements and were disqualified, while the remaining dots represent the detections selected. Of the 6,000 simulations, 1,303 have $p$-value$>$0.1. Hence, their $Q$ factors were not calculated and the corresponding points are not shown in this plot.} 
    \label{fig:pval_Q_valid_test}
\end{figure}

Figure~\ref{fig:pval_Q_valid_test} shows the results for the ($p$, $Q$) pairs of 6,000 time series simulated. We found that, as expected, for low values of $S/N_{\rm base}$ the parameters $p$ and $Q$ tend to be more scattered and present less correlation between themselves. But for higher $S/N_{\rm base}$ values, the couples ($p$, $Q$) tend to cluster around the ``best conditions'' represented by very low $p$-values and high $Q$.
    
Intuitively, the larger the $S/N_{\rm base}$ of a signal the easier it would be for the method to detect it. It is also expected that very noisy signals hamper the period fitting and provide incorrect detections more often, and our results follow that behaviour. For time series with $S/N_{\rm base}$ close to $0.5$, the method achieves 11.3\% of detections, of which  69\% have incorrect periods. For the $S/N_{\rm base}$ class of 1.0, the detections represent 56\% and about half present incorrect periods. For the  $S/N_{\rm base}$ classes larger than 2.0, the method detects more than 93\% of the signals with a low rate of incorrect detections.

\begin{figure*}[htpb]
    \centering
    \includegraphics[width=0.9\linewidth]{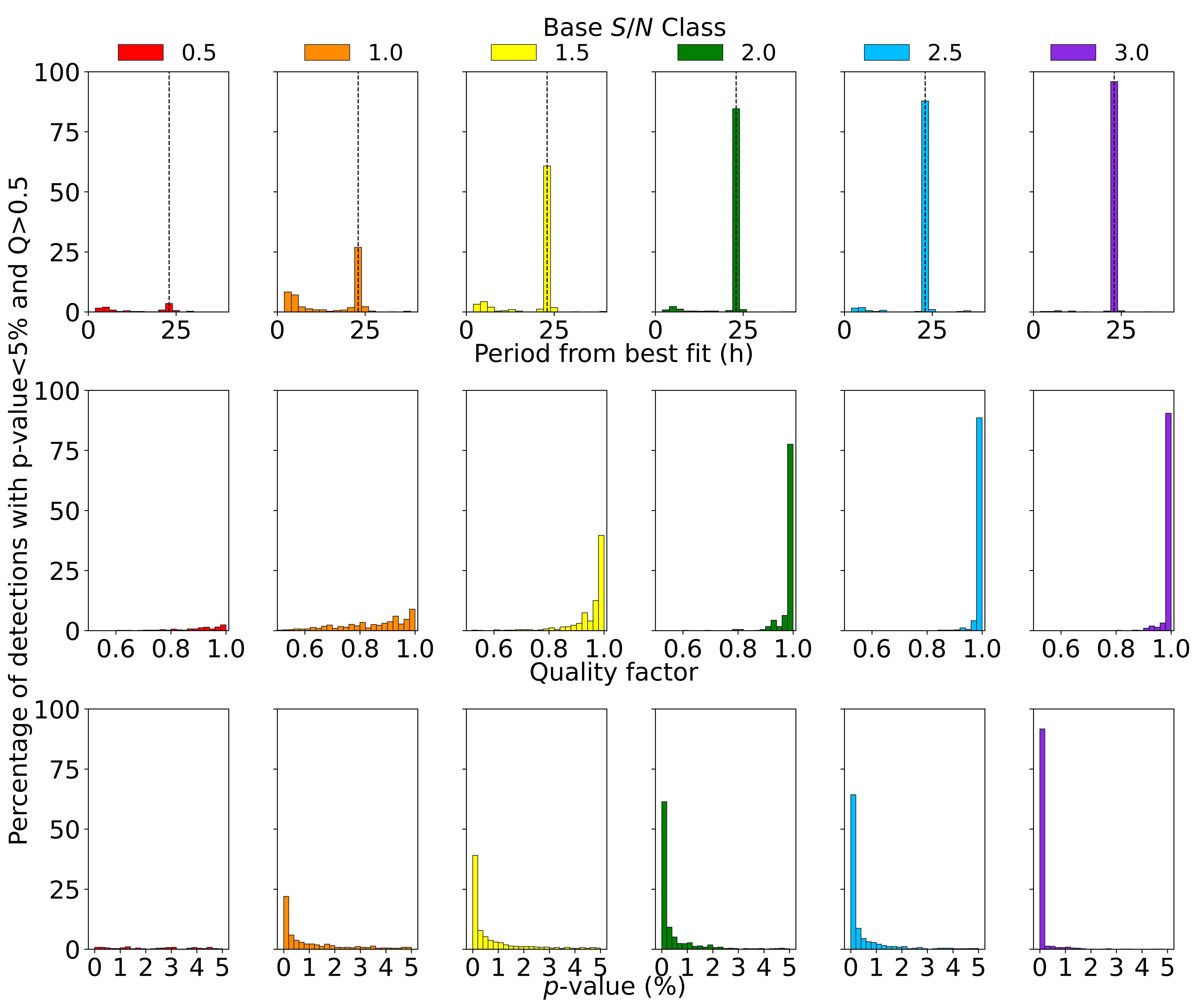}
    \caption{Distribution of periods from best fit (top row), quality factors (centre row), and $p$-values (bottom row) as percentages of the detections selected with $p$-value$<5\%$ and $Q>0.5$. The different coloured bars represent the different classes of $S/N_{\rm base}$ values explored. The dotted line at 23 hours represents the nominal (true) period of the injected signal.}
    \label{fig:valid_test}
\end{figure*}

It is also interesting to check the distribution of the quality factor ($Q$) and the $p$-value depending on the $S/N_{\rm base}$, as presented in Fig.~\ref{fig:valid_test}. It shows the distribution of best periods fitted on the top plots, the distribution of $Q$ in the plots in the middle and the $p$-value distributions on the bottom plots for each different class of $S/N_{\rm base}$ studied. As discussed before, we notice that for small $S/N_{\rm base}$ there is a significant amount of detections with periods far from the real value, represented by the dotted black line. In addition, the corresponding distribution of $Q$ shows that the quality of the detection tends to be mostly below 0.8, much smaller than for larger values of $S/N_{\rm base}$ where $Q$ is usually closer to 1.0. Similar behaviour can also be seen in the $p$-value distributions where the higher the $S/N_{\rm base}$, the larger the concentration of results with small $p$-values.

\begin{table*}[htbp]
    \caption{Results for each class of $S/N_{\rm base}$ studied.}
    \centering
    \begin{tabular}{|c|c|c|c|c|c|}\hline

    \multirow{2}{*}{$S/N_{\rm base}$} &  \multirow{2}{*}{\parbox{1.8cm}{~Detections}}  &  \multirow{2}{*}{\parbox{2.6cm}{Detections around\\correct period}} &  \multirow{2}{*}{\parbox{2.4cm}{Detections with \\ incorrect period}}   &   \multirow{2}{*}{Missed detections} &  \multirow{2}{*}{\parbox{3cm}{Incorrect/Correct \\ period detection rate}}  \\
    &&&&&\\\hline
    0.5 &  11.30 & 3.50 & 7.80 & 19.60 & 69.0 \\
    1.0 &  56.10 & 26.90 & 29.20 & 13.80 & 52.0 \\
    1.5 &  76.88 & 60.76 & 16.12 & 11.81 & 21.0 \\
    2.0 &  93.30 & 84.60 & 8.70 & 5.10 & 9.3 \\
    2.5 &  95.20 & 87.80 & 7.40 & 3.00 & 7.8\\ 
    3.0 &  98.40 & 95.90 & 2.50 & 1.30 & 2.5\\\hline
    \end{tabular}
    
    \begin{minipage}{\linewidth}
   {\small Notes: The first column represents the base signal-to-noise ratio used. The second column shows the total percentage of detections in the sample. The third column shows the percentage of detections in which the best period fit is up to one hour away from the nominal period chosen. The fourth column presents the percentage of detections where the period is further than one hour from the nominal value, complementary to the previous column. The fifth column shows the percentage of missed detections because they did not pass the selection tests, but the period found is close to the nominal period. The last column shows the fraction of the detections with incorrect periods concerning the total number of detections, representing the probability of incorrect detection.}
   \end{minipage}
    \label{tab:validation_baseSNR}
\end{table*}

The $p$-value is usually the only parameter used in the selection of detections. In fact, for $S/N_{\rm base}>1$, we find that almost all missed detections have $Q>0.5$, so the $p$-value is the parameter ruling the selection and $Q$ has almost no influence in the selection process. However, for $S/N_{\rm base}\leq1$ there are many cases with $p$-values smaller than 5\% where $Q$ is used to filter the weak detections, as seen in Fig.~\ref{fig:pval_Q_valid_test}.

\begin{figure*}[htpb]
    \centering
    \includegraphics[width=0.9\linewidth]{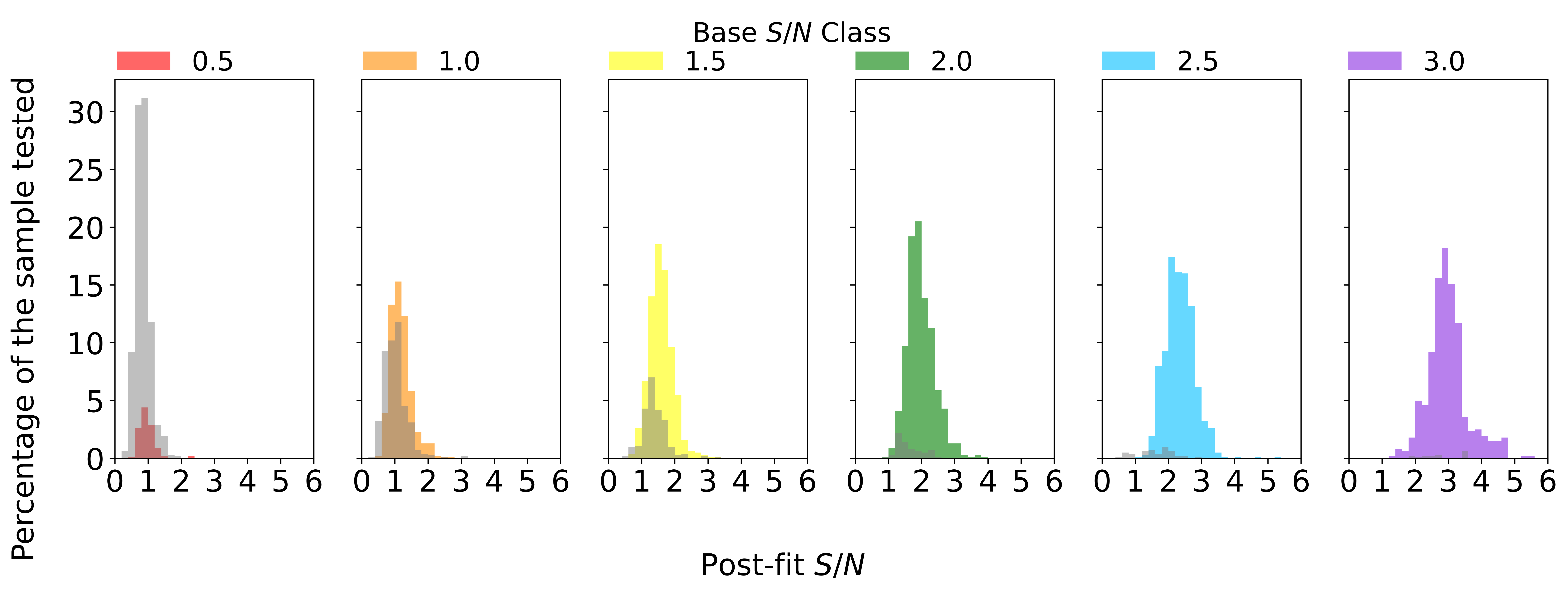}
    \caption{ Distribution of time series selected as detections (coloured bars) and discarded (grey) as a function of the post-fit $\widehat{S/N}$, for the six different $S/N_{\rm base}$ classes explored.}
    \label{fig:base_postfit}
\end{figure*}

Of course, when applying the detection method to real \gaia data, we can only estimate the signal-to-noise ratio based on the sinusoidal fit from GLSP. This is the reason why, in addition to the analysis of the detection performance based on the base $S/N_{\rm base}$ values, it is also interesting to study the results based on the post-fit S/N ($\widehat{S/N}$), which is indeed different from the $S/N_{\rm base}$. As shown in Fig.~\ref{fig:base_postfit} we see that the number of detections increases for the largest values of $S/N_{\rm base}$, as in Table~\ref{tab:validation_baseSNR}.

By combining all the results from the 6,000 time series, we can simulate the situation where we do not know the original parameters of the time series and have only the post-fit signal-to-noise ratio. Then, we separate them in intervals of mean $\widehat{S/N}\pm0.1$ and check the statistics, as it was made previously for the $S/N_{\rm base}$ classes. The results are presented in Table~\ref{tab:validation_postfitSNR}. From the comparison between the analysis based on $S/N_{\rm base}$ classes presented previously, and the one based on the post-fit $\widehat{S/N}$, these results indicate that there are no major differences in the statistics of detections between both approaches, especially for $\widehat{S/N}\geq1.5$.

\begin{table*}[htpb]
    \centering
    \caption{Same as Table \ref{tab:validation_baseSNR}, but with the analysis based on the $\widehat{S/N}$.}
    \begin{tabular}{|c|c|c|c|c|c|}\hline
    \multirow{2}{*}{$\widehat{S/N}$} &  \multirow{2}{*}{\parbox{1.8cm}{~Detections}}  &  \multirow{2}{*}{\parbox{2.6cm}{Detections around\\correct period}} &  \multirow{2}{*}{\parbox{2.4cm}{Detections with \\ incorrect period}}   &   \multirow{2}{*}{Missed detections} &  \multirow{2}{*}{\parbox{3cm}{Incorrect/Correct \\ period detection rate}}  \\
    &&&&&\\\hline
    0.5 &  11.61 & 2.90 & 8.71 & 17.10 & 75.0\\
    1.0 &  43.25 & 20.68 & 22.56 & 16.75 & 52.2\\
    1.5 &  76.56 & 56.64 & 19.92 & 11.62 & 26.0\\
    2.0 &  94.58 & 85.85 & 8.73 & 2.83 & 9.2\\
    2.5 &  98.15 & 94.14 & 4.01 & 1.54 & 4.1\\
    3.0 &  99.15 & 94.87 & 4.27 & 0.00 & 4.3\\\hline
    \end{tabular}
    \label{tab:validation_postfitSNR}
\end{table*}

It is interesting to note from these results that in the scenario where there are signals in all of the time series analysed, but we do not know the $S/N_{\rm base}$, $\widehat{S/N}\leq1$ lead to $p$-values and $Q$ factors that often fail to pass the selection test. Meanwhile, for $\widehat{S/N}\geq2$ we find more than 85\% of the detections around the correct period with a rate of incorrect/correct lower than 10\% and we miss less than 3\% of the detections.

We conclude that a $\widehat{S/N}$ of about one appears to be a limit for the detection regime in this setting.  For increasingly stronger signals, with larger $\widehat{S/N}$, we have an increasingly smaller rate of incorrect detections. The method succeeds accordingly at classifying the sample as detection, at recovering the correct period in the data, and at providing a good approximation of the real $S/N$. 


\subsubsection{Noise-only scenario}

The purpose of this analysis is to understand how the method performs when there is no signal in the data and to assess its robustness in handling data with no wobble at all. For that purpose, we generated 6,000 sets of simulated AL residuals by replicating the error bars and time sampling of original AL residuals from \gaia asteroids. These synthetic noisy time series were generated by drawing random samples with Gaussian noise perturbation corresponding to the target error bars and we added in each time series a random offset consistent with the \gaia time series, cf Appendix \ref{app:offset}.

As discussed in Sec.~\ref{Ss:selection_statistic}, with a $p$-value threshold of 0.05 it is by definition expected that 5\% of the time series are selected (on average) if none of the data present a real signal. Our results showed that in the considered noise-only scenario, the statistical selection process selected as detections about 4.8\% of the time series analysed. 

\begin{figure*}[htpb]
    \centering  \includegraphics[width=0.45\linewidth]{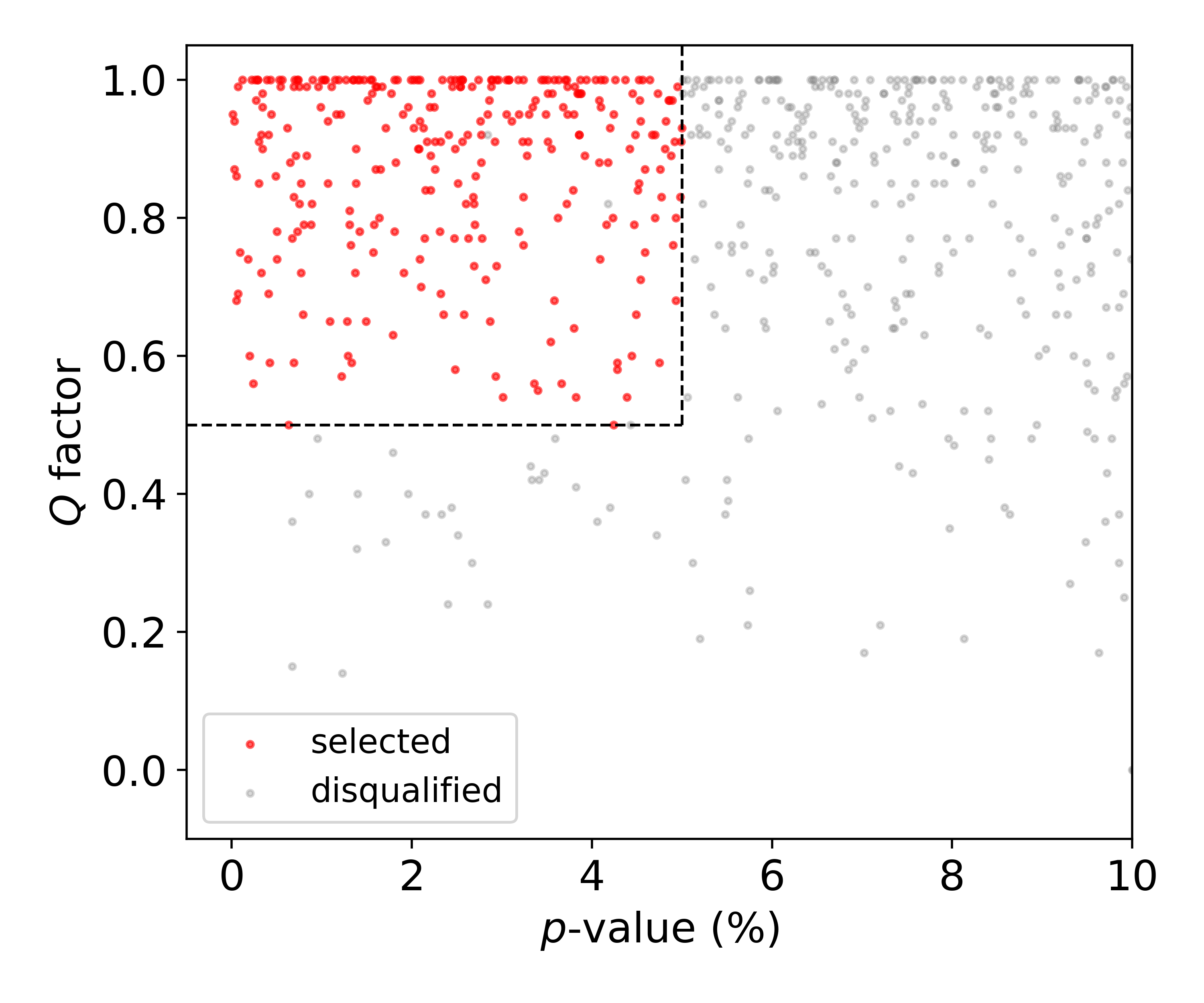}
    \includegraphics[width=0.45\linewidth]{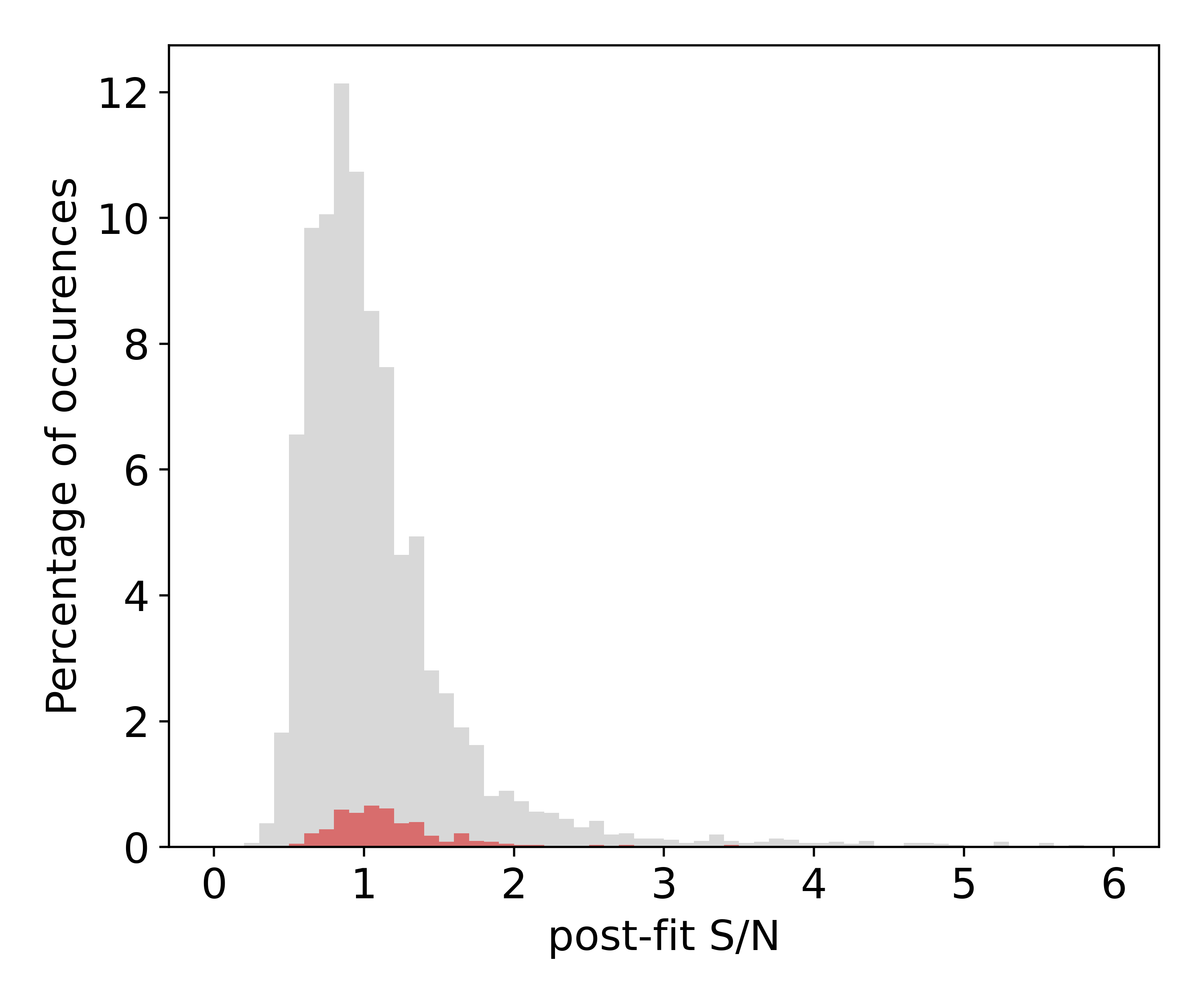}
    \caption{Resulting distribution of parameters from the 6,000 simulations in the noise-only scenario.  Left: Distribution of pairs ($p$, $Q$) for the time series simulated. 5,391 of the cases have $p$-value$>0.1$ (for these cases the Q factors were not calculated and the corresponding points are thus not shown). The grey dots represent the cases where the data was disqualified as a detection (cases for which $p>0.05$, $Q<0.5$, or both) and the red dots represent the detections. The two grey points in the top left square are detections discarded due to $\hat{A}>20$ mas. Right: Post-fit S/N distribution for the pairs in the left plot.}
    \label{fig:pval_Q_noise}
\end{figure*}

The left plot in Fig.~\ref{fig:pval_Q_noise} shows the distribution of ($p$, $Q$) pairs for the 609 time series with $p$-value$<0.1$. The remaining 5,391 time series present $p$-value$>0.1$  and therefore their $Q$ was not calculated. We see that the resulting dispersion in the ($p$, $Q$) plane is similar to that observed in Fig.~\ref{fig:pval_Q_valid_test} for classes of $S/N_{\rm base}\leq1$. This is in contrast with classes of larger S/N values for which the points tend to cluster around the top left corner. \\

Turning to the right panel of Fig.~\ref{fig:pval_Q_noise}, we see that noise-only data lead to post-fit S/N values distributed around $\widehat{S/N}\approx 1$, with few cases having $\widehat{S/N}\geq  2$. As expected, this distribution is compatible with the lowest S/N classes of Fig. \ref{fig:base_postfit}.

In a nutshell, we conclude from this two-fold performance study that, when a wobble is indeed present, our statistical selection process behaves as expected and increasingly succeeds at retrieving this wobble as its amplitude increases. In the situation when there is no wobble, the incorrect detection rate is controlled and does not exceed the prescribed value (set to 5\% here).

Finally, the S/N value of one appears to be a turning point in the detection regime, post-fit S/N values larger than $\approx 2$ being a serious hint of the presence of a population of real wobbles.\footnote{Indeed, the detectability of a signal depends also on the time sampling. Two wobbles with the same S/N but sampled differently will in general have different p-values and probabilities of detection. In our simulations, the time sampling of the simulations is generated in agreement with the \gaia sampling of the considered AL residuals. Our results are averaged over a large number of different sampling grids. Hence, the frontier of a unit S/N should not be taken at face value but reflects a global behaviour corresponding to the considered sampling conditions.}


\subsection{Period search results overview}
\label{Ss:results_overview}

\begin{figure*}[htbp]
    \centering
    \includegraphics[width=0.48\textwidth]{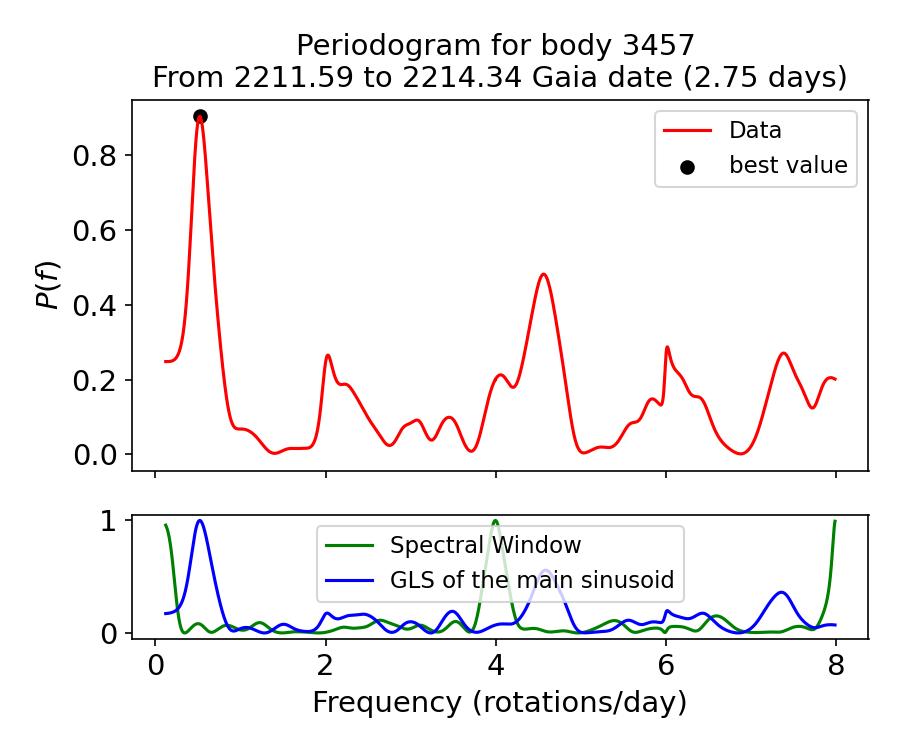}
    \includegraphics[width=0.48\textwidth]{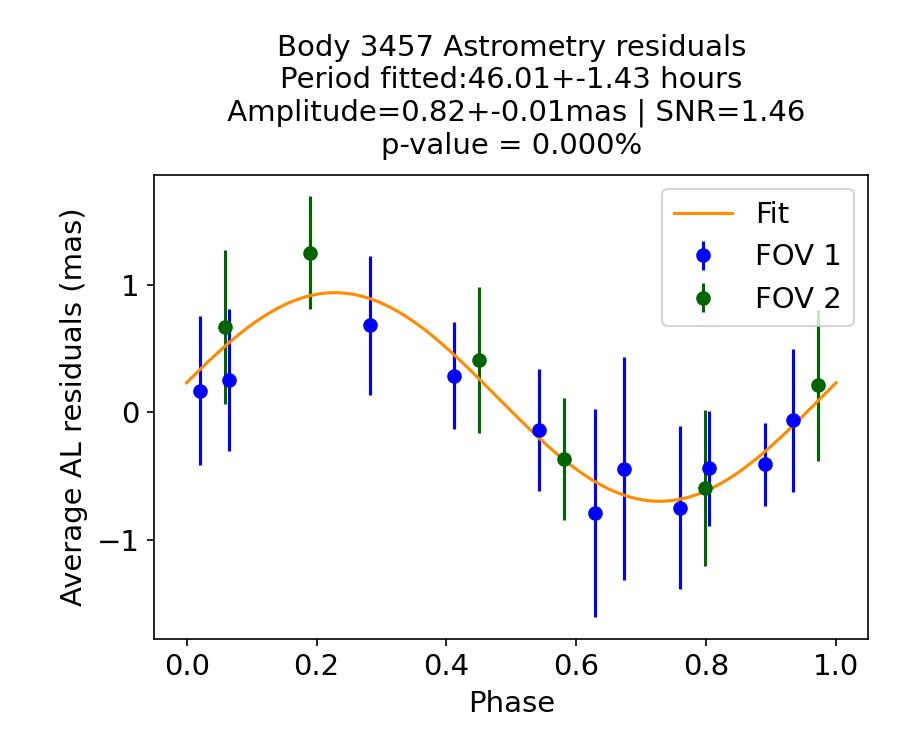}
    \\
    \includegraphics[width=0.48\textwidth]{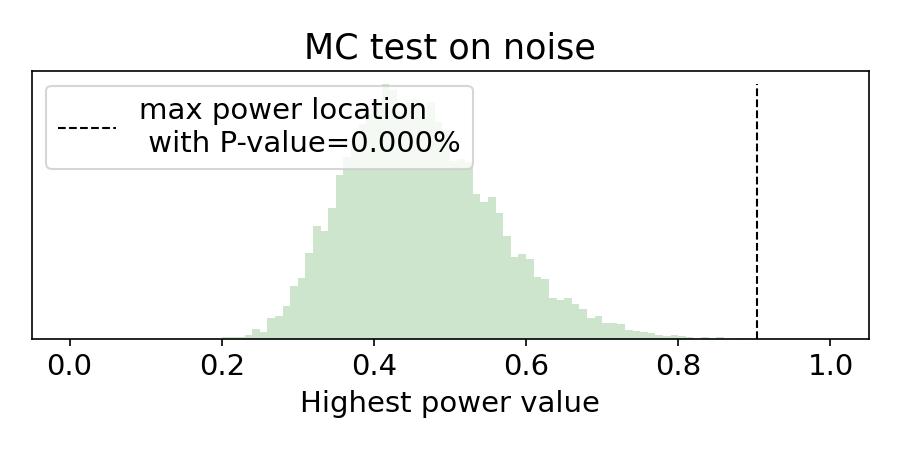}
    \includegraphics[width=0.48\textwidth, height=4.4cm]{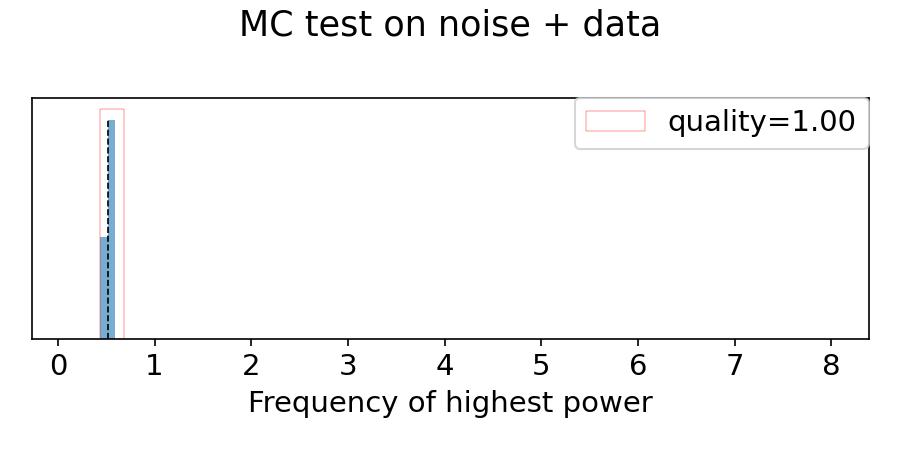}
    \caption{Visualisation of the output of the main steps of the selection process in the period search for one candidate example ((3457) Arnenordheim). The top left plot shows the periodogram from the fitting of the 17 observations in the window. The peak represents a period of 45.95h and the best fit is shown in the phased data in the top right plot. The distribution of data points not phased is shown in the right plot in Fig.~\ref{fig:windows}. The bottom left plot shows the empirical distribution of GLSP power $V$ (maximum value of the GLSP under $ {{\mathcal{H}}}_0) $ and the vertical dashed line shows the position of the value of $v$, the power of the peak from the periodogram with the real data, with $p(v) < 0.001$\%. Finally, the bottom right plot shows the distribution of best frequencies from the 5,000 MC simulations under $\widehat{{\mathcal{H}}}_1)$. The distribution tends to accumulate around the originally estimated frequency, showing that this detection is robust (here $Q$ is close to $1$.)}
    \label{fig:stat_valid_3457}
\end{figure*}

In this section, we apply the detection method to the AL residuals data set. Figure~\ref{fig:stat_valid_3457} visually illustrates the results of the processes of selection and parameters determination for the case of an object that satisfies the requirements (at least one window with ten observations in less than ten days, $p$-value$<$0.05, $Q>$0.5 and estimated wobble amplitude $\hat{A}$ smaller than $20~$mas) and was selected as a possible candidate by the period search. In this case, for (3457) Arnenordheim, we have a window with 17 observations spread over 2.75 days (as shown in Fig.~\ref{fig:windows}, right panel). The periodogram shows a large peak (Fig.~\ref{fig:stat_valid_3457} top left) that provides the best fit in the observations with a period of 45.95h (Fig.~\ref{fig:stat_valid_3457} top right). The MC analysis leads to a very low $p$-value ($<0.001\%$, Fig.~\ref{fig:stat_valid_3457} bottom left) and a quality factor $Q$ of one  (Fig.~\ref{fig:stat_valid_3457} bottom right).

\begin{figure*}[htbp]
    \centering
    \includegraphics[width=13cm]{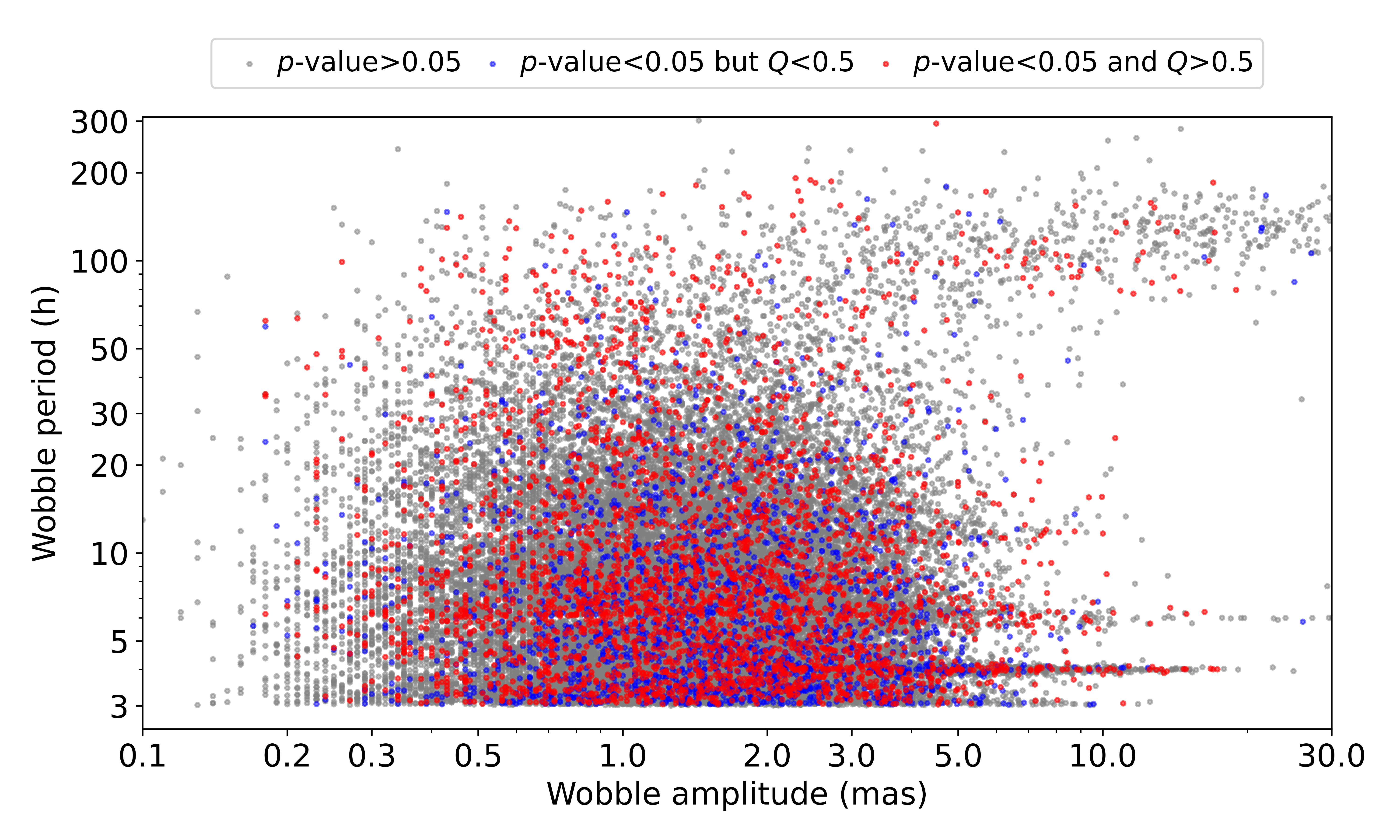}
    \caption{Distribution of estimated wobble amplitudes and periods obtained from the period search for all objects. In grey are the time series disqualified due to a $p$-value$>0.05$; in blue are the cases disqualified because $Q<0.5$, but with $p$-value within the limit of 0.05; and in red are the cases selected as binary candidates from the period search for complying with $p$-value$<0.05$ and $Q>0.5$.}
    \label{fig:hist_periods_wobs}
\end{figure*}

The distribution of estimated periods and wobble amplitudes $\hat{A}$ for the 30,030 candidates (Fig.~\ref{fig:hist_periods_wobs}) shows that small amplitudes are preferentially discarded, as expected since the signal-to-noise ratio tends to be small, but the general distribution remains overall the same. We notice a kind of ``tail'' in the distribution for the large period and amplitude values, and they are mostly not selected as detections. This phenomenon might be caused by unmodelled low-frequency trends (see Sec.~\ref{Ss:selection_statistic}).

It is also noticeable in Fig.~\ref{fig:hist_periods_wobs} a concentration of detections around 3, 4, 6, and 12 hours with larger amplitudes, even though there is no over-representation of such periods. That happens due to the typical frequencies present in the \gaia scanning law (time interval between the two FOVs, rotation period of 6h, precession of about 2 months, and combination of them). In some cases, the sampling of the data set has a dominant effect on the period search causing the GLSP to find the best period at, or very close to, values of 2, 4, 6, and 8 cycles/day. However, some of the detections found at such values might be real, so we leave at the next selection steps the task of cleaning the spurious detections that are not physically meaningful. In future improvements of the method, we intend to account for this effect. Finally, we note that the apparent vertical alignment of points for small values of $\hat{A}$ is due to the bin size in the determination of the mean value estimate, which is more noticeable due to the logarithmic scale of the plot.

We find that 83.8\% of the windows of data analysed lead to non-detections due to a $p$-value larger than the 5\% threshold and about 6.6\% are disqualified because $Q<0.5$ even though the $p$-value is within the limit. Eventually, only 9.6\% of the sets of AL residuals comply with the requirements and are selected as possible binary candidates by the statistical selection, which represents 3,038 asteroids, of which 40 have estimates from two (or more) different windows of observations.

\begin{figure}
    \centering
    \includegraphics[width=\linewidth]{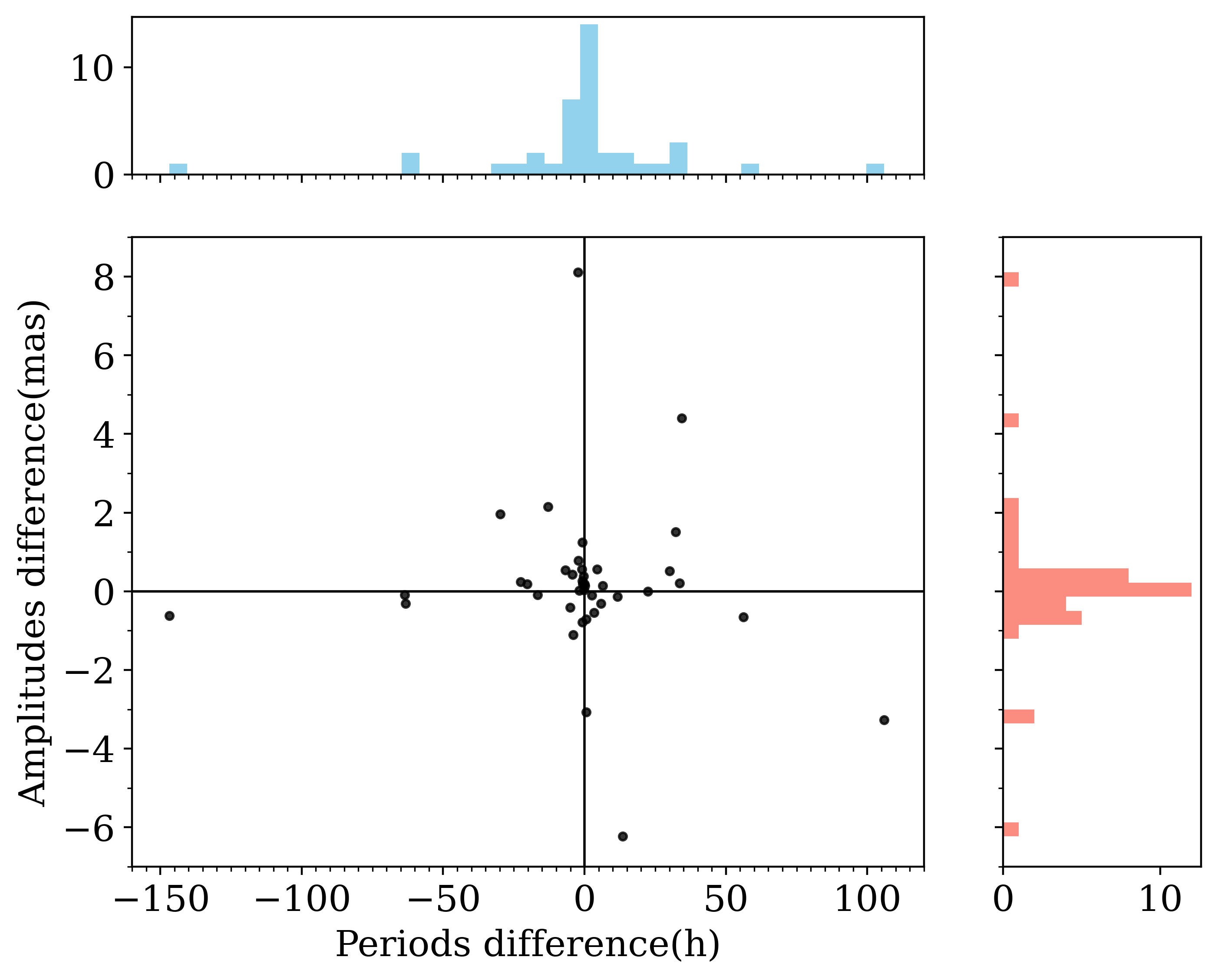}
    \caption{Distribution of differences between the solutions for the 40 objects with multiple windows of consecutive observations.}
    \label{fig:mult_window}
\end{figure}

Some results from different windows are very similar, but most of them show considerably divergent values.  The difference between the solutions is typically up to $\sim$35 h in the estimated period and up to $\sim$2 mas in amplitude as can be seen in Fig.~\ref{fig:mult_window}. At this point, we are not able to decide which window has the most physically meaningful results. Therefore, an asteroid that has been estimated with different periods and amplitudes in different windows at this stage will each have a pair of possible solutions (period, amplitude) considered in the next stage, where the least physically consistent results will be rejected.


\subsection{Selection based on physical properties}
\label{Ss:physical_selection}
For the 3,038 \gaia asteroids whose data passed the statistical detection test above, we move forward to check if the measured wobbling is compatible with the fundamental physical properties expected for an orbiting satellite. Under the hypothesis that the periodic signals correspond to the orbital period, we can exploit known properties measured by other surveys, such as the object size, to infer the range of size ratios and densities that could generate the observed signal. Density, in particular, should be within acceptable ranges, and compatible with the expected mineralogy for a corresponding spectral type.

We start with Kepler's third law of planetary motion
\begin{equation}
    \frac{T^{2}}{a^{3}}=\frac{4\pi^{2}}{M_{\rm total}G} \quad ,
    \label{eq:kepler}
\end{equation}
where $M_{\rm total}$ is the total mass of the studied system ($M_1+M_2$), $T$ is the orbital period of the secondary body around the primary, which we assume to be the same as the wobbling period, and $G$ is the gravitational constant. We manipulate the equation to express it as a function of the mass ratio of the system $q=\frac{M_2}{M_1}=\frac{D_2^3}{D_1^3}$, the diameters of the primary and secondary component ($D_1$ and $D_2$, respectively), and the bulk density $\rho$. 

Furthermore, we also consider that the currently observed diameter corresponds to a sphere of surface brightness equivalent to the sum of the binary components, $D^2=D_1^2+D_2^2$. Hence, the total mass of the system can be written as
\begin{equation}
    M_{\rm total}=\frac{\rho \pi}{6}\frac{D^3}{(1+q^{2/3})^{3/2}}(1+q)\quad .
    \label{eq:mass}
\end{equation}

Finally, substituting Eqs.~\eqref{eq:mass} and \eqref{eq:sep} into Eq.~\eqref{eq:kepler} and with some manipulations, we obtain a relation describing the density $\rho$ as a function of the mass ratio $q$:
\begin{equation}
\rho=\frac{24 \pi \alpha^3  }{G T^2 D^3}  \left|  \frac{(1 + q^\frac{2}{3})^\frac{9}{2} (1 + q)^2}
{q^2(q^\frac{1}{3} -1)^3 } \right| \quad , \quad 0< q < 1\quad .
\label{eq:density}
\end{equation}

By this expression, the dependency of the density from the mass ratio q can be evaluated, by assuming an estimation of D, once $T$ and $\alpha$ are measured from the astrometric wobbling. We retrieve $D$ from literature using the SsODNet service\footnote{https://ssp.imcce.fr/webservices/ssodnet, queried through its rocks python-interface (https://github.com/maxmahlke/rocks) \label{foot_ssodnet}.} \citep{berthier2022ssodnet}, along with the associated uncertainty. If SsODNet does not
provide the diameter for a given
object, we estimate it from the absolute magnitude $H$ by taking the mean diameter for the geometric albedos $p_v$ spanning the interval [0.06,~0.2]. We assume a standard deviation of 15\% for convenience, considering that without this assumption the errors in the diameter estimates are larger than 50\%. This restrictive choice is partially compensated by the generous density interval that is allowed, as explained in the following. 

While $T$ is a direct result of our signal analysis, the measured amplitude $\alpha$ is only a minimum value, due to the unknown projection factor of the real wobble in the AL direction. In this respect then, having fixed all other parameters, Eq.~\eqref{eq:density} provides a {\sl minimum} density as a function of $q$. 

\begin{figure*}[htbp]
    \centering
    \includegraphics[width=0.495\textwidth]{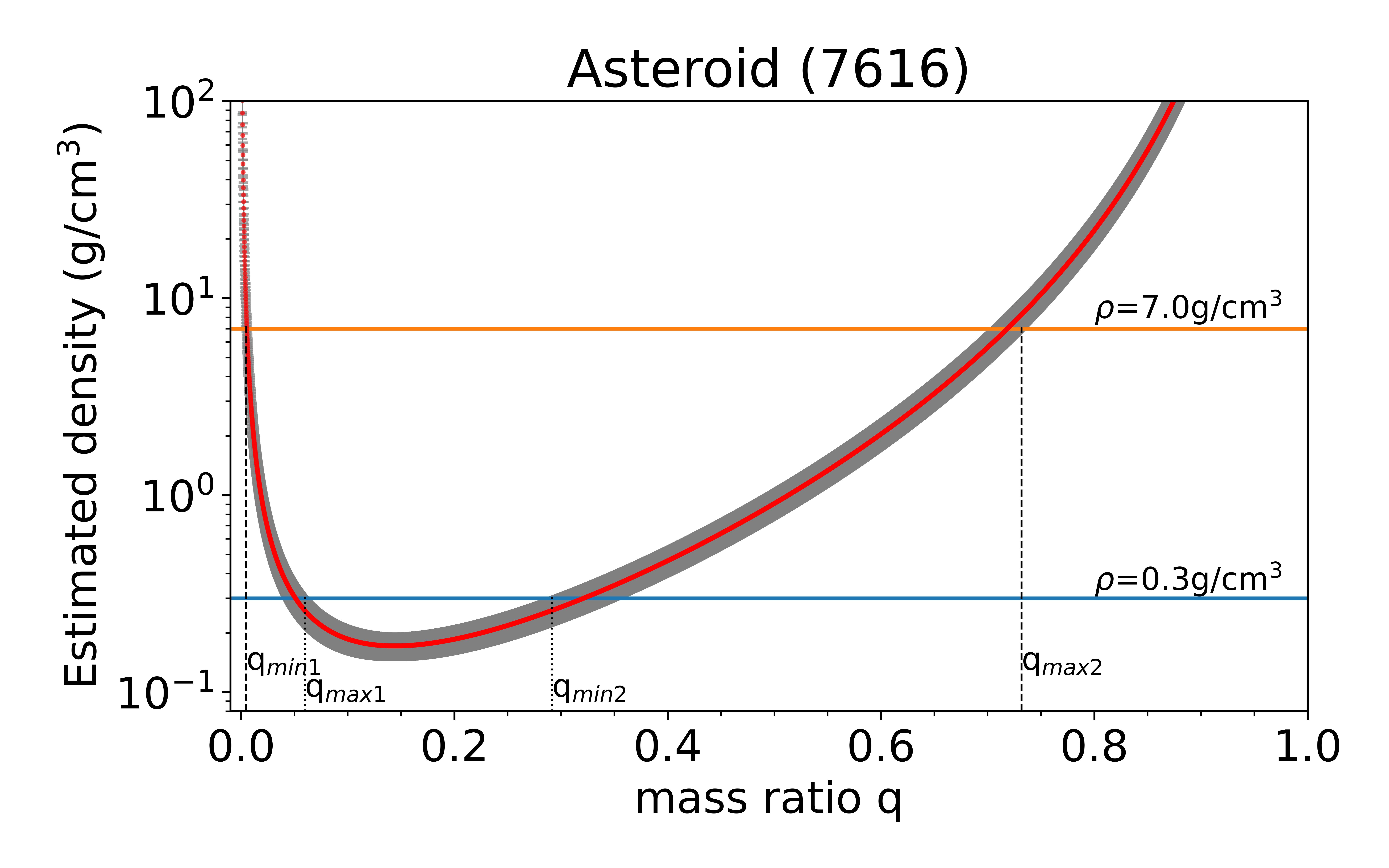}
    \includegraphics[width=0.495\textwidth]{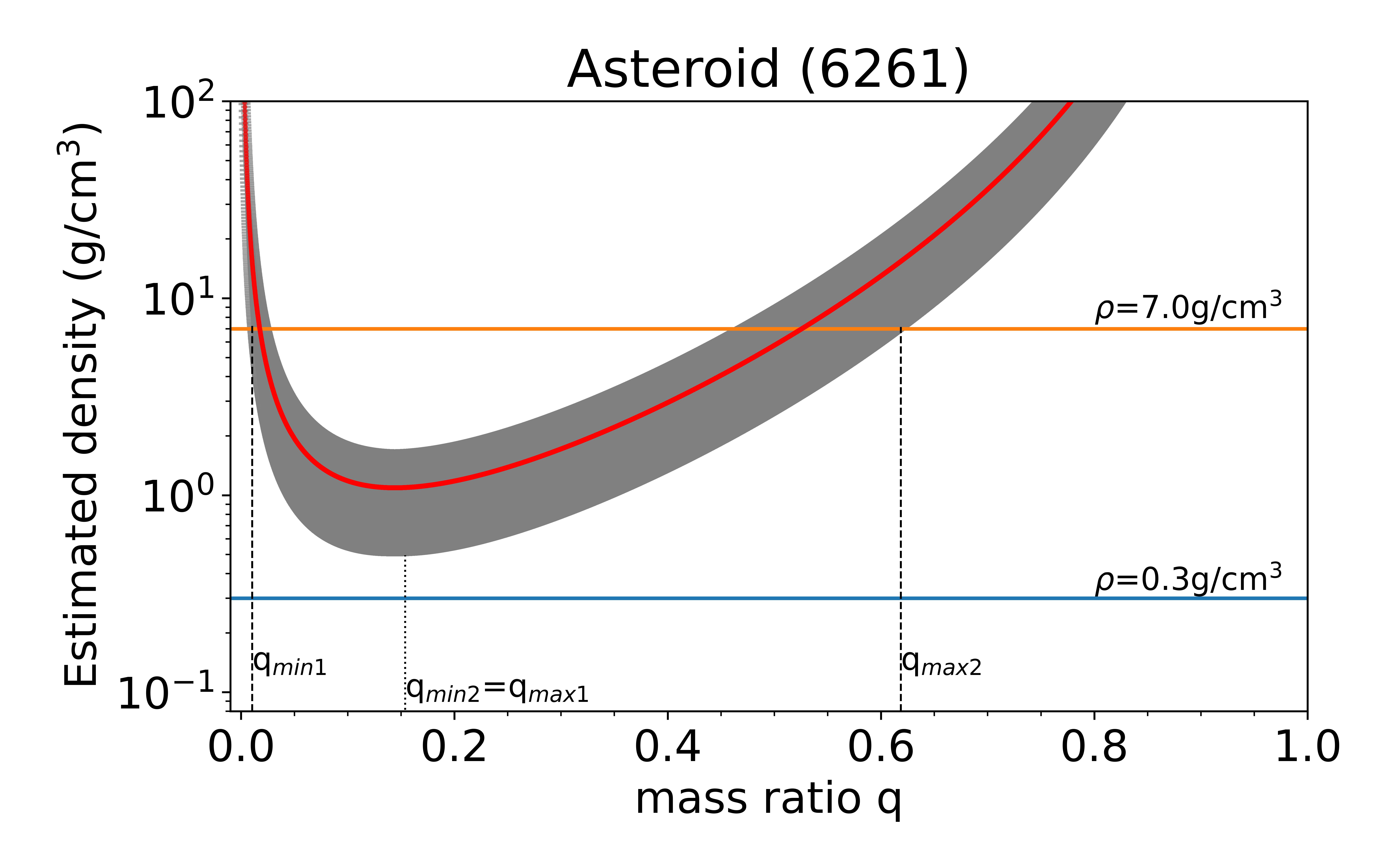}
    \includegraphics[width=0.495\textwidth]{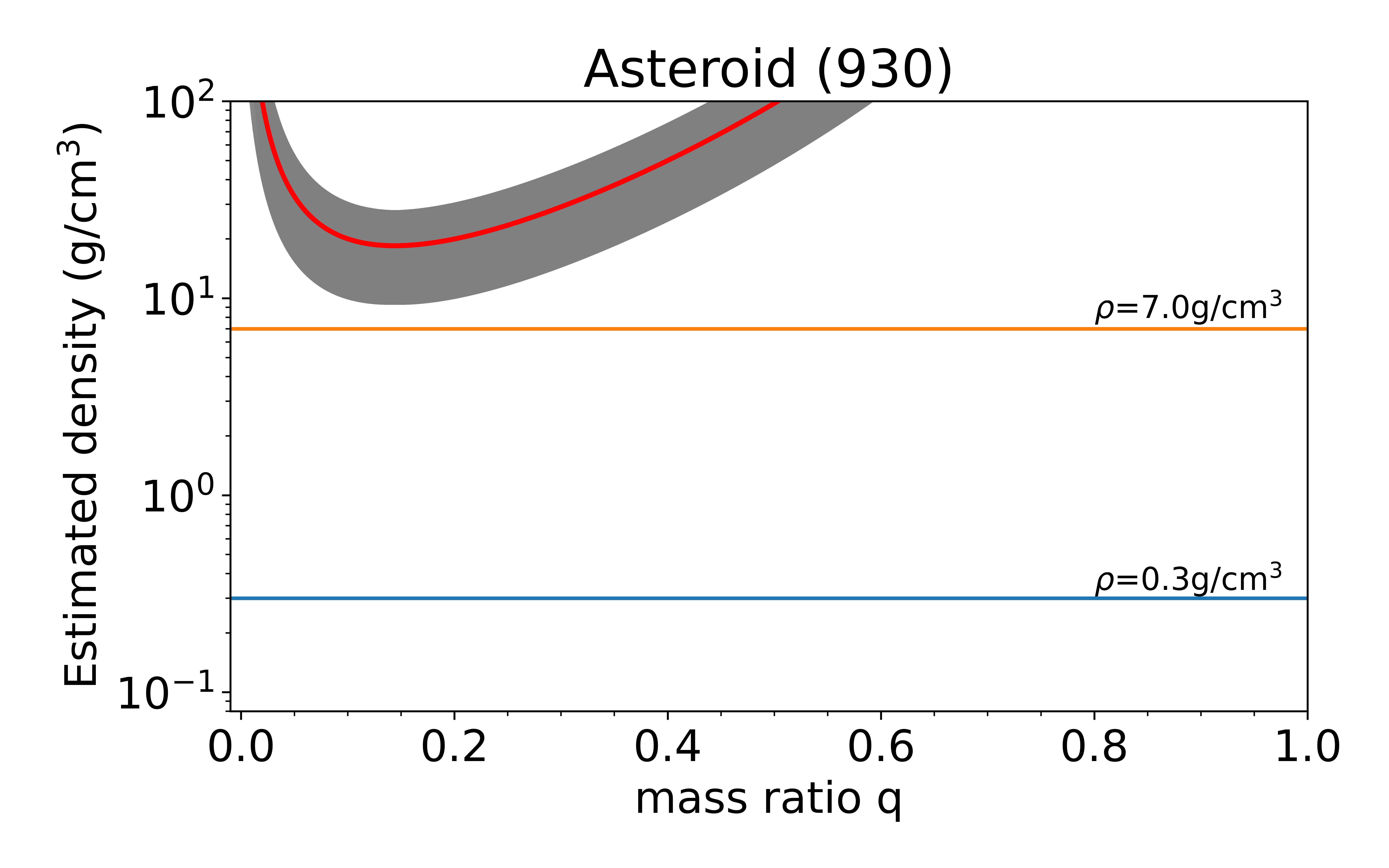}
    \caption{Examples of the bulk density $\rho$ as a function of the mass ratio $q$ for two asteroids, (7616) Sadako (top left) and (6261) Chione (top right). The red line indicates the nominal value. Shown in grey is the error range on the density. The blue and orange horizontal lines represent physically reasonable limits. $q_{min1}$, $q_{max1}$, $q_{min2}$ and $q_{max2}$ mark the corresponding extremes of the mass ratio values. The bottom plot shows the case for (930) Westphalia, eliminated in the density filtering because the minimum estimated density is higher than the chosen range of bulk density.}
    \label{fig:densities_profile}
\end{figure*}

The Eq.~\eqref{eq:density}, as shown in Fig.~\ref{fig:densities_profile}, is a concave function with a minimum corresponding to the maximum wobbling amplitude in Eq.~\eqref{eq:sep}. This minimum represents the smallest $\rho$ of the system compatible with the observed wobbling. The uncertainties are obtained with the linear propagation of the errors in $D$, $T$ and $\alpha$.

For a first selection, we consider a large density range $\rho \in [0.3,7]$~g/cm$^3$. 
We are interested in selecting the possible ranges of mass ratio $q$ corresponding to densities in this interval. Therefore, when the minimum of the density function is $<$0.3~g/cm$^3$, we obtain two separated intervals delimited by 4 different values of mass ratio ($(q_{min1},q_{max1})$ and $(q_{min2},q_{max2})$). From them, and the associated values of $\rho$, the corresponding values of the total mass can be computed and used in Eq.~\ref{eq:kepler} to derive two different intervals of possible separation of the binary ($(s_{max1},s_{min1})$ and $(s_{min2},s_{max2})$). This is the case of (7616) Sadako in Fig.~\ref{fig:densities_profile}. 

In cases where the minimum density estimate is never smaller than the inferior limit, as for (6261) Chione, we get a single interval of $q$ and possible separations. Eventually, if the function has a minimum value larger than the adopted upper threshold, the wobble signature is considered not compatible with a binary object as this would require an excessive density. By adopting the density threshold mentioned above, 934 objects are selected and their possible ranges of $q$ and $s$ have been computed.

A further step consists in considering the smallest separation, for each candidate,  compatible with an orbiting companion. We find that for some of the objects, the range of separations partially overlaps with the Roche limit,\footnote{$d_{Roche}=2.44~R_{primary}(\rho_{primary}/\rho_{secondary})^{1/3}$} computed by assuming the same density for the two components. Binary systems inside the Roche limit are unlikely to exist, and the detected wobble could instead be caused by a very irregular shape. Of course, more accurate physical data, in particular diameter measurements, could change the results for some objects and put them beyond the Roche limit. To be conservative, for this first search, we have decided to reject candidates that have separation intervals overlapping, even partially, with the Roche limit. With this selection, we remain with 362 wobble measurements for \allcand objects.

 As an attempt for a further verification, we considered the largest selected objects in the table having a known shape in the DAMIT\footnote{https://astro.troja.mff.cuni.cz/projects/damit/} database \citep{DAMIT2010A&A...513A..46D} to compute the distance between photocentre and centre of mass, namely for the asteroids (476) Hedwig \citep{marciniak2023scaling}, (487) Venetia \citep{marciniak2018}, (516) Amherstia \citep{durech2009,hanus2011}, (532) Herculina \citep{kaasalainen2002,hanus2017}, (542) Susanna \citep{hanus2021}, (556) Phyllis \citep{marciniak2007}, (699) Hela \citep{marciniak2012}, (1200) Imperatrix \citep{durech2020}, (1639) Bower \citep{durech2019inversion}, (2219) Mannucci \citep{durech2020}, (70045) 1999 DA5 \citep{durech2023DR3photometry}. We simulated the photocentre shift due to the illuminated shape alone, at all epochs used for our period search, by using Lambert's emmition and Lommel-Seeliger's scattering laws \citep{lambert1760photometria,pedrotti2017introduction,Fairbairn2005} and the corresponding illumination geometries. Then, its projection on the AL direction was used to look for correlations with the observed residuals. This test reproduces the one reported in \citet{tanga2022gaia} for (21)~Lutetia.

\begin{figure*}
        \centering
 \includegraphics[width=0.32\linewidth]{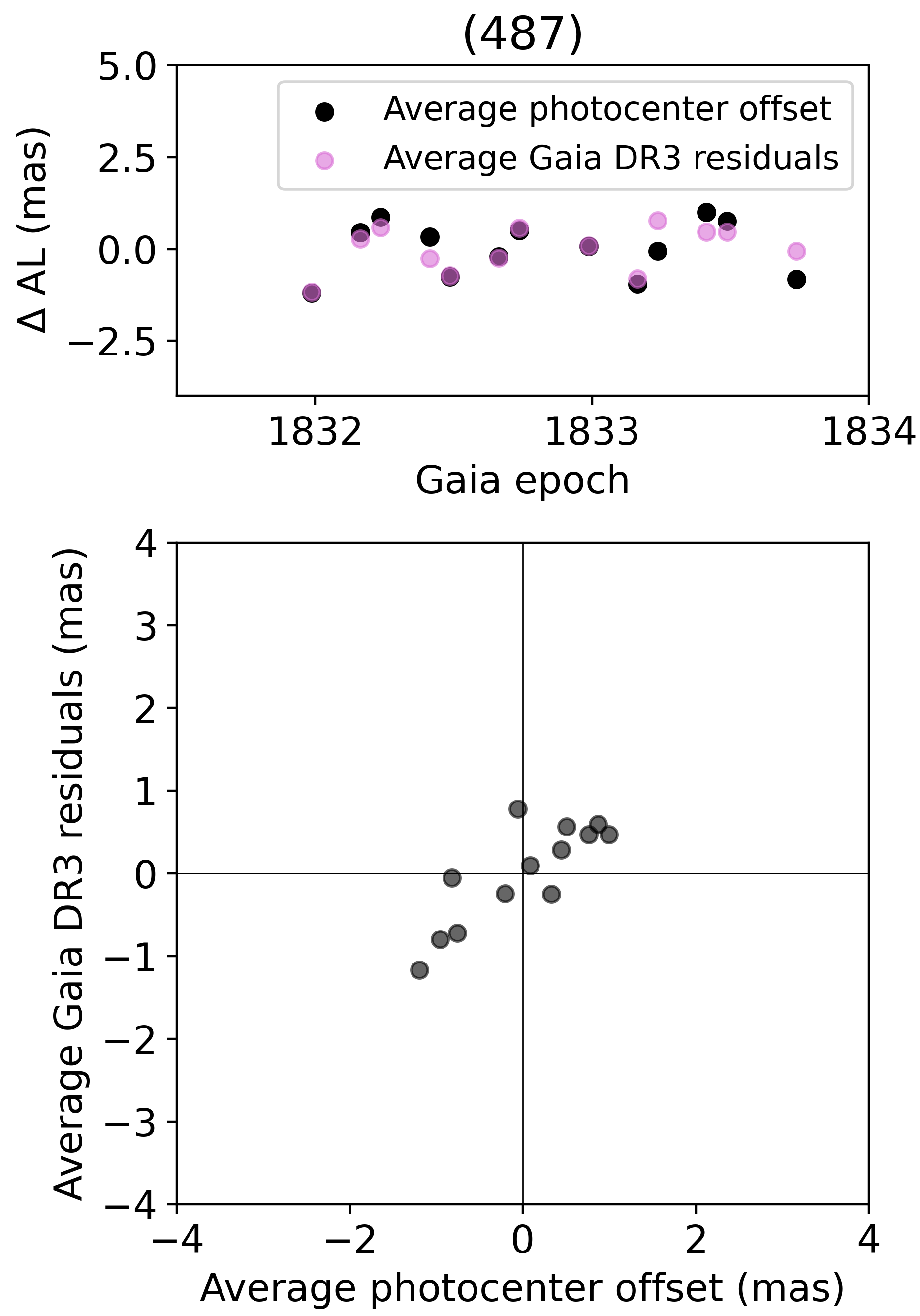}
   \includegraphics[width=0.32\linewidth]{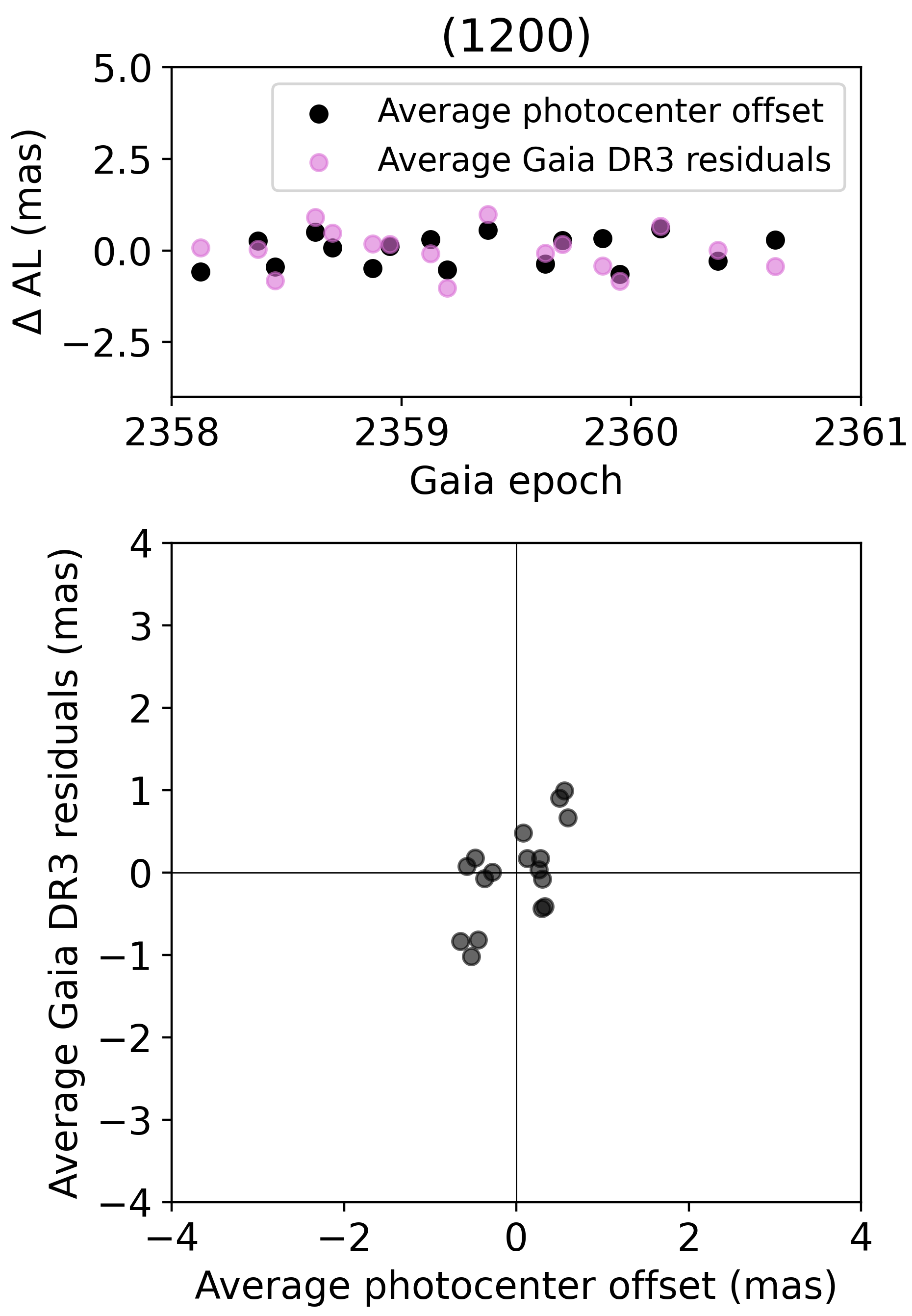}
   \includegraphics[width=0.32\linewidth]{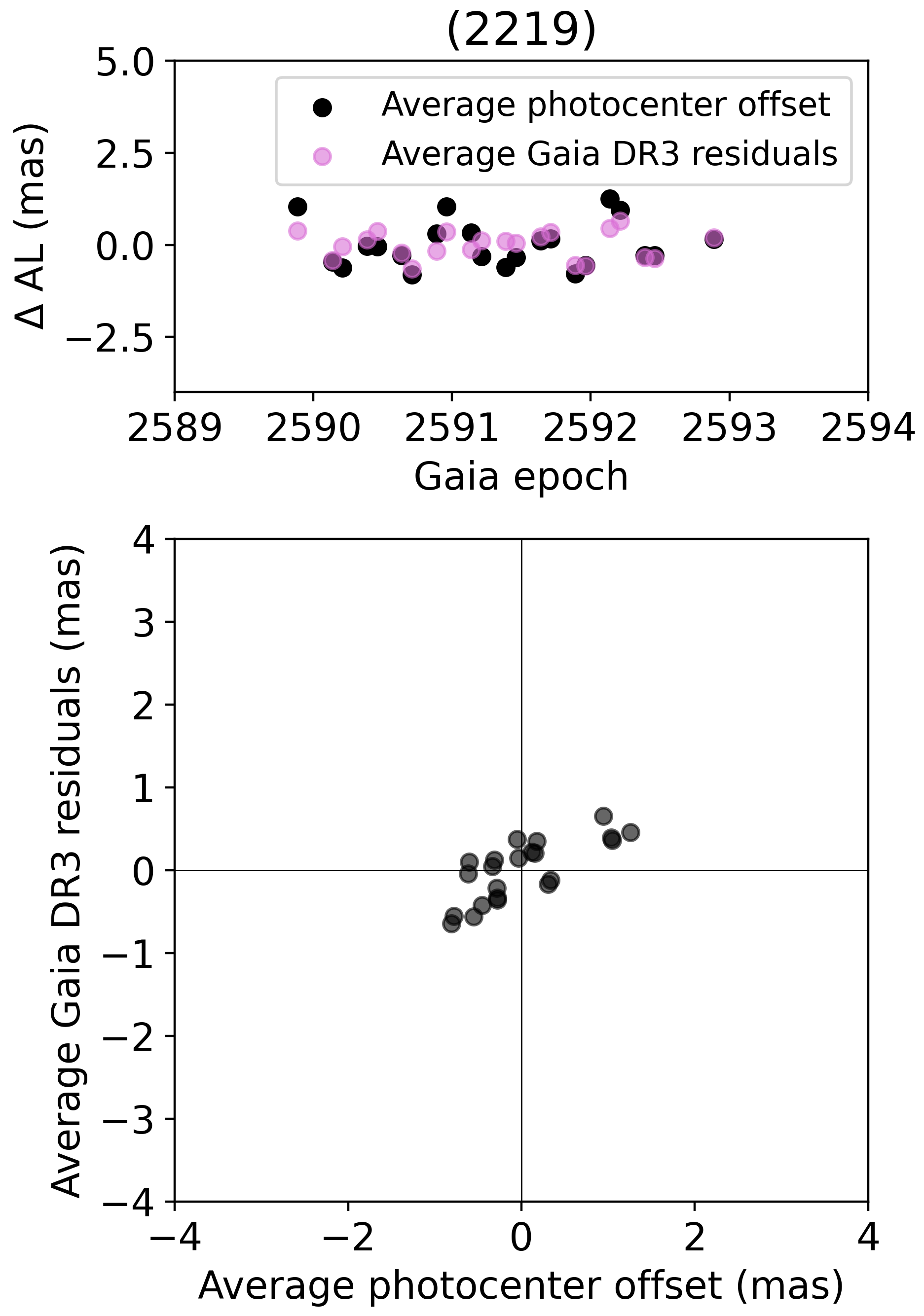}

        \caption{Comparison between the average DR3 astrometric residuals and estimated average photocentre offset for candidates (487) Venetia, (1200) Imperatrix and (2219) Mannucci. The top plots show the variation of the data over time in the along-scan direction, while the bottom plots show the correlation between the two sets of data.}
        \label{fig:photocenter_offset}
    \end{figure*}

We repeated the test with the multiple pole and shape solutions available for each object, also checking the possible role of a rotation phase shift concerning the nominal rotation origin. Our results show that the amplitude of the photocentre displacement could be correlated to the residuals for (487) Venetia, (1200) Imperatrix (for one of the two pole solutions available) and (2219) Mannucci (in one of the two observation windows that we consider), as shown in Fig. \ref{fig:photocenter_offset}. It is also anti-correlated for (532) Herculina and (556) Phyllis. In the other cases, no correlation shows up.

This preliminary test is very sensitive to several details such as shape and pole coordinate errors, scattering law, presence of concavities, etc, and should be taken with caution. However, we notice that the order of magnitude of the wobble amplitude due to a (single) shape would be about the same as the signature produced by a satellite for large asteroids (separations of $\sim$200~mas or more, spatially resolved by \gaia). However, we notice that the amplitude of the photocentre shift is not dissimilar from our residuals. In a synchronous system (primary rotation equal to secondary orbital period) they could even add up with the same or opposite sign, depending on the phase angle of illumination. Only in this case, when the sign is opposite, this could result in the observed anti-correlation which would then be evidence supporting the satellite's presence. 

All these considerations point in the direction of not excluding the largest objects remaining on our list. They certainly deserve more accurate characterisation by future observations and more extensive, complex modelling efforts when more data becomes available.


\subsection{Constraints from taxonomy}

The average density value for each taxonomic class can help to further constrain the physical properties of objects in our selection. We retrieved the taxonomic class for all objects in the list of preliminary candidates using SsODNet. Of the \allcand candidates selected from the previous steps, 264 have their classes determined, corresponding to one of the complexes for which an average density is provided in \cite{carry2012density}. 
 
By using the average density for the taxonomic class, we can repeat the analysis illustrated in Sect.\ref{Ss:physical_selection}, recompute the permitted intervals of possible physical parameters, and refine the selection of the sample. We reached a total of \besttaxocand candidates. In addition, since large real wobbles tend to produce high $\widehat{S/N}$ (cf Sect.~\ref{Ss: validation}) we flag as best candidates those with $\widehat{S/N}>1$.

\begin{figure*}
  \begin{minipage}[c]{0.75\textwidth}
    \includegraphics[width=\linewidth]{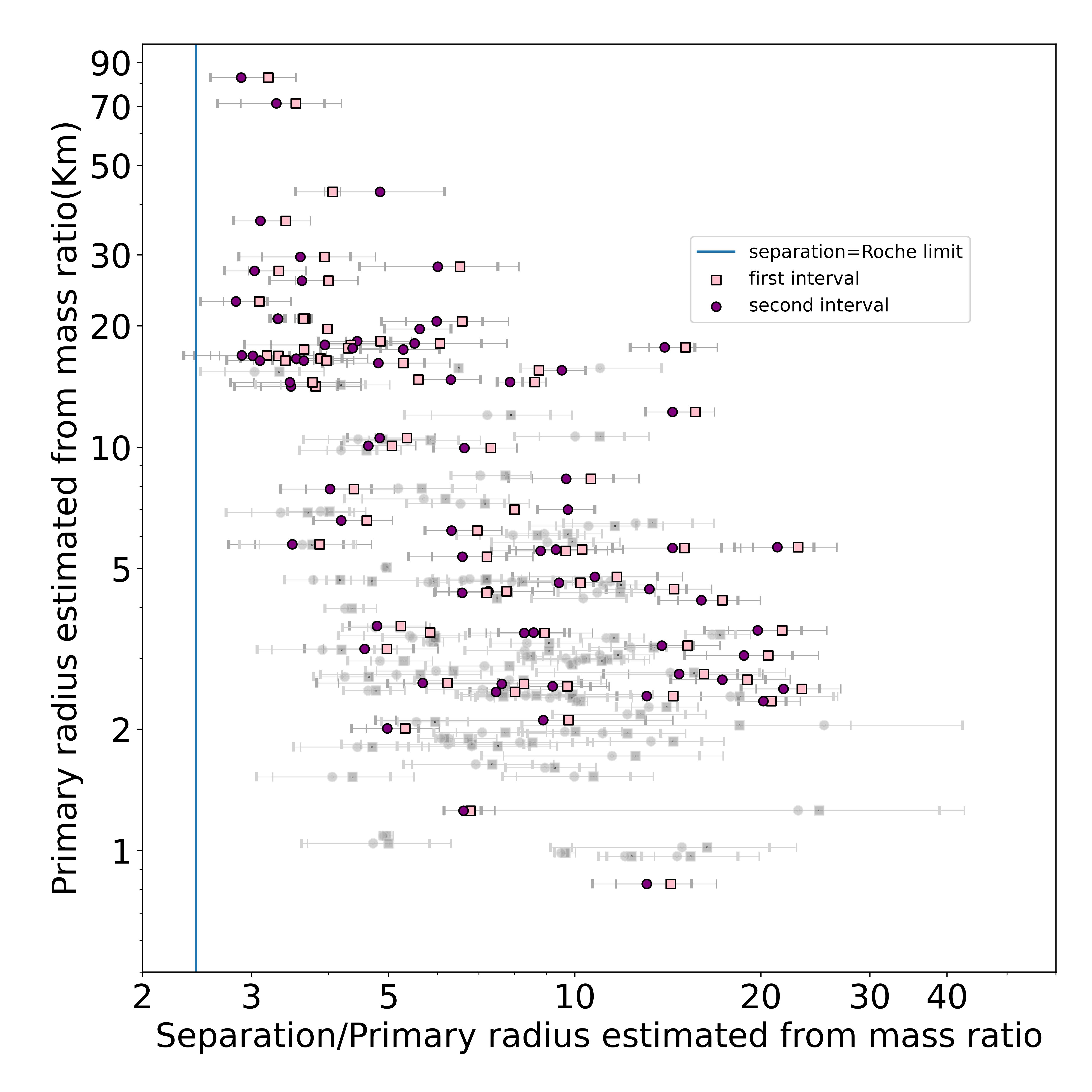}
  \end{minipage}\hfill
  \begin{minipage}[c]{0.25\textwidth}
    \caption{Plot of the radius of the primary as a function of the separation intervals normalised with the Roche limit for fluid satellites, for the sample of \besttaxocand asteroids with known taxonomy. The squares represent the mean value of the first mass ratio interval estimated and the circles represent the second interval. The coloured symbols with dark borders are the objects from the taxonomic filtering with $\widehat{S/N}>1$, and the grey symbols are those with $\widehat{S/N}<1$.  } \label{fig:separation_roche}

  \end{minipage}
\end{figure*}

The objects of our sample on which taxonomy-derived density constraints are applied are presented in Fig.~\ref{fig:separation_roche}, which shows the estimated radius of the primary as a function of the allowed separations (in units of the radius of the primary) for the \besttaxocand objects. The purple and pink points represent the first and second intervals of possible mass ratio values. Those with dark borders are the best candidates with $\widehat{S/N}>1$, and the points with grey borders are the candidates with $\widehat{S/N}<1$ that are not in the list of the best cases but still passed through all the filtering steps.


\section{The final sample of candidate binary asteroids}
\label{S:final}

The total outcome of \allcand objects from this first search of astrometric binary candidates appears in the full version of Table~\ref{tab:all_candidates} in CDS. It includes 6 known binaries, \bestcand of which (highlighted) present results with the density constrained by taxonomy and $\widehat{S/N}>1$.

\begin{table*}[ht]
\caption{List of estimated parameters for the \gaia astrometric asteroid binary candidates.}
\centering\setlength{\aboverulesep}{0pt}\setlength{\belowrulesep}{0pt}

\setlength{\extrarowheight}{2pt}

{\footnotesize
    \centering
    \begin{tabular}{p{0.04\textwidth} l l l l l l c c} \toprule
\bf{ID} &\makecell[b]{\textbf{Wobble}\\ \textbf{period(h)}} & \makecell[b]{\textbf{Wobble} \\\textbf{amplitude}\\\textbf{(mas)}} & \makecell[b]{\textbf{Separation (km)}\\\textbf{1st interval}} & \makecell[b]{\textbf{Mass ratio}\\\textbf{1st interval}} & \makecell[b]{\textbf{Separation (km)}\\\textbf{2nd interval}}&\makecell[b]{\textbf{Mass ratio}\\\textbf{2nd interval}} & \makecell[b]{\textbf{Taxonomy}\\\textbf{correction}} & $\mathbf{\widehat{S/N}>1}$\\\bottomrule

\multicolumn{9}{c}{\textsc{Known binaries}} \\
  317  &  127.56$\pm$64.57  &  2.38$\pm$1.69  &  (5.08$\pm$4.79)e+02  &  (1.21$\pm$1.11)e-02  &  (5.27$\pm$5.27)e+04  &  (7.36$\pm$2.63)e-01  &    &  *\\
  \cellcolor{orange!25}   1509  & \cellcolor{magenta!25} $>$ 84.06  &\cellcolor{magenta!25}--&\cellcolor{magenta!25}--&\cellcolor{magenta!25}--&\cellcolor{magenta!25}--&\cellcolor{magenta!25}--&\cellcolor{magenta!25}--&\cellcolor{magenta!25}--\\
  \rowcolor{gray!25}  2871  & 20.69$\pm$4.63  &  0.45$\pm$0.07  &  (1.56$\pm$0.18)e+01  &  (1.18$\pm$0.25)e-02  &  (1.43$\pm$0.16)e+01  &  (6.00$\pm$0.38)e-01  &  *  &  \\  

 \rowcolor{gray!25}  4337  &  35.93$\pm$6.36  &  0.61$\pm$0.07  &  (6.57$\pm$0.66)e+01  &  (2.40$\pm$0.40)e-03  &  (5.99$\pm$0.62)e+01  &  (8.18$\pm$0.17)e-01  &  *  &  \\
 
\cellcolor{orange!25}  5817  & \cellcolor{magenta!25}  $>$ 144.11  &\cellcolor{magenta!25}--&\cellcolor{magenta!25}--&\cellcolor{magenta!25}--&\cellcolor{magenta!25}--&\cellcolor{magenta!25}--&\cellcolor{magenta!25}--&\cellcolor{magenta!25}--\\
\rowcolor{magenta!25}  18301  &  139.82$\pm$70.02  &  3.10$\pm$1.92  &  (7.61$\pm$1.38)e+01  &  (9.78$\pm$3.19)e-03  &  (6.95$\pm$1.24)e+01  &  (6.35$\pm$0.56)e-01  &  *  &  *\\
\multicolumn{9}{c}{\textsc{Objects with two solutions}} \\

  \rowcolor{magenta!25}   &  10.02$\pm$0.02  &  0.74$\pm$0  &   &  &   &   &  &  \\
\rowcolor{magenta!25}  542  &  10.04$\pm$0.04  &  0.56$\pm$0.01  &  (7.64$\pm$0.20)e+01  &  (2.30$\pm$0.10)e-03  &  (6.90$\pm$0.19)e+01  &  (8.20$\pm$0.05)e-01  &  *  &  *\\
\rowcolor{magenta!25} &  14.37$\pm$0.19  &  0.48$\pm$0.01  &  & &  &   &    &  \\
\rowcolor{magenta!25}  2219  &  13.56$\pm$0.36  &  0.74$\pm$0.06  &  (5.59$\pm$1.10)e+01  &  (2.50$\pm$0.80)e-03  &  (5.08$\pm$0.99)e+01  &  (8.16$\pm$0.33)e-01  &  *  &  *\\
\rowcolor{gray!25}   &  63.47$\pm$13.54  &  0.68$\pm$0.05  &   &  &  &  &    &  \\
\rowcolor{gray!25}  9573  &  57.59$\pm$3.56  &  0.99$\pm$0.05  &  (3.37$\pm$0.57)e+01  &  (1.08$\pm$0.33)e-02  &  (3.08$\pm$0.50)e+01  &  (6.18$\pm$0.53)e-01  &  *  &  \\
\rowcolor{gray!25}  &  10.33$\pm$0.13  &  0.56$\pm$0.05  &  &  &   &  &    &  \\
\rowcolor{gray!25}  10766  &  73.57$\pm$9.51  &  0.87$\pm$0.55  &  (1.17$\pm$0.24)e+02  &  (4.99$\pm$1.70)e-03  &  (1.07$\pm$0.22)e+02  &  (7.36$\pm$0.48)e-01  &  *  &  \\

\multicolumn{9}{c}{\textsc{Objects with one solution}} \\
  53  &  9.08$\pm$0.01  &  1.42$\pm$0.01  &  (2.39$\pm$0.96)e+02  &  (2.50$\pm$1.50)e-03  &  (2.54$\pm$1.25)e+02  &  (8.34$\pm$0.76)e-01  &    &  *\\
  203  &  23.89$\pm$0.27  &  0.88$\pm$0.01  &  (2.32$\pm$0.03)e+02  &  (1.00$\pm$0)e-03  &  (5.94$\pm$3.71)e+02  &  (9.35$\pm$0.39)e-01  &    &  *\\
  217  &  12.70$\pm$0.04  &  0.88$\pm$0.02  &  (1.84$\pm$0.82)e+02  &  (2.75$\pm$1.75)e-03  &  (2.19$\pm$1.25)e+02  &  (8.33$\pm$0.88)e-01  &    &  *\\
  238  &  9.85$\pm$0.01  &  2.23$\pm$0.02  &  (3.23$\pm$1.07)e+02  &  (2.00$\pm$1.00)e-03  &  (3.92$\pm$1.96)e+02  &  (8.59$\pm$0.66)e-01  &    &  *\\
  247  &  12.10$\pm$0.06  &  0.55$\pm$0.02  &  (1.46$\pm$0.05)e+02  &  (1.00$\pm$0)e-03  &  (4.41$\pm$2.38)e+02  &  (9.53$\pm$0.24)e-01  &    &  *\\
    \vdots& \vdots  & \vdots &  \vdots & \vdots &\vdots & \vdots &  \vdots & \vdots\\

  \end{tabular}\label{tab:all_candidates}
}
  \begin{minipage}{0.95\linewidth}
    {\small Notes: The grey lines represent the objects with density constrained by taxonomy, and the magenta lines indicate the best candidates accounting for the taxonomic correction and a strong detection with $\widehat{S/N}>1$. The objects with orange ID represent those that were selected as binary candidates but with periods larger than the length of the observation window, therefore the period given as the minimum is the length of the observation window used. This is an extract of the full table, which is available at the CDS.}
\end{minipage}
  \end{table*}

\begin{figure}[htpb]
    \centering
    \includegraphics[width=0.9\linewidth]{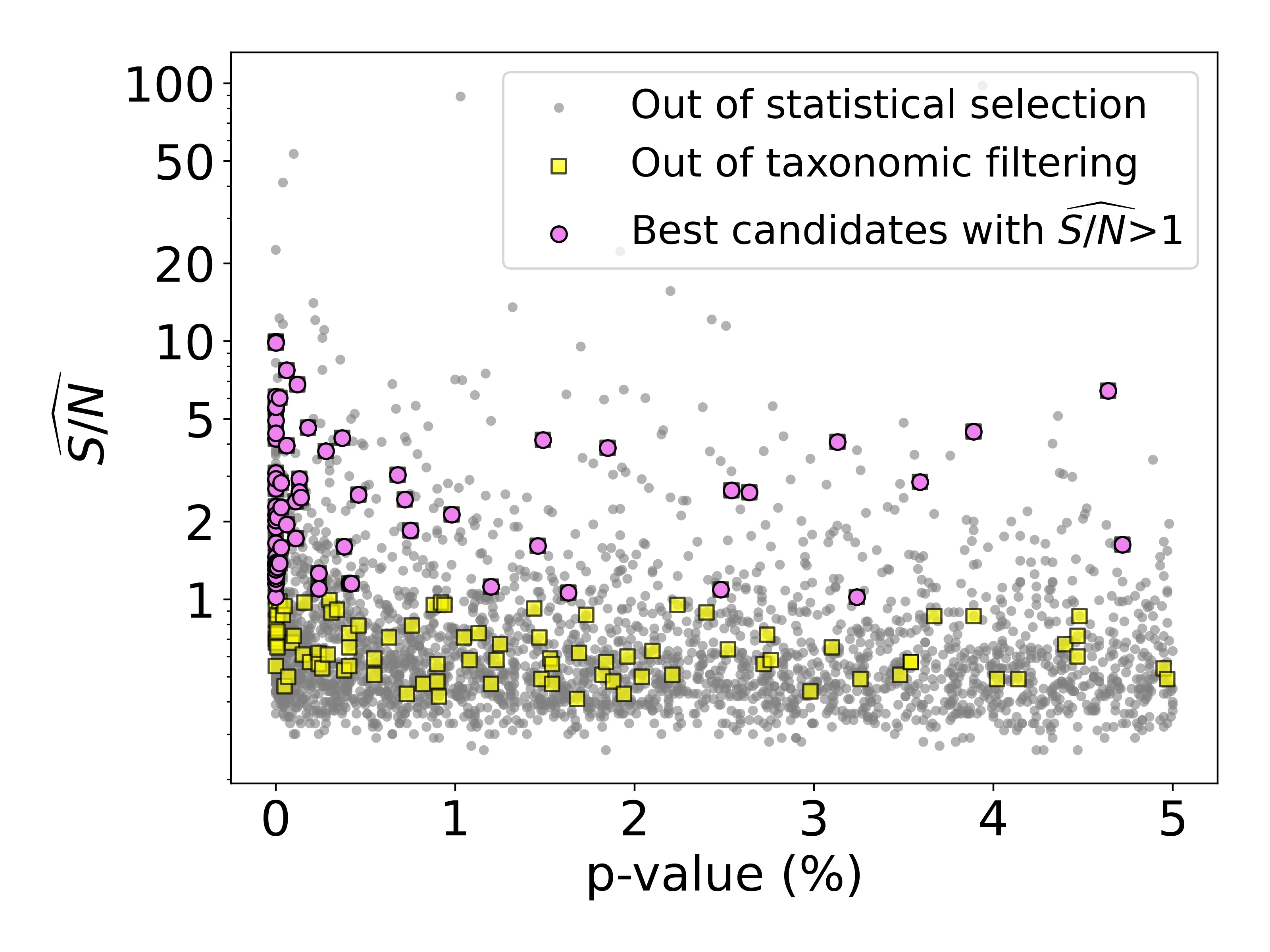}\\
    \includegraphics[width=0.9\linewidth]{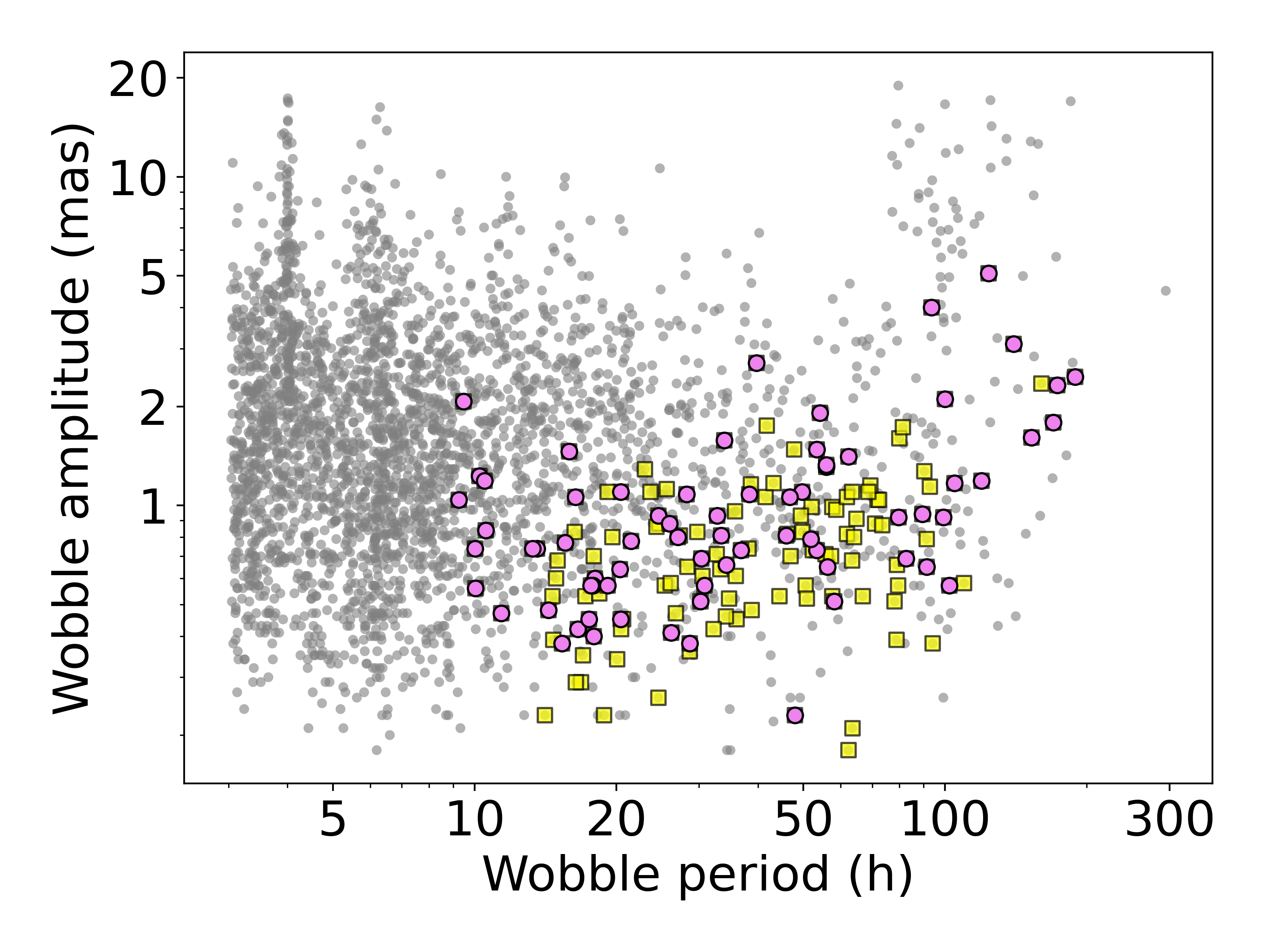}\\
    \includegraphics[width=0.9\linewidth]{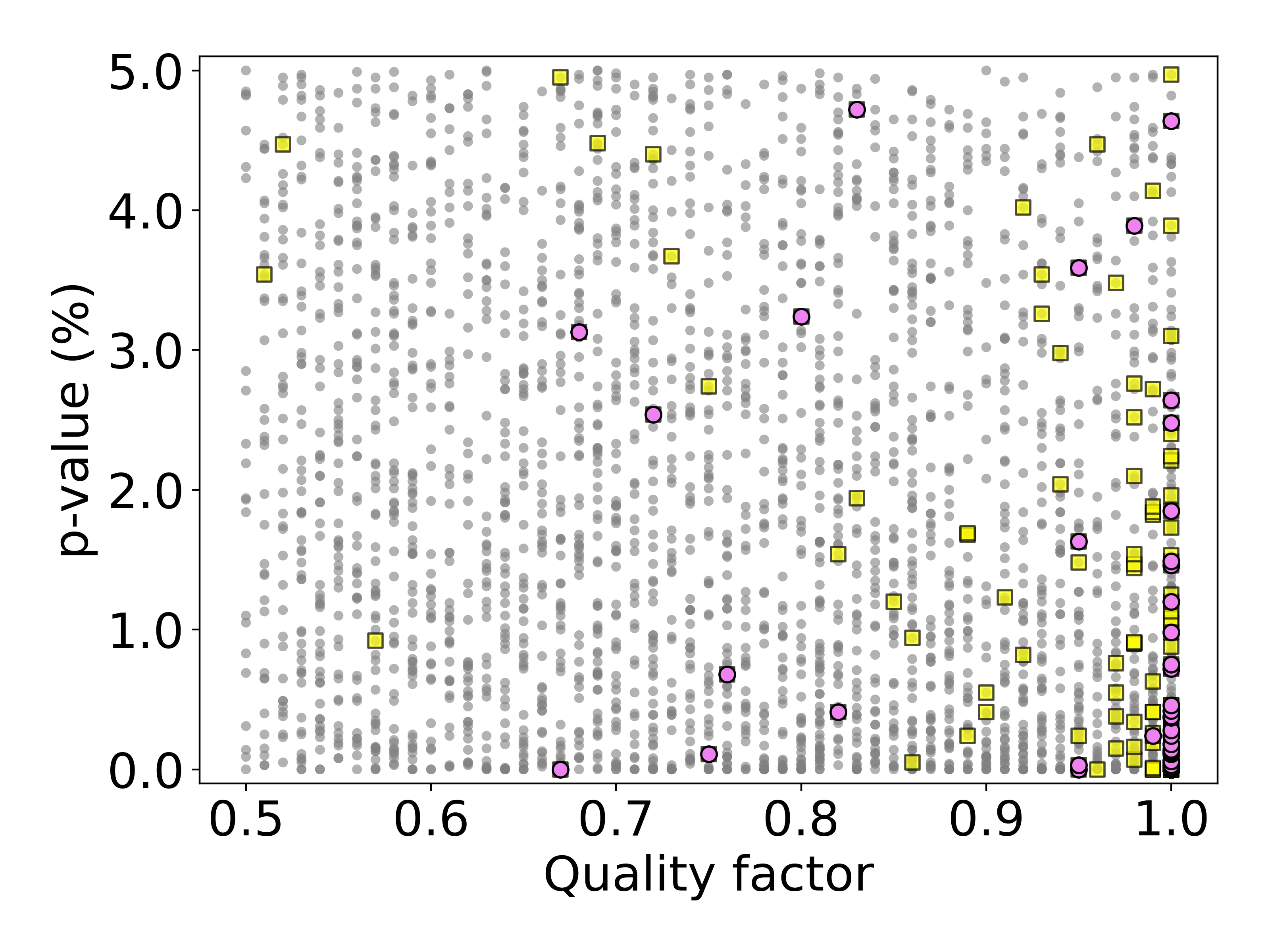}
    \caption{Distribution of $\widehat{S/N}$ vs $p$-value (top); estimated wobble period vs wobble amplitude (center); and $p$-value vs quality factor (bottom) for the 3,038 objects obtained from the statistical selection (in grey), the \besttaxocand objects that survived the taxonomic filtering (in yellow), and the \bestcand best candidates that passed all tests and have $\widehat{S/N}>1$ (in pink).}
    \label{fig:final_dist_cand}
\end{figure}

Table \ref{tab:candidates_numbers} summarises the number of candidates after each step of our selection process. We select the 30,030 asteroids in DR3 with consecutive observations in a short time. From the statistical selection, we obtained 3,038 preliminary candidates that presented $p$-value$<0.05$ and $Q>0.5$. The validation on the physical ground and the selection based on known taxonomic type further restricted this list to \besttaxocand objects presenting characteristics compatible with binary systems. Most of them have an estimated $\widehat{S/N}<1$, as the regime of the scarceness of the data and potentially low wobble amplitudes make detectability difficult. Therefore, we can consider that those with $\widehat{S/N}>1$ among the \besttaxocand, are the best characterised. The distribution of their parameters concerning the statistical thresholds of period determination is shown in Fig.~\ref{fig:final_dist_cand}.

 \begin{table*}[htpb]
    \centering
    \caption{Number of binary candidates remaining after each stage of the systematic search.}
    
    \begin{tabular}{c|c}
        & Number of candidates\\\hline
       Asteroids in DR3  &  156,801\\
       Candidates with a reasonably long sequence of observations  & 30,030\\
       Candidates' time series with detection from the statistical selection & 3,038\\
       Candidates' time series with physically meaningful density estimate & 934\\
       Candidates with separation beyond the Roche limit & 358\\
       Candidates from the taxonomic filtering & 156\\
       Best candidates with $\widehat{S/N}>1$ & 67\\
       
    \end{tabular}
    \label{tab:candidates_numbers}
\end{table*}

In the final list of \allcand candidates, there are also 31 objects whose estimated period is not very reliable. In fact, their values reach is towards the highest ones in the allowed range of the GLSP (reaching three times the length of the observation window).

\begin{figure*}
\centering
     \includegraphics[width=0.42\linewidth]{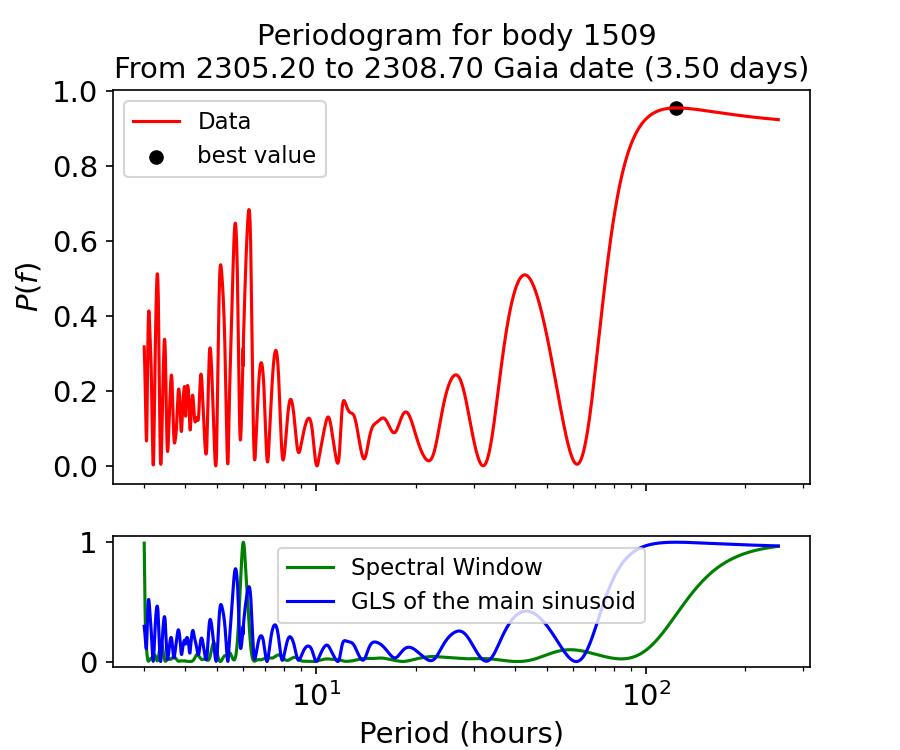}
     \includegraphics[width=0.42\linewidth]{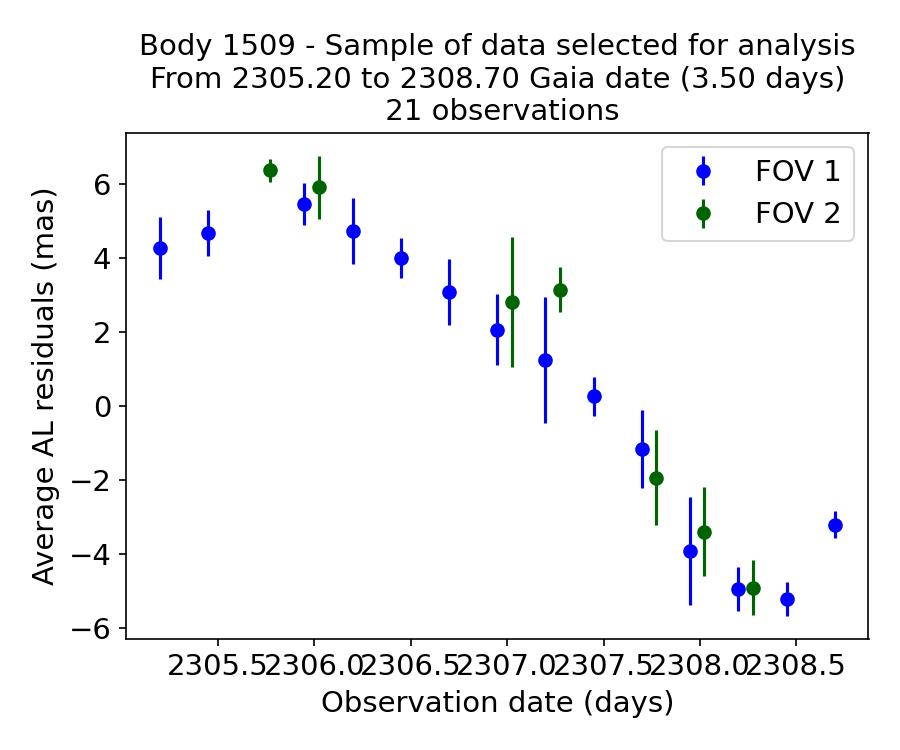}

    \caption{(1509) Esclangona residuals distribution (right) and the periodogram from the fitting of the 21 data points in the window (left).}
 \label{fig:1509}
\end{figure*}

An example is (1509) Esclangona, a known binary with an estimated period of $\sim$ 410$\pm$120 hours \citep[assuming a bulk density of $\rho=2.0\pm1.0~g/cm^3$,][]{marchis2012} well beyond the periodogram range (left panel of Fig.\ref{fig:1509}). A poorly defined best period is found, in a plateau of the periodogram where a whole range of values is equally possible. Going towards the extreme of the highest periods tested, the fitting function to the residuals (covering 84 hours right panel) tends to locally approach a straight line, resulting in a whole range of similar fitting quality.  

It should be noted that the clear trend could indicate a genuine long period. Hence, the data may contain part of a periodic signal that we cannot consistently estimate, as this would require a longer data sequence. In this case, and the other ones resulting in a similar periodogram, we are only able to give a minimum boundary and estimate that the period should be larger than the length of the observation window. Candidate binaries in this category are identified by an orange background in the final list of candidates in Table~\ref{tab:all_candidates}. 

Additionally, we notice that the method seems to favour the selection of objects presenting periods longer than 10 hours, which is perhaps the sign that high frequencies not captured by the constant plus sinusoid model of the GLSP are present in the time series. By filtering the physically meaningful candidates we eliminate most of the cases with $p$-values$>2\%$ and $Q<0.9$ and the best candidates are mostly those with very low $p$-value and $Q$ close to 1.
\begin{figure}
    \centering
    \includegraphics[width=\linewidth]{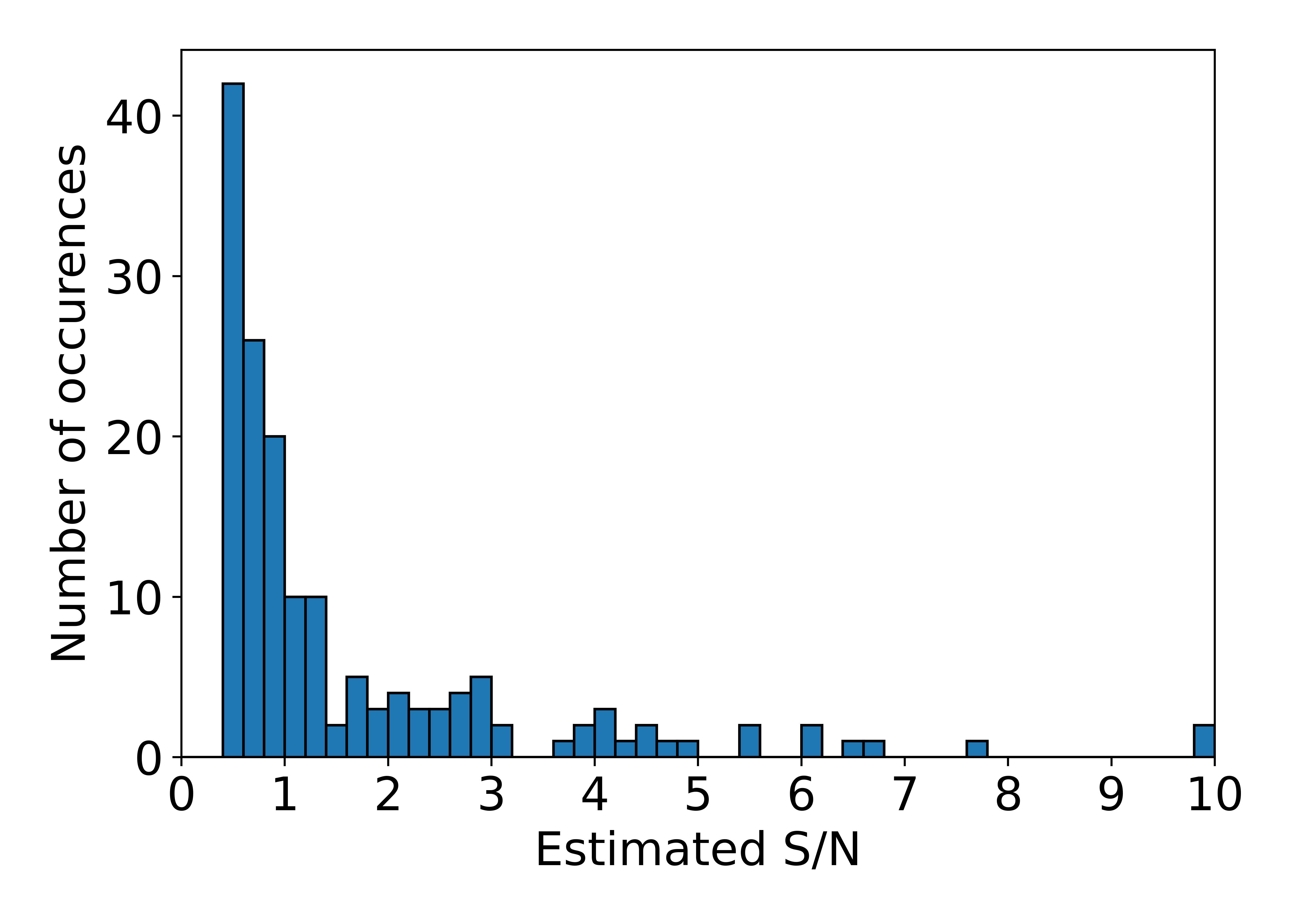}
    \caption{Distribution of $\widehat{S/N}$ for our list of \besttaxocand binary candidates selected from the taxonomic filtering. The \bestcand best candidates are those with $\widehat{S/N}>1$}
    \label{fig:hist_snr_156}
\end{figure}

The comparison between Fig.~\ref{fig:hist_snr_156} and Fig.~\ref{fig:pval_Q_noise}, showing the heavier right tail of the $\widehat{S/N}$ distribution obtained from \gaia data, further strengthen the evidence that our candidates are indeed binary asteroids. Two of the best candidates in the list, (2219)~Mannucci and (542)~Susanna present matching wobble periods from two different time windows, separated in time by weeks or months. At such large intervals, the orientation of the wobbling motion relative to the scan direction can be different (as confirmed by the different wobble amplitude), thus making these measurements independent. The fact that the two estimated wobble periods of each object match is another sign that our method of period search is rather robust.

The bodies selected as best binary candidates, including the known binary asteroids, are in large majority main-belt asteroids (Inner MB: 11, Middle MB: 26, Outer MB: 22), but 3 Phocaea, 3 Mars-Crosser, 2 Hungaria, 1 NEA Amor, and 1 Trojan are also present. All of our best candidates present separations $<200~$mas, below the limit at which the onboard Video Processing Unit of \gaia treats sources separately.


\subsection{Known binaries}
Our results contain 6 known binaries: (1509)~Escanglona, (5817)~Robertfrazer, (2871)~Schober, (4337)~Arecibo, (317) Roxane and (18301) Konyukhov. Such a small number of known binaries is not unexpected, since our filtering and statistical validation reduce a large sample to a very small population. For each binary, the probability of showing an astrometric signature in \gdrthree is not very high, as it requires the appropriate orientation of the wobble motion concerning the scan motion, during the periods of consecutive observations.

\begin{table*}[ht]
    \centering
    \caption{Parameters of the known binaries from the literature.}
    \setlength\tabcolsep{3pt} 
    \begin{tabular}{|c|c|c|c|c|c|c|c|}\hline
        ID & $a$ (km) & $D_1$(km) & $D_2$(km) & Size ratio $k$ & $T$ (h) & Misc. & Ref. \\\hline
        
        \multirow{2}{*}{\parbox{2cm}{(1509)\\ Esclangona}} & 140 & -- & 4 & -- & -- & $D_{eff}$=9$\pm$1 km & \cite{merline2003} \\
        & $<$132 & 8.0 $\pm$ 0.8 & 4.0 $\pm$ 0.7 & 0.5 $\pm$ 0.1 & -- & type Sw &\cite{marchis2012}\\\hline
        \multirow{2}{*}{\parbox{2cm}{(4337) \\Arecibo}} & -- & 24.4 $\pm$ 0.6 & 13.0 $\pm$ 1.5 & 0.533 $\pm$ 0.063 & -- & Occultation &  \cite{gault2022}\\
        & 49.9 $\pm$ 1.0 & -- & -- & 0.186 $\pm$ 0.022 & 32.9728 & $\rho \approx$ 1g/cm$^3$ & \cite{tanga2022gaia} \\\hline
        (2871) Schober & -- & -- & -- & $>$ 0.28 & 42.47 $\pm$ 0.02 & -- & \cite{Benishek2023}\\\hline
        (5817) Robertfrazer & -- & -- & -- & $>$ 0.19 & 28.862 $\pm$ 0.007 & -- & \cite{stephens2020}\\\hline
        (317) Roxane & 245 & 19 & 5 & -- & -- & -- & \cite{Merline2009}\\\hline
        \multirow{2}{*}{\parbox{2cm}{\centering (18301)\\ Konyukhov}} & \multirow{2}{*}{\parbox{1cm}{\centering --}} & \multirow{2}{*}{\parbox{1cm}{\centering --}} & \multirow{2}{*}{\parbox{1cm}{\centering --}} & \multirow{2}{*}{\parbox{1.5cm}{\centering 0.26 $\pm$ 0.02}} & \multirow{2}{*}{\parbox{1.5cm}{\centering 35.7 $\pm$ 0.1\\or twice}}  & \multirow{2}{*}{\parbox{1cm}{\centering --}} & \multirow{2}{*}{\parbox{2cm}{\centering \cite{konyukhov_bin}}}\\
        &&&&&&&\\  \hline

    \end{tabular}
\centering
    \begin{minipage}{\linewidth}
        
   {\small Notes: $a$ is the binary separation; $D_1$ and $D_2$ are the primary and secondary bodies diameters, respectively; $k$ is the size ratio $D_2/D_1$; $T$ is the orbital period of the secondary around the primary body.}
    \end{minipage}
    \label{tab:Known_binaries_param}
    \end{table*}

Table \ref{tab:Known_binaries_param} contains the parameters described in the literature and obtained from observations of the six mentioned known binary systems. These six known binaries offer in principle the possibility of validating our approach, but in practice, other limitations appear. As mentioned previously, (1509) Esclangona cannot be exploited for the comparison, as it has a small satellite with an orbital period well beyond the reach of our data sample. The most probable periodicity given by our periodogram, 123.71$\pm$11.09 hours, is roughly between 2 and 4 times smaller and should be considered as a lower limit (Sect.~\ref{S:final}).

In the case of (5817) Robertfrazer the data from \gdrthree presents two windows of observation that we use, but the result obtained from one of them does not pass the physical validation process and is therefore discarded. So, for the remaining window, we also find a matching size ratio ($k=D_2/D_1$), however, our period is more than 5 times larger than reported ($\sim$28.862 hours), and again close to the upper limit of the frequency range of the GLSP.

Interestingly, however, for both these incorrect matches the apparent wobble signal is strong, with a $\widehat{S/N}$ of 6.09 for Esclangona, and 2.64 for Robertfrazer, and the wobble amplitudes are significantly high ($5.09\pm0.54$~mas for Esclangona and $1.61\pm1.19$~mas for Robertfrazer), especially for Esclangona. A possible alternative interpretation could be the presence of another satellite in the system, and future \gaia astrometry will probably help to clarify which scenario is more probable. In both cases, we estimate that the interest in selecting these objects by our procedure remains.

For (2871) Schober, the size ratio from the literature matches our results. We find solutions from two different time windows of observation, with wobble periods of 20$\pm$4 hours and 50$\pm$11 hours: the first can be considered an alias emerging from the data at about double the frequency due to the irregular sampling from Gaia, while the second results in intervals of separation that overlap partially with the Roche limit and, therefore, it was rejected. Both windows are compatible with the value reported in the literature, obtained by photometry \citep{Benishek2023}. 

In the case of (317) Roxane, we find that our results also match well with the size ratio from the literature. We also estimated the orbital period of the secondary based on the separation, as was made for Esclangona, and we found that our period estimate is about half of the expected value. The observation window from Gaia DR3 that was selected to perform the search is shorter than the expected period of the secondary, which once again limits the extent of our period search and invisibilises the method to find the correct period.

For (18301) Konyukhov we find that the observed size ratio of 0.26 \citep[$\approx$0.017 in mass ratio,][]{konyukhov_bin} is very close to the upper limit of the first interval of mass ratio of 0.013, translated into an upper value for size ratio of approximately 0.24.
Regarding the period, our Monte Carlo procedure to estimate the confidence interval on this value reveals that the distribution of the estimated periods (cf Sec. \ref{Ss:selection_statistic}) is nearly bimodal, with two modes on 70.1 h and 140.6 h. As a consequence, the procedure provides a very large confidence interval ($\pm 70.02$h), which reflects the possibility that the true period can actually be close to either of the two values.

\begin{figure}
    \centering
    \includegraphics[width=\linewidth]{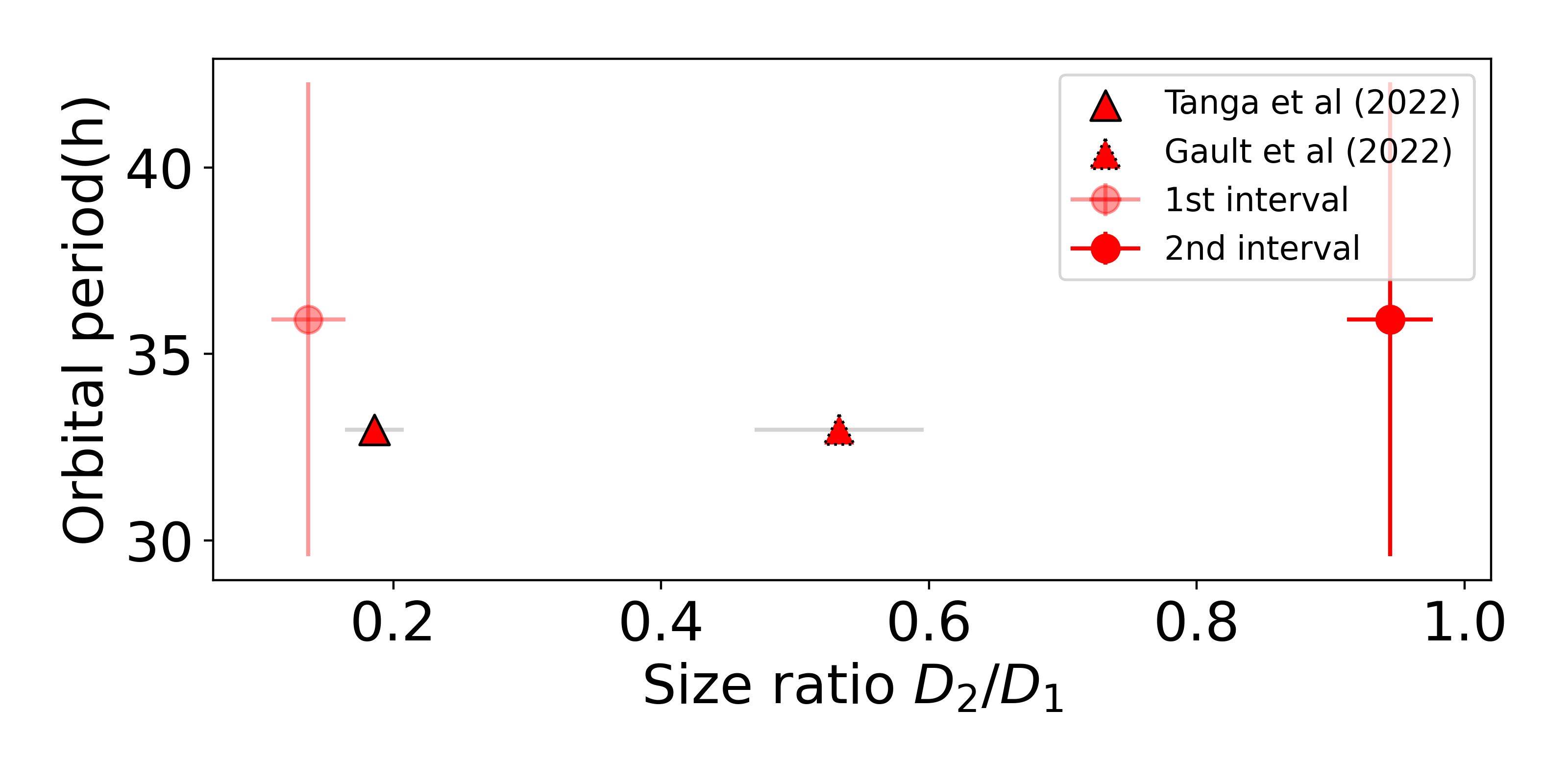}
    \caption{Comparison between the parameters for the Arecibo system. The triangle with black edges and grey error bars represents the values from literature (Table \ref{tab:Known_binaries_param}), while the circles represent the first (transparent) and second intervals of size ratio estimated from our method using the same wobble amplitude as in \cite{tanga2022gaia}.}
    \label{fig:are}
\end{figure}

For (4337) Arecibo we find that our estimated period is equivalent within the uncertainty to other sources \citep{tanga2022gaia, durech2023DR3photometry}. This binary has two different size ratio values from the literature. Stellar occultations \cite{gault2022} provide $k_{occ}=0.533\pm0.063$, while the fit to both \gaia astrometry and occultations \cite{tanga2022gaia} results in $k_{Tanga}=0.35~k_{occ}$, with indications of a probable ellipsoidal shape of the components. Using the best value of surface--equivalent diameter from SsODNet (19.746$\pm$0.324 km) and our estimates for wobble amplitude (0.61$\pm$0.07 mas) and wobble period (35.9284$\pm$6.3600 h), we obtain results close to the parameters of the system previously published (lower size ratio solution in Fig.~\ref{fig:are}). The remaining difference with respect to \cite{tanga2022gaia} arises from the different estimates of the wobble amplitude and period, and from the fact that we do not consider here additional constraints such as those coming from the observation of stellar occultations.

\subsection{Comparison to the known population}

Our candidate binaries can be compared to the existing population. As our observational biases are different from other techniques, the goal is not to reproduce the same distribution. In principle, we can discover asteroids with satellites whose dynamical properties are unusual, in particular in a domain of intermediate sizes (that we can conventionally define between 20 and 100 km) where traditional techniques have very limited capabilities of discovery. On the other hand, we can also reasonably expect that some of our candidates are similar, in physical and/or dynamical properties, to the binaries known so far. 

A first comparison between the astrometric binary candidates and the known binary population is based on taxonomy. Our goal is to verify if an excess of S-type binaries is present in our sample, by an approach similar to \citet{minker2023deficit}.
We start by grouping the taxonomic classes in large complexes, as shown in Table~\ref{tab:reducedtaxo}. Notably, P-type asteroids are in the C complex, not the X complex. 
The reference sample of asteroids was constructed based on a magnitude limit of absolute magnitude $H < 16$ and within ranges of a$\,\in$~[2,~3.5]~au, i$\,\in$~[0$^{\circ}$, 30$^{\circ}$] and e$\,\in$~[0, 0.5] roughly matching the distribution of the sample of binary candidates.

\begin{table}[htbp]
\caption{Taxonomic classes grouped as complexes ($\Sigma$).} 
\centering
\begin{tabular}{l|l}

 Classes & $\Sigma$ \\\hline

 S & S \\
 Q  & Q \\
 V  & V \\
 C, Ch, B, D, P, Z & C \\
 X   & X, E \\
 K, L, M  & M \\
 
\end{tabular}
\label{tab:reducedtaxo}
\end{table}
The reference sample is selected by choosing 30 asteroids around each binary candidate, from a partition of the 4D ($a$,$e$,$i$,$D$) parameter space (see \cite{minker2023deficit} for details). 

\begin{figure}[htpb]
\centering
\includegraphics[width=\linewidth]{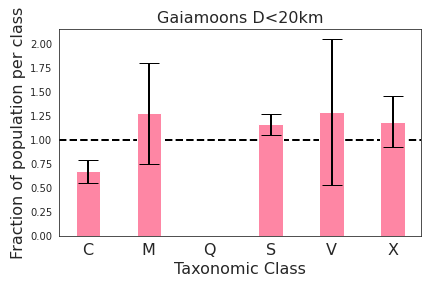}
\caption{\label{fig:small_tax} Comparison of the fraction of small objects with $D<$20 km in each taxonomic complex between the reference sample of background asteroids and the population of our astrometric binary candidates. A taxonomic type with no column indicates that asteroids of that type are present in the reference
population but not in the group of binary candidates.}
\end{figure}
Figure~\ref{fig:small_tax} shows the comparison between the reference sample and our binary candidates with $D<$20 km, which represent most of our candidates. The S-complex taxonomic types present a fraction above 1, which means that there is an over-representation of objects of this complex in our sample of small binary candidates. On the other hand, our small candidates show a deficit of C-complex taxonomic types in comparison with the background asteroid population.

The results from this analysis are in great agreement with \cite{minker2023deficit}, where it is shown that even though chondritic asteroids are the most common type in the Solar System, there is a preferential presence of S-type over C-type, among binary objects at small diameters. Thus, we can conclude that the taxonomic distribution of our candidates is comparable to the one of the previously known population.

\begin{figure}[htpb]
\centering
\includegraphics[width=\linewidth]{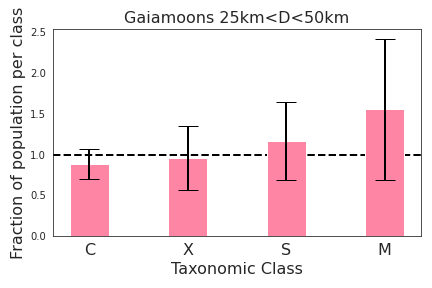}
\caption{\label{fig:medium_tax} Fraction of binary candidates with 30$<D<$50 km in each taxonomic complex, with respect to the reference sample of background asteroids.}
\end{figure}

Taking a closer look at the sub-sample with 30$<$D$<$50~km, we find no dominance of any taxonomic type (Fig.~\ref{fig:medium_tax}), suggesting a disconnect between the small-- and medium--sized binary asteroid populations. However, statistical comparisons to a known population are not possible, as the only known binary asteroid in this size range is the system (243) Ida -- Dactyl. Interestingly, it was discovered by a spacecraft during a fly-by \citep{chapman1995discovery}. This could indicate that un--conventional techniques such as the astrometric method are necessary to find similar binaries.
The limited presence of large (D$>$50~km) asteroids in our sample of binary candidates eliminates the possibility of continuing this taxonomic analysis at larger sizes.

\begin{figure*}
  \begin{minipage}[c]{0.75\textwidth}
    \includegraphics[width=\linewidth]{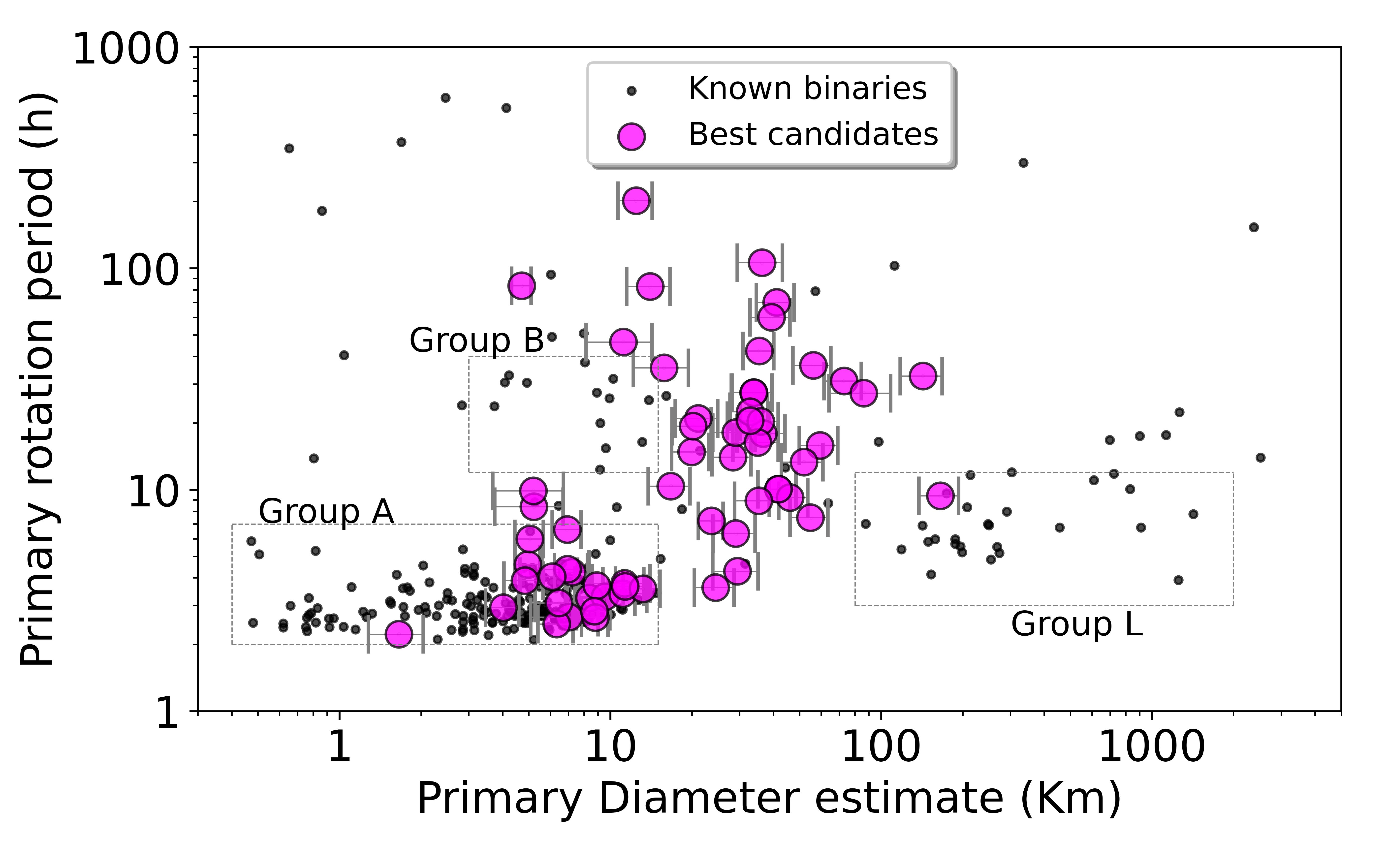}
  \end{minipage}\hfill
  \begin{minipage}[c]{0.25\textwidth}
    \caption{Plot of the primary rotation period vs the estimated primary diameter for the best candidates. The black dots are known binaries in the shown intervals and the magenta circles are the best binary candidates obtained.} \label{fig:best_candidates}

  \end{minipage}
\end{figure*}

In order to compare the dynamical properties of our candidates to the known population, we first present (Fig.~\ref{fig:best_candidates}) the distribution of the primary rotation period as a function of the estimated primary diameter. 
The rectangles in the plot represent a raw approximation of the known binary asteroid groups \citep{pravec2007binary,pravec2010formation,margot2015asteroid}. Group L is for fast-rotating and large objects. Group A represents small primary asteroids with fast rotation; B, systems with small primaries but size ratios $D_2/D_1 \sim 1$ that rotate slower as the primary diameter increases. 

The first striking feature of our sample is the partial overlap with the pre-existing population discovered by photometry (Group A), which clearly appears below $\sim$20~km. Some candidates also partially overlap group B, but their mass ratios are small in general (of the order of 0.1). At the other extreme of the size range, binary candidates from our search do not match group L (besides one). This is expected, as the wobbling signal from large objects should not be dominated by a small satellite, but by the photocentre displacements on the shape of the primary alone. Coherently with this interpretation, some potential candidates in this region are excluded due to the separation being too small relative to the Roche radius. Eventually, intermediate sizes (20-100~km), clearly underrepresented in the known sample, are present for our astrometric candidates, directly confirming the new potential of this technique.

\begin{figure*}
  \begin{minipage}[c]{0.75\textwidth}
    \includegraphics[width=\linewidth]{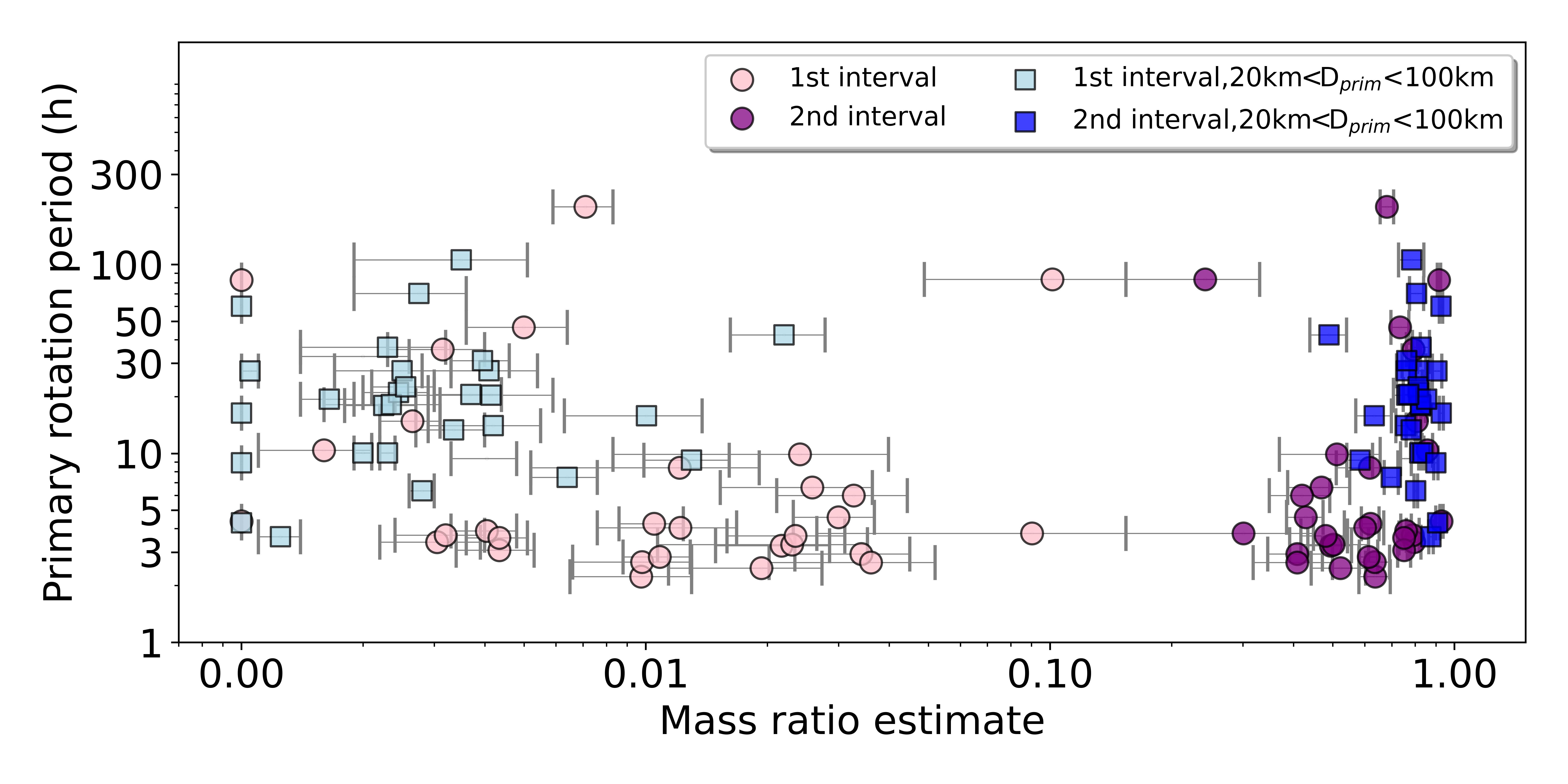}
  \end{minipage}\hfill
  \begin{minipage}[c]{0.25\textwidth}
    \caption{Plot of the primary rotation period vs the estimated mass ratio intervals for the \bestcand selected candidates. The lighter and darker coulored circles represent, respectively, the first and second intervals of possible mass ratio values for the best binary candidates.}
    \label{fig:best_candidates_massratio}

  \end{minipage}
\end{figure*}

The comparison of satellite orbits and sizes is more difficult and intrinsically ambiguous, as our analysis introduces two families of solutions, corresponding to two different possible ranges of mass ratio, as shown in Fig.~\ref{fig:densities_profile}. The estimated mass ratio intervals for the \bestcand selected best binary candidates, and their respective primary rotation periods, are shown in Fig.~\ref{fig:best_candidates_massratio}. The two mass ratio intervals are more or less separated depending on the minimum estimated density for the system. At this stage, we have no criteria to favour a solution over the other. However, we notice that objects of intermediate size populate preferentially extreme values of the mass ratio. This could imply that binaries of similar-sized components could exist in that population.

\begin{figure*}
  \begin{minipage}[c]{0.75\textwidth}

    \includegraphics[width=\textwidth]{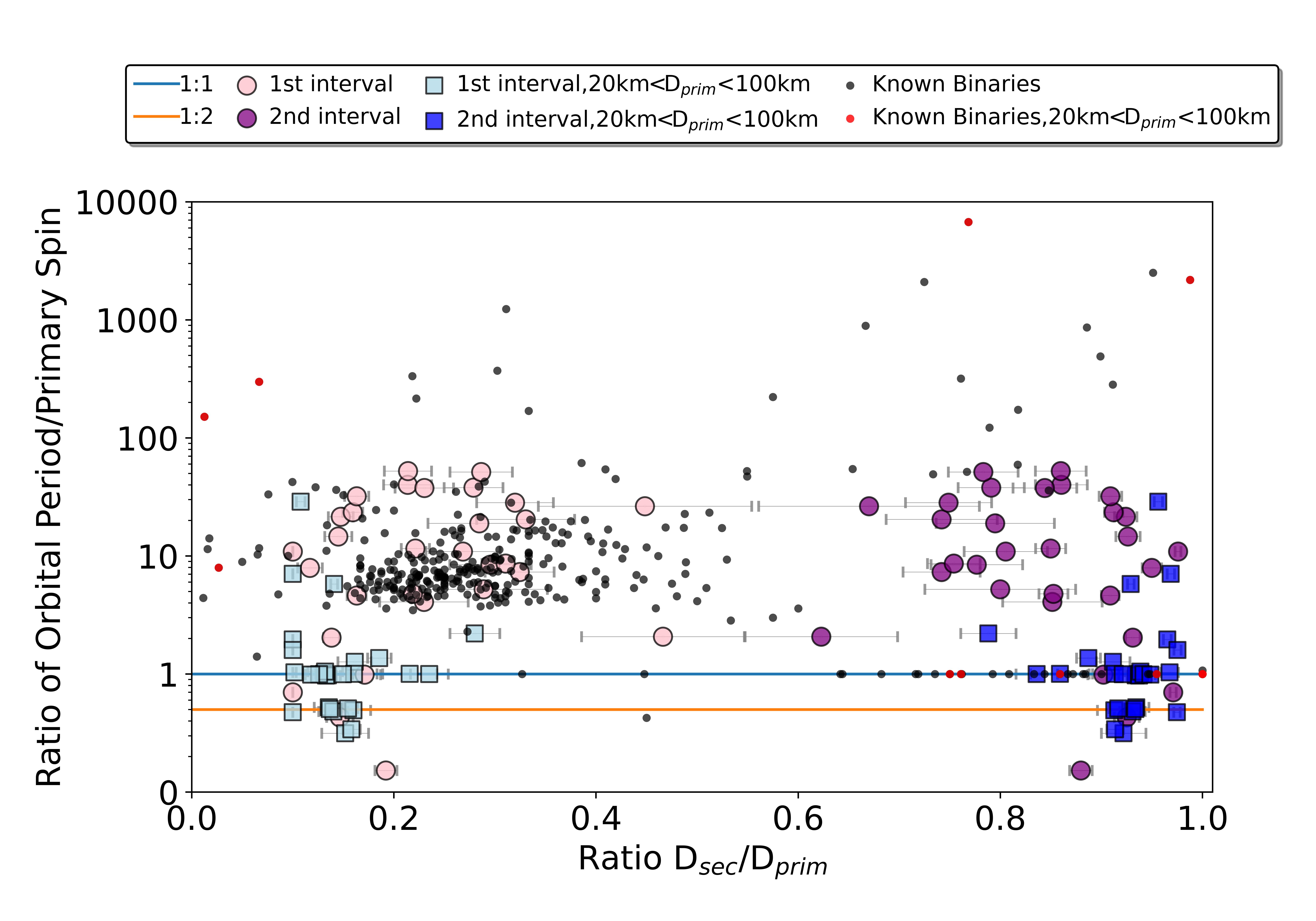}  \end{minipage}\hfill
  \begin{minipage}[c]{0.25\textwidth}
    
    \caption{Plot of the ratio of the estimated orbital period (or wobble period) of the secondary over the photometric rotation period of the primary vs the estimated size ratio of the binary system. The black dots represent all of the known binaries and the red dots highlight those with intermediate sizes (20 km $<~D_{primary}<~$ 100 km). The light and dark purple circles represent the average value of the best-estimated size ratio intervals for objects smaller than 20 km and the light and dark blue circles represent the larger objects. The blue line shows the condition where the estimated orbital period of the secondary is almost equal to the rotation period of the primary object, representing a 1:1 period ratio, while the orange line represents the 1:2 period ratio.}
    \label{fig:best_candidates_ratio}

  \end{minipage}
\end{figure*}

The analysis of the ratio of the wobble (orbital) period to primary rotation period, reveals the presence of synchronous objects and, to a minor extent, of objects in a 1:2 period ratio (Fig.~\ref{fig:best_candidates_ratio}). Among the known main belt binaries, almost all those with $D_{s}/D_{p}>0.5$ are synchronous, as the tidal coupling is more efficient for large satellites. Applied to our sample, this evidence would imply that large mass ratios have a higher probability of showing orbit--spin coupling. Arecibo, at $D_{s}/D_{p}>0.53$ and synchronous, clearly falls in this category. Periods on the 1:2 ratio could simply represent an alias and could be considered synchronous too. Interestingly, intermediate-size objects are in large majority in this category.


\section{Discussion}
\label{S:Discussion}

As explained above, this first attempt of binary search in \gdrthree astrometry is performed by adopting some conservative criteria and an approach that does not take into account all the observations available for each asteroid. Other choices are possible, such as those described in \cite{segev2023astrometric}, where a solution of the complete inverse problem leading to satellite orbits, and involving all the AL residuals of each object, is proposed. Their search is not successful and the main explanation given is the rejection mechanism of astrometric outliers implemented by the validation pipeline in DR2 \citep{gaiacollaborationDR2_2018}. 

However, our results show that periodic fluctuations in the residuals are compatible with the expected amplitude and orbital periods of satellites. We thus think that the missed detection in the cited article is due to three main reasons: the strong increase in the time range and number of asteroids of \gdrthree concerning DR2 (by factors 1.5 and 11, respectively); the improved astrometric quality of DR3; and, of lesser importance, the possible presence of systematic errors in isolated astrometric measurements that are included in the analysis. 

Concerning the first argument, by simple statistical scaling we can estimate that, if DR2 astrometry were good enough to reveal our best candidates, only a few of them ($<$10) would have been found. However, the probability of detection is not a linear function of the time range, as a reduction in the observed arc can have a strong adverse impact on the quality of the orbital fit and, in turn, on its residuals.

The fact that DR3 improves astrometric quality in a significant way over DR2, is well visible not only in the overall statistics but also in specific cases such as (4337)~Arecibo, whose residuals show a clear trend in DR3 \citep{tanga2022gaia} but less clearly in DR2 \citep[Fig. 19 in ][]{segev2023astrometric}, over the same time span. Hence, this is the first work that provides a large list of binary candidates obtained from astrometry and crosses dedicated statistical and physical selection criteria. At this stage, without independent confirmation of the binary status and a clear discrimination between the two solutions that we obtain for each candidate, the impact of our possible discoveries on formation scenarios cannot be very specific. However, some general considerations can be attempted.

First, the intermediate size range (20-100 km) where very few binaries are currently known (only $\sim$5\%n of the total population) is probably the most interesting, and the one having a stronger potential impact. If they are synchronous and have large mass ratios, a well-known prototype at large sizes can be considered the asteroid (90) Antiope \citep[size ratio $\sim$0.95, diameter $\sim$88 km,][]{merline2000discovery}. Even though there are several theories for the formation of the Antiope system \citep{pravec2007binary,weidenschilling2001origin}, observations support the hypothesis that the Antiope system objects were formed simultaneously from the disruption of a parent body\citep{descamps2009giant,marchis2011origin}. 

With a similar mechanism, as shown in numerical simulations \citep{doressoundiram1997formation,durda2004formation}, Escaping Ejecta Binaries (EEBs) result from the creation of collisional fragments at low relative velocity, that remain gravitationally bound. They preferentially have highly eccentric orbits, that could be circularised by tidal forces in their secular evolution. They should not have particularly fast primary rotation as they are not created by fission or fragment ejection. However, the current sample of known binaries, if simulations are correct, has a clear under-representation of this category. The only good candidate is (317) Roxane with its satellite Olympias \citep{drummond2021orbit}. The population of EEBs could also be compatible with binary Mars impactors revealed by \citet{vavilov2022evidence}. Once validated, our sample could clearly contribute to revealing many details about the presence, origin and evolution of EEBs. 

We also notice the compatibility of our highest mass ratio solutions with the findings by 
\citet{scheeres2007rotational} and \citet{pravec2010formation}, that determine (theoretically and by observations) the conditions for splitting of pairs created by fission. Their main finding is a general tendency to create separate pairs when the mass ratio is $<$0.2, which is about the inferior threshold for our distribution of solutions for high values of mass ratio when small asteroids $<$20~km are considered (magenta circles in Fig.~\ref{fig:best_candidates_massratio}). Some of the systems close to (and above) that critical splitting threshold, could have been produced by fission. Considering that mass shedding and re-accumulation in orbit is the most likely formation process for small asteroids ($D\lesssim$20 km) \citep{pravec2007binary, walsh2008rotational}, we expect that future observations characterising our sample should find results compatible with the known population in this size range. 


\section{Conclusions}
\label{S:Conclusions}

Our work presented a simple but robust method to detect periodic signals in the residuals of the orbital fit in \gaia astrometry, which is compatible with the presence of binary asteroids. We explored all astrometric data of asteroids available in the DR3 catalogue and, after a careful selection process combining statistical and physical analyses, we obtained a consolidated list of candidates that are likely to be binary asteroids. 

Among the strengths of our results, we can highlight that the presence fraction of our candidates has physical properties similar to the known binary population (small asteroids with companions discovered by photometry), but also spans a parameter range that tends to fill the gap where other techniques have been failing to find binaries, as expected from preliminary studies \citep{mignard2007gaia,pravec2012small}. The distribution of the taxonomy of our candidate binaries also provides interesting matches to the known population and some potentially new evidence. We also find a large fraction of possible synchronous binaries, which are notoriously difficult to discover by photometry, especially if the secondary rotation is also synchronised (doubly synchronous binaries).

Some limitations of our approach are also clear. As we deal with an astrometric accuracy of the order of 1~mas, we cannot exclude that some signals are not due to companions, but to the photocentre displacement occurring during the rotation of irregular, single asteroids. This could be the case, especially for the largest objects approaching the 100~km diameter, although the relatively small phase angles should mitigate the effect. However, our preliminary analysis of the detection of photocentre variation shows that even though for some cases it could be the source of the signal detected, the results are not significant enough to discard the possibility of the astrometric detection of a satellite.

We also note that our comparison with known satellites is limited to a very small sample, but seems to point out that in some cases we detect companions, although not with the parameters derived by previous surveys. The parameters estimated by making use of taxonomy must be considered with caution since they could be misclassified, which would significantly change the separations. Additionally, we face problems due to the limitations of the data. For instance, the length of the observation window can be shorter than the wobble period, the irregular sampling that causes the rise in aliases in the period determination, and the usually small number of data points along with low S/N that complicates the detections. 

Future improvements may involve many aspects: the exploitation of the astrometry could adopt a more refined error model at the transit level; the inclusion of observations more distant in time, with an increase in the accessible period range; a more complex model for the astrometric signal, involving flattening of the components and non-uniform illumination; a more detailed analysis of the spectral properties and the choices of sizes and plausible densities. In addition, a full simulation of the whole process starting from a synthetic binary population and simulated \gaia astrometry is a complex task that should provide a better view of the biases present in our method, and its limitations. However, such extensive investigation is beyond the scope of this work.

Other \gaia data releases, such as the Focused Product Release (which appeared in October 2023) and, in particular, the DR4 (including all the asteroid astrometry from the nominal mission) can certainly offer opportunities to consolidate our approach and refine the results. We expect the astrometric quality to increase further, and the number of consecutive data sequences (also for the same asteroid) to be more frequent. 

Despite these limitations, we think that we have revealed the capabilities of \gaia in discovering asteroid satellites by the astrometric method, and we have set the foundations for future, more refined searches. Stellar occultation campaigns focused on our candidates may pin down the orbits, sizes, and shapes of the companions. In the absence of our \gaia binary candidates, a blind search of asteroid companions by stellar occultation observations would be unlikely to succeed. In addition, without the stellar occultation technique, it is hard to better assess some of the candidate system properties. 

For some of the candidates with similar characteristics to the known binary population discovered by photometry, light curves could be sufficient to determine the presence of a companion. Therefore, we warmly invite the community to collaborate on the campaigns that we are organising, by photometry and stellar occultations, to validate our findings and physically characterise the \gaia astrometric binaries.


\section{Acknowledgements}

This work presents results from the European Space Agency (ESA) space mission \gaia. \gaia data are being processed by the \gaia Data Processing and Analysis Consortium (DPAC). Funding for the DPAC is provided by national institutions, in particular, the institutions participating in the \gaia Multilateral Agreement (MLA). The \gaia mission website is https://www.cosmos.esa.int/\gaia. The \gaia archive website is https://archives.esac.esa.int/\gaia.

This work was supported by the project \gaia~Moons of the Agence Nationale de Recherche (France), grant ANR-22-CE49-0002.

It was financed in part by the Coordenação de Aperfeiçoamento de Pessoal de Nível Superior – Brasil (CAPES) – Finance Code 001, also by CAPES-PRINT Process 88887.570251/2020-00, by the French Programme National de Planetologie, and by the BQR program of Observatoire de la C\^ote d'Azur. 

P.B. acknowledges funding through the Spanish Government retraining plan ’María Zambrano 2021-2023’ at the University of Alicante (ZAMBRANO22-04).

We made use of the software products: SsODNet VO service of IMCCE, Observatoire de Paris\citep{berthier2022ssodnet}; Astropy, a community-developed core Python package for Astronomy \citep{0astropy2013, 1astropy2018, 2astropy2022}; Matplotlib \citep{matplotlib_Hunter:2007}; Multiprocess package \citep{mckerns2010multiprocess,mckerns2012multiprocess}.

\bibliographystyle{aa}
\bibliography{references}

\begin{appendix}
\section{Amplitude offset}
\label{app:offset}

To evaluate the statistical significance of the highest GLSP peak, we explained in Sec.~\ref{Ss:selection_statistic} that we use Monte Carlo simulations. Their purpose is to generate a large number of synthetic times series that are consistent with our model when there is no sinusoid, that is, an unknown constant sampled at the same time sampling grid as the data set under investigation plus noise consistent with the data error bars.

\begin{figure}[htpb]
    \centering
    \includegraphics[width=\linewidth]{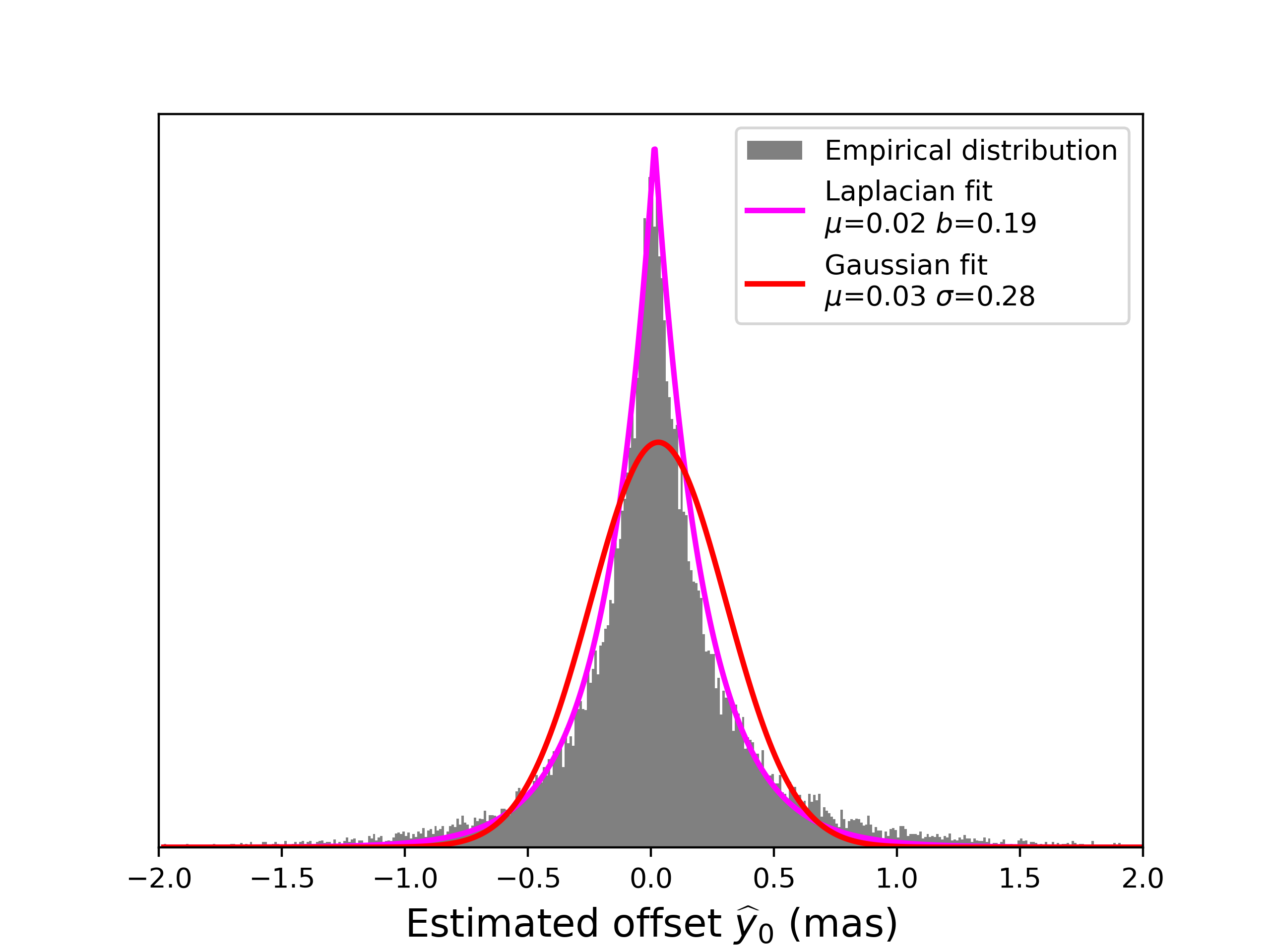}
    \caption{Histogram of distribution of the estimated offset $\widehat{y}_0$ for each window of search explored. The grey bars represent the empirical distribution, while the coloured lines represent different distributions fitted to the data and their respective best-fit parameters.}
    \label{fig:offset}
\end{figure}

Therefore, to generate unknown constants (offsets) that are consistent with our \gaia residual time series, we estimated this offset from a large number of time series and plotted its distribution. The result is in Fig.~\ref{fig:offset}, limited between $-2$ and $2~$mas for visualisation purposes.

The figure compares the empirical distribution with a Laplacian (magenta, parameters $\mu$ and $b$) and Gaussian (red, parameters $\mu$ and $\sigma$) best fit. Clearly, the Laplacian fit is better and we use this distribution to generate offsets in the Monte Carlo simulations.

\section{Detection of eccentric system}
\label{app:ecc}
Highly eccentric binary asteroid systems are rare and not expected to be frequently found in our sample. However, it is interesting to evaluate how efficiently we detect eccentric binaries' astrometric signal. For such analysis, we start by creating samples of data from synthetic eccentric binary asteroid systems in one dimension, which mimics the projection of the residuals from the orbital fit of the \gdrthree astrometric data in the AL direction.

\begin{figure}[ht]
    \centering
    \includegraphics[width=0.85\linewidth]{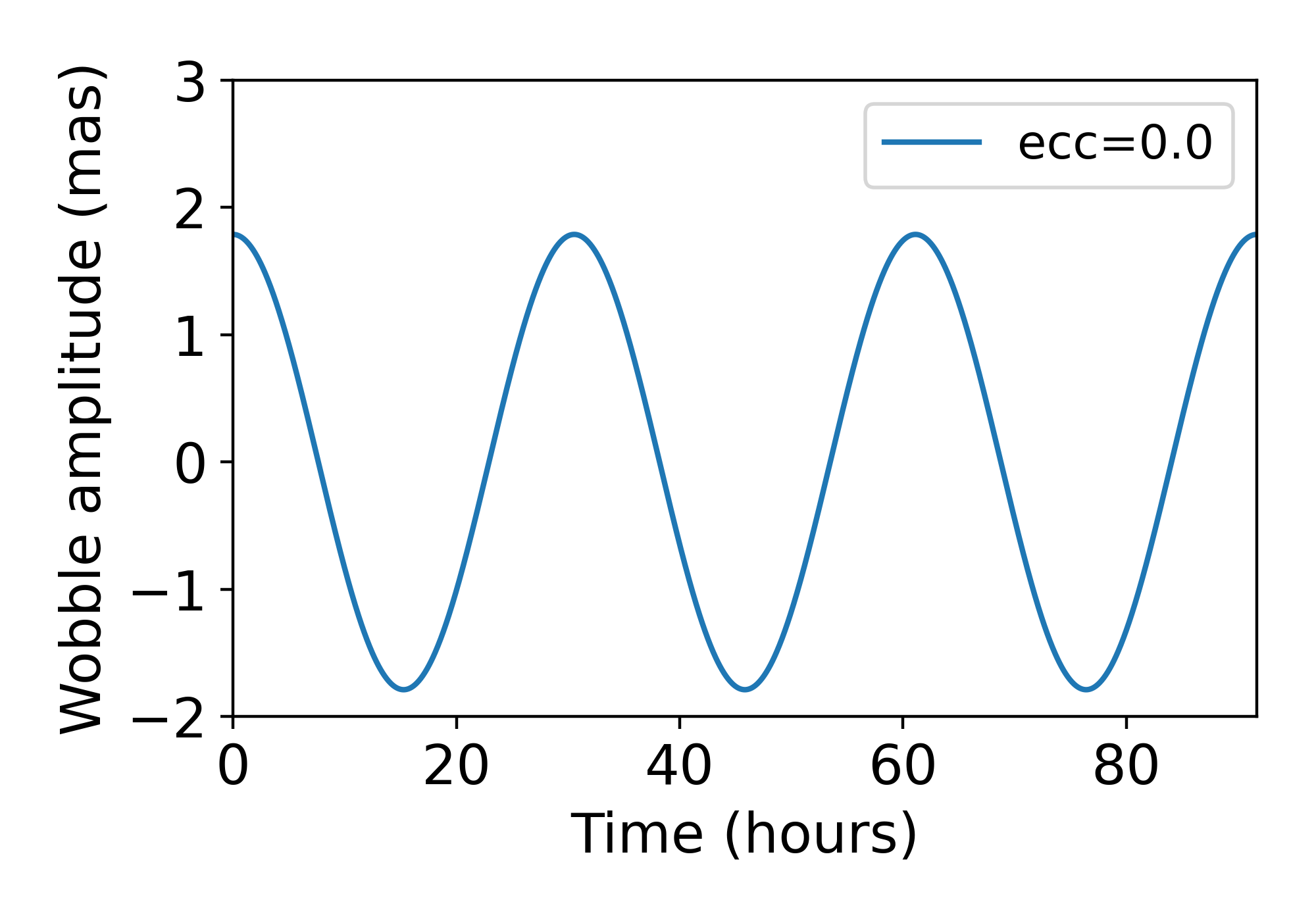}\\
    \includegraphics[width=0.85\linewidth]{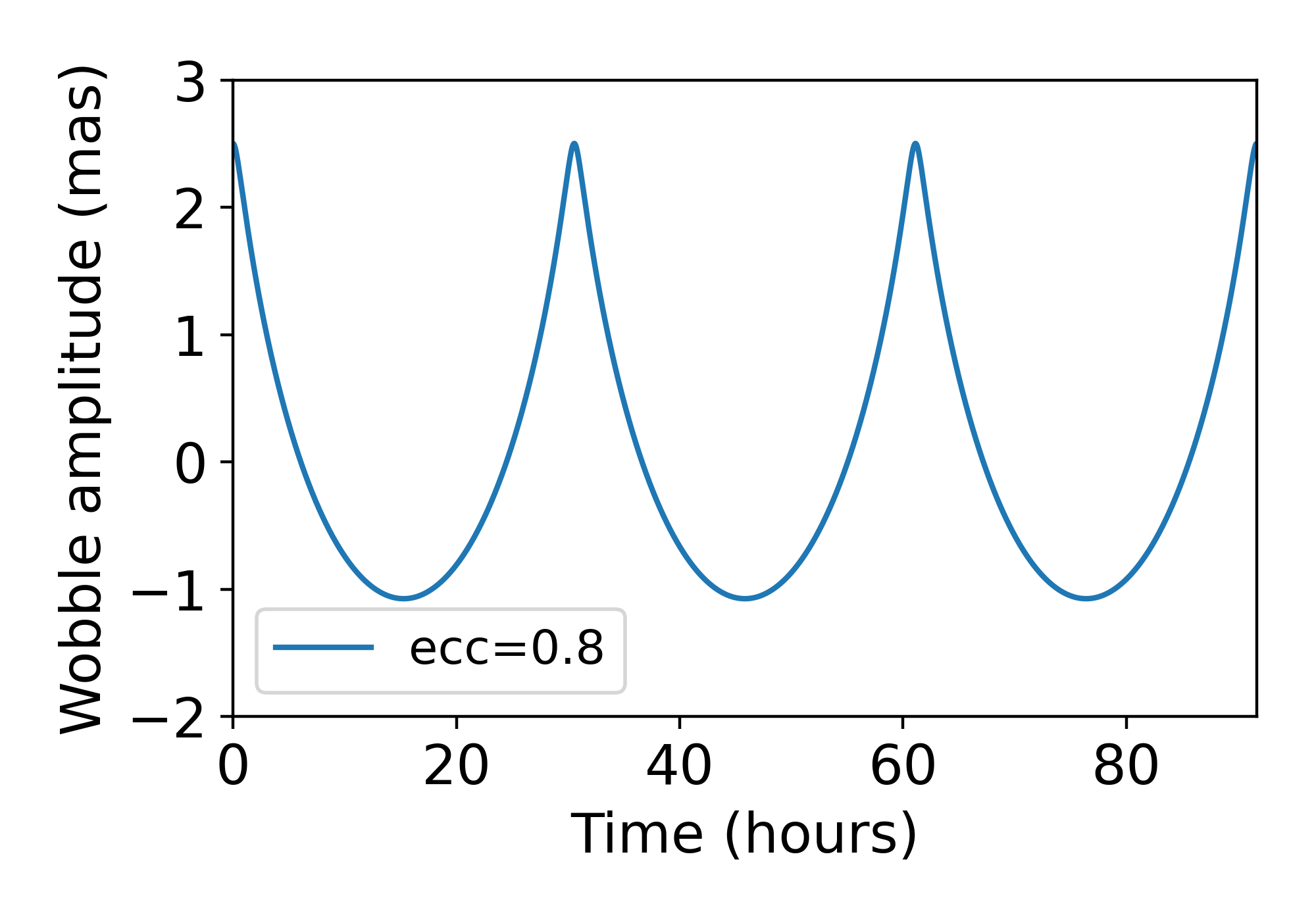}
    \caption{Nominal signal from the generic binary system with extreme eccentricity values. }
    \label{fig:ecc_signals}
\end{figure}

The generic binary system's parameters were chosen arbitrarily but consistent with what we expect to detect. It consists of a primary with a diameter of $D_1$=40 km and a secondary with $D_2$=8km resulting in a mass ratio of $q$=0.008. The bulk density of both objects was chosen to be $\rho$=2.5 g/cm$^3$, a separation of $a$=120 km equivalent to 3 radii of the primary and eccentricities of $e$= 0, 0.2, 0.5, and 0.8. Assuming that it was observed at a distance of 2.8 au, the resulting signal has a nominal wobble amplitude of $\alpha$=1.8 $mas$ and a period of $T$=30.5h, as shown in Fig.~\ref{fig:ecc_signals}.

We randomly select 20 data points from the nominal signal over 3.75 days and create 1000 different samples of data with a process similar to the one described in Sec. \ref{Ss: validation}, as it is also done for the white noise samples.

\begin{figure}[ht]
    \centering
    \includegraphics[width=1\linewidth]{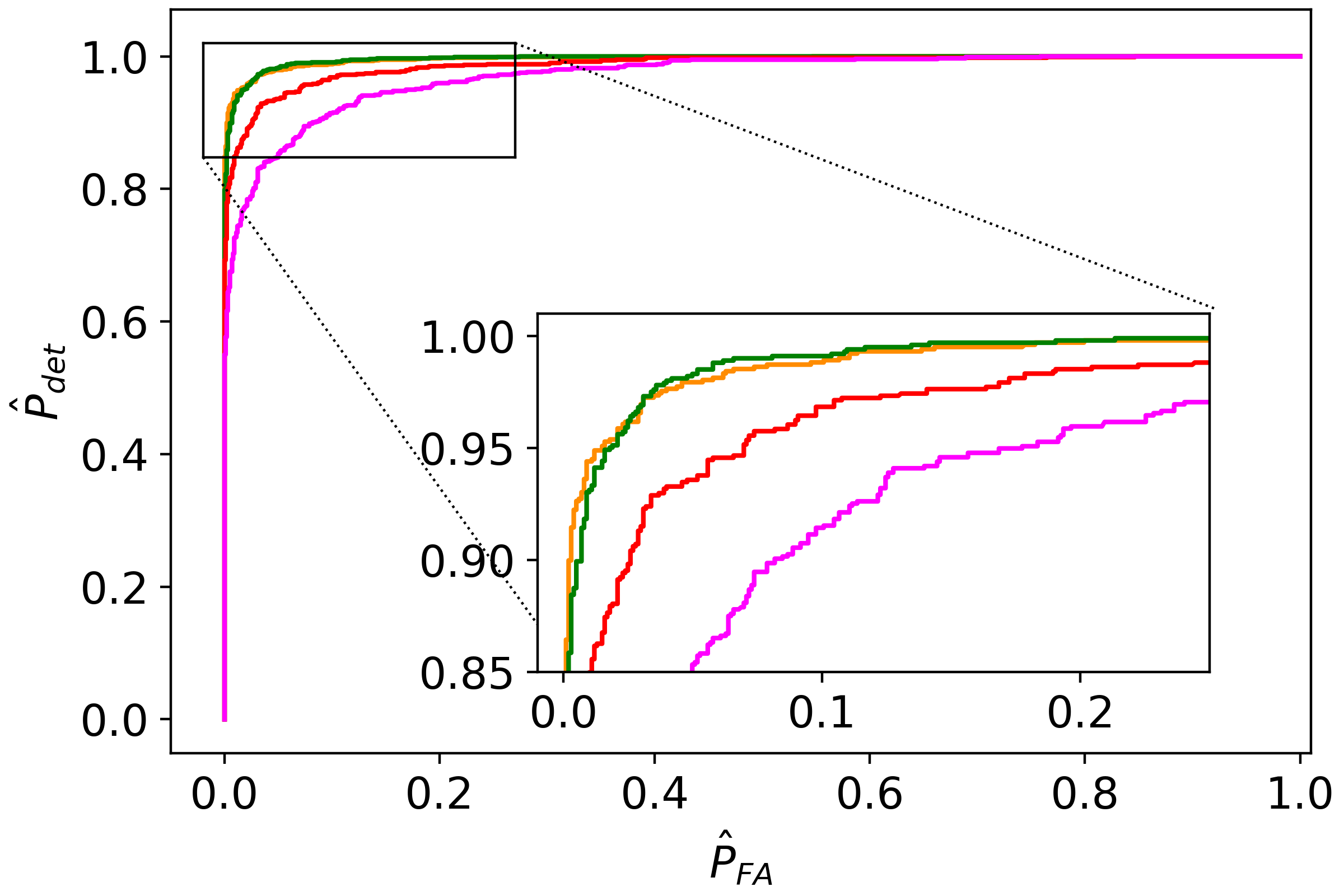}
    \caption{ROC curve for the different eccentricity values considered.}
    \label{fig:probabilities}
\end{figure}

For each one of the 5 situations analysed (4 different values of eccentricity + noise only), we then calculate the {empirical} probability of detection ($\hat{P}_{det}$) and {of false alarm } ($\hat{P}_{FA}$). When plotting $\hat{P}_{det}$ as a function of $\hat{P}_{FA}$
we obtain Fig. \ref{fig:probabilities}, which provides a visual representation of how well our test distinguishes between the two classes (noise only, or noise plus wobble).  The plot shows that the detection method is efficient since all curves are concentrated on the top-left corner.

\begin{figure}[htpb]
    \centering
    \includegraphics[width=\linewidth]{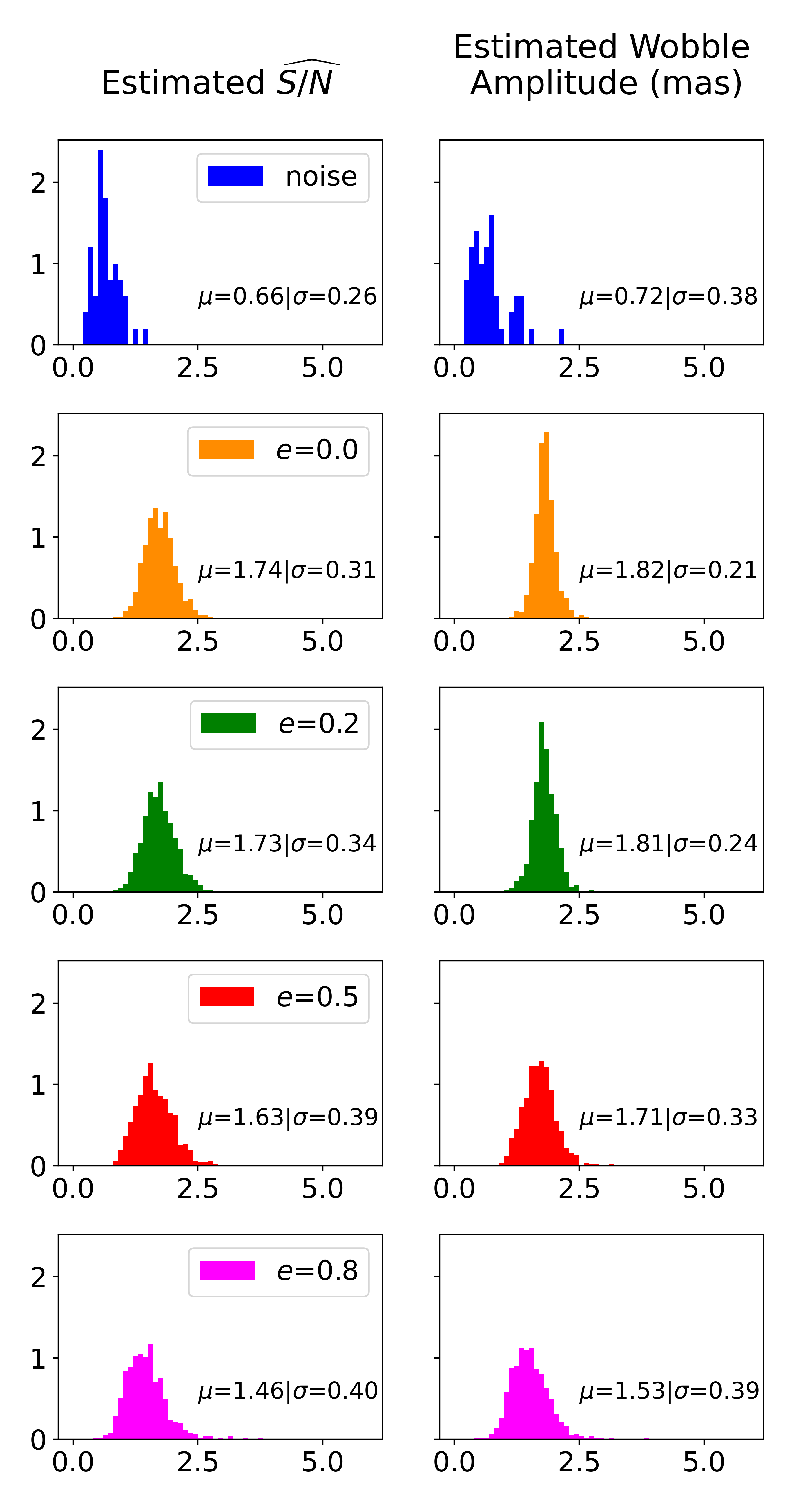}
    \caption{Density histograms for estimated $\widehat{S/N}$ (top) and estimated wobble amplitude (bottom) for the cases with $\hat{P}_{FA} = 5\%$.}
    \label{fig:snr_amp}
\end{figure}

For instance, if we choose to accept a false alarm probability of 5\% we find that about 98\% of the signals from the circular case are detected against approximately 85\% from the highly eccentric systems. Turning to the estimated $\widehat{S/N}$ and estimated wobble amplitude for the case where $\hat{P}_{FA} = 5\%$, we obtained the results of   Fig. \ref{fig:snr_amp}. Here, the considered wobble amplitudes correspond to a nominal S/N of approximately 1.7. We see that the estimated S/N are consistent with this value. We notice that as we increase the eccentricity, the estimated amplitude decreases and the uncertainty grows. Since the periodogram uses a sinusoidal model to fit the signal from the eccentric systems, we lose some precision on the amplitude estimate when the signal comes from an eccentric system, which also impacts the $\widehat{S/N}$ estimate as shown. Still, the statistical detection process succeeds at selecting at least 90\% of the samples of data from the most extreme eccentricity case. Note finally that when there is no wobble in the data, the estimated S/N tends to be much smaller (below) $1$ than in the wobble-present case.

Furthermore, when analysing the frequencies at which the maximum of the periodogram is found, we find that for the circular case about 97\% of the detections are within the {correct} signal frequency $\pm$ 10\% against and 79\% for the case with $e$=0.8.  Even in extreme eccentricity cases, when a signal is correctly classified as a detection, the period is correctly estimated 63\% of the time. We can conclude that the detection of eccentric binaries is indeed more difficult, but this study shows that the method is still efficient under these conditions. We finally note that highly eccentric binaries are not commonly found and they are also not expected to survive for long periods of time under these extreme configurations \citep{MARCHIS2008,WALSH2008,walsh2015formation}, so they must be exceptions and thus rarely found in our data.
\end{appendix}

\end{document}